\documentclass[11pt,a4paper]{article}
\usepackage{jheppub}
\usepackage[utf8]{inputenc}
\usepackage{amsmath}
\usepackage{mathtools}
\usepackage{tikz}
\usepackage{empheq}
\usepackage{enumitem}
\usepackage{graphics}
\usepackage{wrapfig}
\usepackage{caption}
\usepackage{graphicx}
\usepackage{subcaption}

\definecolor{light-gray}{gray}{0.85}

\usepackage{xcolor}
\usepackage{framed}
\definecolor{shadecolor}{gray}{0.925}

\newcommand{\bi}{\begin{itemize}}
\newcommand{\ei}{\end{itemize}}
\newcommand{\bea}{\begin{align}}
\newcommand{\eea}{\end{align}}
\newcommand{\be}{\begin{equation}}
\newcommand{\ee}{\end{equation}}

\newcommand{\tcr}{\textcolor{red}}

\newcommand{\tcb}{\textcolor{blue}}

\makeatletter
\renewcommand*\env@matrix[1][\arraystretch]{%
  \edef\arraystretch{#1}%
  \hskip -\arraycolsep
  \let\@ifnextchar\new@ifnextchar
  \array{*\c@MaxMatrixCols c}}
\makeatother

\author{Charlotte SLEIGHT\footnote{Also at the Universit\'e Libre de Bruxelles and International Solvay Institutes, Belgium.}}

\affiliation{School of Natural Sciences, Institute for Advanced Study,\\
1 Einstein Drive, Princeton, NJ 08540}

\emailAdd{csleight@ias.edu}

\title{\centering
\LARGE{A Mellin Space Approach to Cosmological Correlators}}

\abstract{We propose a Mellin space approach to the evaluation of late-time momentum-space correlation functions of quantum fields in $\left(d+1\right)$-dimensional de Sitter space. The Mellin-Barnes representation makes manifest the analytic structure of late-time correlators and, more generally, provides a convenient general $d$ framework for the study of conformal correlators in momentum space. In this work we focus on tree-level correlation functions of general scalars as a prototype, including $n$-point contact diagrams and $4$-point exchanges. For generic scalars, both the contact and exchange diagrams are given by (generalised) Hypergeometric functions, which reduce to existing expressions available in the literature for $d=3$ and external scalars which are either simultaneously conformally coupled or massless.  This approach can also be used for the perturbative bulk evaluation of momentum space boundary correlators in $\left(d+1\right)$-dimensional anti-de Sitter space (Witten diagrams).}

\begin{document}
\maketitle

\newpage

\section{Introduction}

The inflationary paradigm \cite{Guth:1980zm,Linde:1981mu,Albrecht:1982wi,Starobinsky:1982ee} has emerged as a leading candidate for the origin of primordial fluctuations, which are postulated to arise from quantum fluctuations during a phase of (approximately) de Sitter expansion in the early universe. Cosmological correlations can be re-wound back to primordial correlation functions at the end of inflation, which provide a remarkably successful description of the current data from cosmological observations. But in spite of this, the physics of inflation has remained elusive. This is due to both theoretical and experimental limitations. In recent years, a significant theoretical effort has been dedicated to refining our understanding of inflationary correlation functions. Non-Gaussianities in primordial correlators encode information about interactions and field content during inflation \cite{Chen:2010xka}, so it is imperative to develop systematic approaches to classify the possible shapes of non-Gaussianities for comparison with upcoming observational data. See e.g. \cite{Maldacena:2011nz,Mata:2012bx,Anninos:2014lwa,Ghosh:2014kba,Kehagias:2015jha,Arkani-Hamed:2015bza,Lee:2016vti,Arkani-Hamed:2017fdk,Benincasa:2018ssx,Li:2018wkt,Farrow:2018yni,Arkani-Hamed:2018kmz,Goon:2018fyu} for some recent efforts.

We propose (see also \cite{ToAppear}) an approach to the perturbative evaluation of late-time correlators in de Sitter space based on the Mellin-Barnes representation in Fourier space. This has various advantages. Computationally, the bulk integrals are trivialised since the dependence of the propagators on conformal time is a simple power-law at the level of the Mellin-Barnes representation. This feature straightforwardly gives rise to analytic expressions for boundary correlators with any number of legs. The Mellin-Barnes representation of the boundary correlators makes manifest their analytic structure, not only in the momenta but also in the boundary dimension $d$ and the scaling dimensions of the fields. Asymptotic expansions of the correlators in the momenta/mass can moreover be systematically derived using well-established methods in the Mellin-Barnes literature.

These features of the Mellin-Barnes representation could make it a convenient framework to explore the basic principles that must be satisfied by late-time de Sitter correlators, which may eventually be used to constrain (or ``Bootstrap") such observables without any reference to bulk time-evolution (see e.g. \cite{Arkani-Hamed:2015bza,Arkani-Hamed:2018kmz} for some initial works on the Bootstrap of Cosmological Observables). Indeed, as we shall see, the location of the poles in the Mellin-Barnes integrand are fixed by conformal symmetry,\footnote{The constraints of Conformal Ward identities on Conformal Structures in Fourier space have been studied in \cite{Maldacena:2011nz,Coriano:2013jba,Bzowski:2013sza,Bzowski:2015pba,Bzowski:2017poo,Coriano:2018zdo,Coriano:2018bbe,Isono:2018rrb,Bzowski:2018fql,Coriano:2018bsy,Maglio:2019grh,Isono:2019ihz}.} while the zeros may be fixed by imposing the correct boundary conditions at possible singularities. This is the focus of the companion work \cite{ToAppear}.

In this work we shall focus on using the Mellin framework for the bulk computation of tree-level correlation functions of general scalars on $\left(d+1\right)$-dimensional de Sitter space, including $n$-point contact diagrams and four-point exchange diagrams. For generic scalars, the Mellin framework naturally identifies both types of diagrams with (generalised) Hypergeometric functions, which simplify for special values of the scaling dimensions. The exchange diagrams in particular have an appealing Mellin-Barnes representation, where they are given as a product of the corresponding three-point structures that are sewn together by a simple factor with poles that encode the Effective Field Theory Expansion and a second factor containing only zeros which encode the boundary condition. The exchange four-point function moreover factorises on the poles associated to the exchanged single particle state.

When $d=3$ our results for generic external scalars are new, and reduce to existing expressions available in the literature for when the external scalars are either simultaneously conformally coupled or massless. To the best of our knowledge, the results for general $d$ were not previously available, even for the simplest case of external conformally coupled scalars. The ease at which these general results are obtained in the Mellin formalism is testament to its efficacy. From the results for generic external scalars, it is straightforward to extract inflationary corrections at leading order in slow roll by taking one of the external legs to have soft momentum and a small mass.

Before concluding the introduction let us note that the Mellin formalism under consideration can also be used in the perturbative bulk computation of boundary correlators in anti-de Sitter space (i.e. Witten diagrams). As we shall demonstrate, the Mellin-Barnes representations of de Sitter and anti-de Sitter propagators in momentum space differ only by a simple phase. In position space, the Mellin representation for Witten diagrams \cite{Liu:1998th,Mack:2009gy,Mack:2009mi} has already stood out as an indispensable tool which moreover makes manifest the analogy between AdS correlators and flat space scattering amplitudes \cite{Penedones:2010ue,Paulos:2011ie,Fitzpatrick:2011ia,Fitzpatrick:2011hu}. It would be interesting to investigate whether the techniques presented in this work could facilitate similar progress in momentum space.\footnote{Momentum space techniques for Witten diagrams have been relatively little explored to date except for a handful of works, see e.g. \cite{Chalmers:1998wu,Raju:2010by,Raju:2012zr,Ghosh:2014kba,Anninos:2014lwa,Albayrak:2018tam,Albayrak:2019asr}.} 

\paragraph{Outline.} This paper is organised as follows. In section \ref{sec::propagators} we discuss propagators of scalar fields in dS$_{d+1}$. After reviewing the pertinent aspects of the Wightman two-point functions and the Schwinger-Keldysh formalism, in section \ref{subsec::splitrep} we show that the Wightman two-point function can be obtained from the corresponding Harmonic function in Euclidean anti-de Sitter space (EAdS$_{d+1}$) by analytic continuation. This allows us to establish a ``split representation" for propagators in dS$_{d+1}$. In section \ref{subsec::mellinfourierspace} we introduce the Mellin-Barnes representation for propagators in Fourier space, where the analytic continuation from EAdS$_{d+1}$ is encoded in a simple phase. In section \ref{sec::contactdiagrams} we consider the computation of late-time contact diagrams in dS$_{d+1}$. We start off in section \ref{subsec::3ptscalargen} with three-point functions of generic scalars, before presenting the extension to $n$-point functions in section \ref{subsec::nptcontact}. In section \ref{subsec::3pttwocc} we discuss simplifications which occur when one or more of the scalars is conformally coupled, together with some subtleties which arise when the Mellin integration contour becomes pinched. This is also discussed in section \ref{subsecc::mlesssc} for the special case of three-point functions of massless scalars for $d=3$. In section \ref{subsec::3ptlimits} we show how kinematic limits in the phase space of momenta can be studied using the Mellin-Barnes representation. In section \ref{sec::exch} we consider exchange four-point functions of general scalars, deriving its Mellin-Barnes representation in section \ref{subsec::genresult4pt}. In sections \ref{subsec::opelimit} and \ref{subsec::seriesexpfrommellin} we detail how the Operator Product Expansion and Effective Field Theory expansion respectively are encoded in the Mellin-Barnes representation.

Our notations and conventions are given in section \ref{subsec::notationsconventions}. Various technical details are relegated to the appendices, where we also review relevant aspects of Mellin-Barnes integrals

\subsection{Notation and conventions}
\label{subsec::notationsconventions}

We study quantum scalar fields $\phi$ on a fixed $(d+1)$-dimensional de-Sitter background, which we denote by dS$_{d+1}$. This can be viewed as a time-like hyperbola embedded in an ambient $(d+2)$-dimensional Minkowski space-time $\mathbb{R}^{1,d+1}$,
\begin{equation}\label{tlhypcond}
  X \cdot X := \eta_{AB}X^A X^B=L^2, \qquad \eta_{AB}=\text{diag}\left(-\,+\,+\,...\,+\right), \quad A,B=0,...,d+1,
\end{equation}
where $L$ is the de Sitter radius.
We shall consider a flat slicing $x^\mu=\left(\eta,y^i\right)$, $i=1,...,d$, where \begin{subequations}
\begin{align}
    X^0&=\frac{\eta^2-L^2-y^iy^i}{2\eta},\\
    X^i&=-\frac{L y^i}{\eta},\\
    X^{d+1}&=\frac{-\eta^2-L^2+y^iy^i}{2\eta},
\end{align}
\end{subequations}
and the metric reads
\begin{equation}\label{flatslice}
    ds^2 = \frac{L^2}{\eta^2}\left(-d\eta^2+d\vec{y}^2\right).
\end{equation}
The conformal time $\eta$ is related to the proper time $t$ by
\begin{equation}
    d\eta=\frac{dt}{a(t)}, \qquad a(t)=e^{t/L}.
\end{equation}

We shall be interested in the late-time correlation functions of $\phi$, which are evaluated on the future boundary of de Sitter by taking the late-time limit $\eta \to 0$. Spatial slices of de Sitter, including the future boundary, are parametrised by the spatial vectors $y^i$. The spatial momentum is represented by $k^i$ or $\vec{k}$, with magnitude $k=|\vec{k}|$.

\section{Propagators}
\label{sec::propagators}

We will begin in section \ref{subsec::reviewprop} with a brief review on the relevant aspects of freely propagating scalar fields on a fixed background de Sitter space-time, including the Wightman function and Keldysh propagators. For more complete and detailed pedagogical reviews see e.g. \cite{Spradlin:2001pw,Baumann:2009ds,Anninos:2012qw,Akhmedov:2013vka,Chen:2017ryl}. In section \ref{subsec::splitrep} we present a ``split-representation" for de Sitter two-point functions in position space, which are given as an integrated product of bulk-to-boundary propagators. This is obtained as an analytic continuation of the split representation for Harmonic functions in $(d+1)$-dimensional Euclidean anti-de Sitter space (EAdS$_{d+1}$). In section \ref{subsec::mellinfourierspace} we introduce a Mellin-Barnes representation for the propagators in Fourier space, where the dependence on the conformal time is a simple power-law and the analytic continuation from EAdS$_{d+1}$ is encoded in a simple phase. 

\subsection{Review: Wightman two-point functions and Keldysh propagators}
\label{subsec::reviewprop}

Let us consider the free propagation of a scalar field $\phi$ of mass $m$, which satisfies the Klein-Gordan equation
\begin{equation}\label{kgeq}
    \left(\nabla^2-m^2\right)\phi=0.
\end{equation}
At late times $\eta \to 0$, the scalar field behaves as\footnote{The behaviour \eqref{ltl} can be derived by considering the asymptotic form of the equation of motion
\begin{equation}
    0=\left(\nabla^2-m^2\right)\phi\sim L^{-2}\left[-\left(\eta\partial_\eta\right)^2\phi+(d-1)\left(\eta\partial_\eta\right) \phi\right]-m^2\phi,
\end{equation}
and searching for solutions of the form $  \phi\left(\eta,\vec{x}\right)=A\left(\vec{x}\right)\eta^{\Delta}$.}
\begin{align}\label{ltl}
     \phi\left(\eta,\vec{x}\right)\,&\sim\,{\cal O}_{\Delta_+}\left(\vec{x}\right)\eta^{\Delta_+}+{\cal O}_{\Delta_-}\left(\vec{x}\right)\eta^{\Delta_-},
\end{align}
where the scaling dimensions of the boundary operators ${\cal O}_{\Delta_\pm}\left(\vec{x}\right)$ are related to the mass via 
\begin{subequations}
\begin{align}
\Delta_{\pm}&=\frac{d}{2}\pm i\nu,\\
    \left(mL\right)^2&=\Delta_+\Delta_-.
\end{align}
\end{subequations}

Particles in de Sitter space are classified according to Unitary Irreducible Representations (UIRs) of the isometry group $SO\left(1,d+1\right)$ \cite{doi:10.1063/1.1665471,Dobrev:1977qv}.\footnote{See \cite{Joung:2006gj,Joung:2007je,Basile:2016aen} for the complete dictionary between UIRs of the de Sitter isometry algebra $\mathfrak{so}\left(1,d+1\right)$ and fields on dS$_{d+1}$.} The non-tachyonic representations for scalar particles fall into two categories:
\begin{align}\label{PS}
  &\hspace*{-0.5cm}\bullet \:\: \text{Principal Series:} \:\:\: \text{Massive Particles},\:\:\: \nu \in \mathbb{R}, \:\:\: m^2 \geq \left(\tfrac{d}{2}\right)^2. \\ \label{CS}
  &\hspace*{-0.5cm}\bullet \:\: \text{Complementary Series:} \:\:\: \text{Light Particles}, \:\:\: \nu \to i\mu, \:\:\: |\mu| \in \left(0,\tfrac{d}{2}\right), \:\:\: 0 < m^2 < \left(\tfrac{d}{2}\right)^2.
\end{align}
Massless particles correspond to $|\mu|=\frac{d}{2}$ and lie on the boundary of the complementary series (which is sometimes referred to in the literature as the exceptional series). 

In the following we will consider the Wightman two-point function,
\begin{equation}\label{wfsc}
    G\left(x_1,x_2\right)=\langle 0 | \phi\left(x_1\right)\phi\left(x_2\right)|0\rangle,
\end{equation}
which obeys the homogeneous Klein-Gordon equation \eqref{kgeq}. This is the basic object from which other two-point functions (e.g. retarded, advanced and Feynman Green's functions) can be obtained, as we shall review below. de Sitter invariant two-point functions are functions of the invariant length between the two points,
\begin{equation}
    P\left(x_1,x_2\right)=\frac{\eta_{AB}X^{A}_1\left(x_1\right)X^{B}_2\left(x_2\right)}{L^2},
\end{equation}
the dependence on which is convenient to express through the variable
\begin{align}
    \sigma(x_1,x_2)=\frac{L^2+X_1\left(x_1\right) \cdot X_2\left(x_2\right)}{2L^2},
\end{align}
which in the flat slicing \eqref{flatslice} reads
\begin{equation}
    \sigma=1+\frac{\left(\eta_1-\eta_2\right)^2-\left(\vec{y_1}-\vec{y_2}\right)^2}{4\eta_1\eta_2}.
\end{equation}
As a function of $\sigma$ the equation for the Wightman function takes the form (for $\sigma \ne 1$) 
\begin{equation}\label{eqgeobubu}
   L^{-2}\left[ \sigma\left(1-\sigma\right)\partial^2_\sigma G\left(\sigma\right)-\left(\tfrac{d+1}{2}\right)\left(2\sigma-1\right)\partial_\sigma G\left(\sigma\right)\right]-m^2G\left(\sigma\right)=0,
\end{equation}
which is Euler's Hypergeometric differential equation. This has two independent solutions:
\begin{equation}
    G\left(\sigma\right)=A\,{}_2F_1\left(\tfrac{d}{2}+i\nu,\tfrac{d}{2}-i\nu;\tfrac{d+1}{2};\sigma\right)+B\,{}_2F_1\left(\tfrac{d}{2}+i\nu,\tfrac{d}{2}-i\nu;\tfrac{d+1}{2};1-\sigma\right),
\end{equation}
linear combinations of which correspond to the one-parameter family of de Sitter invariant vacua, known as $\alpha$-vacua \cite{Burges:1984qm,Mottola:1984ar,Allen:1985ux}. The solution with $B=0$ corresponds to the standard Bunch-Davies de Sitter vacuum \cite{Gibbons:1977mu}. This solution has a singularity at $\sigma=1$, which is a short-distance singularity.\footnote{This can be understood by noting that $\sigma$ is related to the geodesic distance $D$ as
\begin{align}\label{siggeo}
    \sigma(x_1,x_2)=\frac{1+\cos\left(D\left(x_1,x_2\right)/L\right)}{2}.
\end{align}}
This allows us to fix the coefficient $A$ by requiring that it has the same strength as the short distance singularity in flat space, which is:
\begin{equation}
        G_{\text{flat}}\left(x_1,x_2\right) \approx \frac{1}{D(x_1,x_2)^{d-1}} \frac{\Gamma\left(\frac{d+1}{2}\right)}{2(d-1)\pi^{(d+1)/2}},
\end{equation}
while in de Sitter we have that\footnote{Note that the expansion of the Gauss Hypergeometric function around $z=1$ is \begin{multline}
      {}_2F_1\left(a,b;c;z\right) = \left[\frac{\Gamma\left(c-a-b\right)\Gamma (c)}{\Gamma (c-a) \Gamma (c-b)}+O\left(z-1\right)\right]\\-(1-z)^{c-a-b} e^{2 i \pi  (c-a-b) \left\lfloor \frac{\arg (z-1)}{2 \pi }\right\rfloor } \left[\frac{\Gamma\left(a+b-c\right) \Gamma (c)}{\Gamma (a) \Gamma (b)}+O\left(z-1\right)\right].
  \end{multline}} 
\begin{equation}
        {}_2F_1\left(\tfrac{d}{2}+i\nu,\tfrac{d}{2}-i\nu;\tfrac{d+1}{2};\sigma\right) \approx \frac{\Gamma\left(\tfrac{d+1}{2}\right)\Gamma\left(\tfrac{d-1}{2}\right)}{\Gamma\left(\tfrac{d}{2}+i\nu\right)\Gamma\left(\tfrac{d}{2}-i\nu\right)}\frac{2^{d-1}}{\left(D(x_1,x_2)/L\right)^{d-1}},
    \end{equation}
using the relation \eqref{siggeo} between $\sigma$ and the geodesic distance. This gives the following expression for the Bunch-Davies solution: 
\begin{equation}\label{wightprop}
G\left(\sigma\right)= \frac{1}{L^{d-1}} \frac{\Gamma\left(\tfrac{d}{2}+i\nu\right)\Gamma\left(\tfrac{d}{2}-i\nu\right)}{(4\pi)^{(d+1)/2}\Gamma\left(\tfrac{d+1}{2}\right)}{}_2F_1\left(\tfrac{d}{2}+i\nu,\tfrac{d}{2}-i\nu;\tfrac{d+1}{2};\sigma\right).
\end{equation}
The Hypergeometric function moreover has a branch cut for $\sigma \in \left(1,\infty\right)$, where the two points become time-like separated. The possible $i \epsilon$ prescriptions for going around the singularity in the complex plane, which are
\begin{equation}\label{sigmaeps}
    \sigma_{\pm}=1-\frac{\left(\vec{y_1}-\vec{y_2}\right)^2-\left(\eta_1-\eta_2\right)^2\mp i\text{sgn}\left(\eta_1-\eta_2\right)\epsilon}{4\eta_1\eta_2},
\end{equation}
correspond to the two possible Euclidean orderings of the operators,\footnote{In particular,
\begin{equation}
    \langle 0 | \phi\left(t_1,\vec{y}_1\right)\phi\left(t_2,\vec{y}_2\right)|0\rangle = \lim_{\epsilon_i \to 0} \langle 0 | \phi\left(t_1-i\epsilon,\vec{y}_1\right)\phi\left(t_2+i\epsilon,\vec{y}_2\right)|0\rangle,
\end{equation}
where $\epsilon>0$. See e.g. \cite{Streater:1989vi}.} 
\begin{subequations}\label{pmWM2pt}
\begin{align}
 G_{-+}\left(x_1,x_2\right)&=\langle 0 | {\hat \phi}\left(x_1\right){\hat \phi}\left(x_2\right)|0\rangle=G\left(\sigma_-\right),\\
   G_{+-}\left(x_1,x_2\right)&=\langle 0 | {\hat \phi}\left(x_2\right){\hat \phi}\left(x_1\right)|0\rangle=G\left(\sigma_+\right).
\end{align}
\end{subequations}
See e.g. \cite{PhysRevLett.73.1746,Bros:1995js} for detailed considerations of this point. This will also be important when we derive the Wightman function \eqref{wightprop} as an analytic continuation from Euclidean anti-de Sitter space.

In this way the Bunch-Davies time-ordered and anti-time-ordered two-point functions are given in terms of the Wightman two-point function \eqref{pmWM2pt} as
\begin{subequations}
\begin{align}
   \hspace*{-0.4cm}  \langle 0 |T {\hat \phi}\left(x_1\right){\hat \phi}\left(x_2\right)|0\rangle &= \theta\left(\eta_1-\eta_2\right) G_{-+}\left(x_1,x_2\right) + \theta\left(\eta_2-\eta_1\right) G_{+-}\left(x_1,x_2\right), \\
   \hspace*{-0.4cm}  \langle 0 |{\bar T} {\hat \phi}\left(x_1\right){\hat \phi}\left(x_2\right)|0\rangle &= \theta\left(\eta_1-\eta_2\right) G_{+-}\left(x_1,x_2\right) + \theta\left(\eta_2-\eta_1\right)G_{-+}\left(x_1,x_2\right),
\end{align}
\end{subequations}
where $T$ and ${\bar T}$ denote time and anti-time-ordered products.

\paragraph{Schwinger-Keldysh formalism.} In time-dependent backgrounds it is useful to employ the Schwinger-Keldysh (or ``in-in'') formalism \cite{doi:10.1063/1.1703727,kadanoff1962quantum,Keldysh:1964ud} for perturbative evaluations of expectation values. In this formalism, to compute fixed-time expectation values one performs a time-ordered integral which goes from the initial time to the time of interest $\eta=\eta_0$, and then performs an anti-time-ordered integral back to the initial time. This is called the ``in-in contour''. The propagators with points along the different parts of the contour are 
\begin{subequations}\label{SKprop}
\begin{align}
  G_{++}\left(x_1,x_2\right)&=\langle 0 |T {\hat \phi}\left(x_1\right){\hat \phi}\left(x_2\right)|0\rangle, \\
 G_{+-}\left(x_1,x_2\right)&= \langle 0 | {\hat \phi}\left(x_2\right){\hat \phi}\left(x_1\right)|0\rangle,
   \\
  G_{-+}\left(x_1,x_2\right)&= \langle 0 | {\hat \phi}\left(x_1\right){\hat \phi}\left(x_2\right)|0\rangle,
   \\
    G_{--}\left(x_1,x_2\right)&=\langle 0 |{\bar T} {\hat \phi}\left(x_1\right){\hat \phi}\left(x_2\right)|0\rangle,
\end{align}
\end{subequations}
where the $+$ $(-)$ subscripts denote the (anti-)time ordered part of the in-in contour. This formalism was first applied to the evaluation of cosmological correlators in \cite{Maldacena:2002vr,Bernardeau:2003nx,Weinberg:2005vy}.

\subsection{Split representation}
\label{subsec::splitrep}

In this section we introduce a convenient integral representation of de Sitter propagators, where they are given as an integrated product of bulk-to-boundary propagators.\footnote{By boundary here we are referring to the late-time boundary at $\eta=0$.} In Euclidean anti-de Sitter space, such a representation for propagators is often referred to in the AdS/CFT literature as the ``split representation", which has proven to be an invaluable tool in the evaluation of Witten diagrams \cite{Hartman:2006dy,Penedones:2010ue,Giombi:2011ya,Paulos:2011ie,Fitzpatrick:2011ia,Costa:2014kfa,Bekaert:2014cea,Bekaert:2015tva,Sleight:2016hyl,Chen:2017yia,Sleight:2017fpc,Tamaoka:2017jce,Giombi:2017hpr,Yuan:2017vgp,Giombi:2018vtc,Yuan:2018qva,Nishida:2018opl,Costa:2018mcg,Carmi:2018qzm,Jepsen:2019svc}. 

To derive the split representation in de Sitter space, it is useful to re-visit the split representation in Euclidean anti-de Sitter space. In the Poincar\'e patch the EAdS$_{d+1}$ metric reads:
\begin{equation}\label{poincarcoord}
    ds^2=\frac{L^2}{z^2}\left(dz^2+d\vec{y}^2\right).
\end{equation}
This is related to the flat slicing of the de Sitter metric \eqref{flatslice} by the analytic continuation $z=-\eta e^{\pm \frac{\pi i}{2}}$ and changing the sign of the metric \cite{Polyakov:2007mm}. In Euclidean anti-de Sitter space, the counter-part of the Bunch-Davies Wightman function \eqref{wfsc} is the Harmonic function, 
\begin{equation}
    \left(\nabla^2_{\text{AdS}}-m^2\right)\Omega_\nu\left(x_1,x_2\right)=0,
\end{equation}
which admits the split representation (see e.g. \cite{Leonhardt:2003qu,Moschella:2007zza,Costa:2014kfa}): 
 \begin{equation}\label{spliteads}
    \Omega_{\nu}\left(x_1,x_2\right)=\frac{\nu^2}{\pi} \int d^d\vec{y}\, K_{\frac{d}{2}+i\nu,0}\left(x_1;\vec{y}\right)K_{\frac{d}{2}-i\nu,0}\left(x_2;\vec{y}\right).
\end{equation} 
This is a product of bulk-to-boundary propagators in EAdS$_{d+1}$ that are integrated over their common boundary point $\vec{y}$, which in Poincar\'e patch \eqref{poincarcoord} read \cite{Witten:1998qj}:
\begin{align}\label{bubo}
  \hspace*{-0.25cm}  K_{\Delta,0}\left(z_1,\vec{y}_1;\vec{y}\right)=C_{\Delta,0}\left(\frac{z_1}{z^2_1+\left(\vec{y}_1-\vec{y}\right)^2}\right)^\Delta, \quad C_{\Delta,0} =\frac{1}{L^{\left(d-1\right)/2}} \frac{\Gamma\left(\Delta\right)}{2\pi^{d/2}\Gamma\left(\Delta+1-\frac{d}{2}\right)}.
\end{align}

\begin{figure}[t]
    \centering
    \captionsetup{width=0.8\textwidth}
    \includegraphics[scale=0.45]{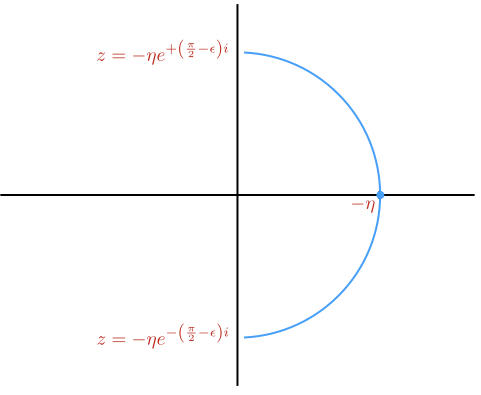}
    \caption{Analytic continuation from EAdS to dS}
    \label{fig:wick}
\end{figure}

That the Harmonic function is related to the Bunch-Davies de Sitter Wightman function \eqref{wightprop} becomes manifest upon evaluating the boundary integral in \eqref{spliteads}, which gives the Gauss Hypergeometric function (see e.g. \cite{Penedones:2007ns}): 
\begin{equation}\label{harm2f1}
       \Omega_{\nu}\left(x_1,x_2\right)=\frac{1}{\Gamma\left(i\nu\right)\Gamma\left(-i\nu\right)}\frac{\Gamma\left(\tfrac{d}{2}+i\nu\right)\Gamma\left(\tfrac{d}{2}-i\nu\right)}{L^{d-1}(4\pi)^{\tfrac{d+1}{2}}\Gamma\left(\tfrac{d+1}{2}\right)}{}_2F_1\left(\tfrac{d}{2}+i\nu,\tfrac{d}{2}-i\nu;\tfrac{d+1}{2};\sigma_{\text{AdS}}\right),
\end{equation}
where, in the Poincar\'e patch \eqref{poincarcoord},
\begin{align}\label{chordaldistancads}
    \sigma_{\text{AdS}}=1
    -\frac{\left(z_1+z_2\right)^2+\left(\vec{y}_1-\vec{y}_2\right)^2}{4z_1z_2}.
\end{align}

\begin{figure}[t]
\centering
\captionsetup{width=0.95\textwidth}
\begin{subfigure}[b]{\textwidth}
\centering
   \includegraphics[width=0.975\linewidth]{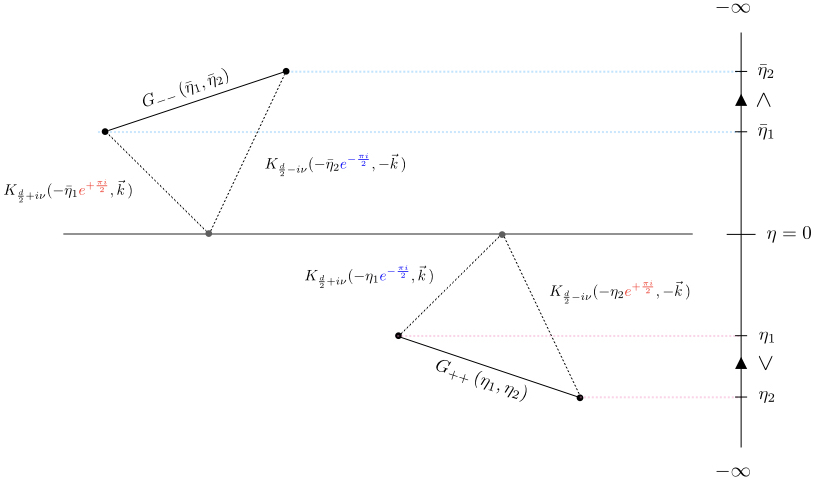}
   \caption{.}
   \label{fig:Ng1} 
\end{subfigure}

\begin{subfigure}[b]{\textwidth}
\centering
   \includegraphics[width=0.975\linewidth]{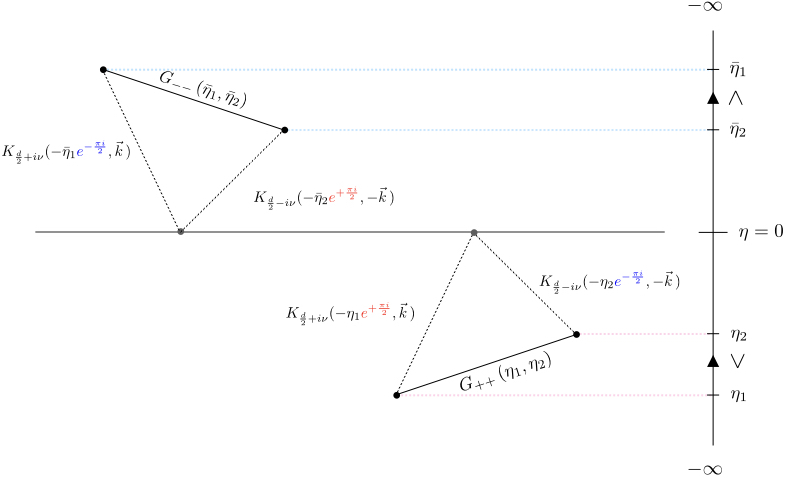}
   \caption{.}
   \label{fig:Ng2}
\end{subfigure}

\caption{Depiction of the split representation for de Sitter propagators on the $++$ and $--$ branches of the in-in contour, for $\eta_1 > \eta_2$, ${\bar \eta}_1 > {\bar \eta}_2$ (figure (a)) and $\eta_2 > \eta_1$, ${\bar \eta}_2 > {\bar \eta}_1$ (figure (b)). The arrows along the vertical axis indicate the path along the in-in contour.}
\label{fig::splitrepdS}
\end{figure}

The Wightman function \eqref{wightprop} can therefore be obtained from the EAdS$_{d+1}$ Harmonic function by Wick rotating $z_{1}$ and $z_2$ in opposite directions: 
\begin{equation}
    z_1=-\eta_1 e^{\tcr{\pm} \left(\frac{\pi }{2}-\epsilon\right)i}, \qquad z_2=-\eta_2 e^{\tcb{\mp}\left(\frac{\pi }{2}-\epsilon\right)i},
\end{equation}
which correspond to the two possible Euclidean orderings \eqref{pmWM2pt}. See figure \ref{fig:wick}. In particular, 
\begin{shaded}
\noindent \emph{Split representation of the Bunch-Davies Wightman two-point function}
\begin{equation}\label{Wightsplit}
    G\left(\sigma_{\pm}\right)  = \Gamma\left(i\nu\right)\Gamma\left(-i\nu\right)\Omega_{\nu}(-\eta_1 e^{\pm\frac{\pi i}{2}},\vec{y}_1;-\eta_2 e^{\mp\frac{\pi i}{2}},\vec{y}_2),
\end{equation}
\end{shaded}
\noindent which, via \eqref{spliteads}, provides the split representation for the Bunch-Davies de Sitter Wightman function. This is equivalent to the integral expressions derived in \cite{Bros:1994dn,Bros:1995js,Joung:2006gj}, though the connection with the split representation of the EAdS$_{d+1}$ Harmonic function was not made manifest. The expression \eqref{Wightsplit} moreover provides a split representation for the Keldysh propagators \eqref{SKprop}. This is depicted in figure \ref{fig::splitrepdS}. \newpage

\subsection{Mellin representation in Fourier space and the late-time limit}
\label{subsec::mellinfourierspace}

Because of translation invariance, it is convenient to study Cosmological Correlators in Fourier space. The Fourier transform of the split representation \eqref{spliteads}, being a convolution, completely factorizes: 
\begin{equation}\label{splitrepharm}
    \Omega_{\nu,\vec{k}}(z_1;z_2) = \frac{\nu^2}{\pi} K_{\frac{d}{2}+i\nu}(z_1,\vec{k})K_{\frac{d}{2}-i \nu}(z_2,-\vec{k}).
\end{equation}
In Fourier space, the bulk-to-boundary propagator in EAdS is given by a Modified Bessel function of the second kind \cite{Gubser:1998bc}, which admits the following convenient representation as a Mellin-Barnes integral:
\begin{multline}\label{mellinK}
     K_{\tfrac{d}{2}+i\nu}(z,\vec{k})=  \frac{z^{\tfrac{d}{2}-i\nu}}{2L^{\left(d-1\right)/2}\Gamma\left(1+i\nu\right)}\int^{i\infty}_{-i\infty}\frac{du}{2\pi i}\, \Gamma\left(u+\tfrac{i\nu}{2}\right)\Gamma\left(u-\tfrac{i\nu}{2}\right)\left(\frac{zk}{2}\right)^{-2u+i\nu},
\end{multline}
where $k=|\vec{k}|$ and at the level of the Mellin integrand the dependence on the co-ordinate $z$ is a simple power-law. Combined with \eqref{Wightsplit}, this gives the following Mellin-Barnes representation of the Bunch-Davies Wightman function \eqref{Wightsplit}:\footnote{To simplify the presentation we introduced the compact notation: 
\begin{equation}
    \int \left[du\right]_2= \int^{i\infty}_{-i\infty}\frac{du_1}{2\pi i}\frac{du_2}{2\pi i}.
    \end{equation}} 
\begin{shaded}
\noindent \emph{Mellin-Barnes representation for the Wightman two-point function in Fourier space}
\begin{multline}\label{MBwightman}
G_{\pm \mp, \vec{k}}\left(\eta_1,\eta_2\right)= \left(-\eta_1\right)^{\tfrac{d}{2}}\left(-\eta_2\right)^{\tfrac{d}{2}} \frac{1}{4\pi L^{d-1}} \int \left[du\right]_2\,e^{\delta_\pm\left(u_1,u_2\right)} \rho_{\nu,\nu}\left(u_1,u_2\right)\\ \times \prod^2_{j=1}\left(-\tfrac{\eta_jk}{2}\right)^{-2u_j},
\end{multline}
\end{shaded}
\noindent where the Wick rotations to de Sitter space introduce a phase:
\begin{align}\label{phasewightmann}
    \delta_\pm\left(u_1,u_2\right)=\mp i\pi \left(u_1-u_2\right),
\end{align}
and we collected the $\Gamma$-functions from each bulk-to-boundary propagator together in the function
\begin{equation}
    \rho_{\nu_1,\nu_2}\left(u_1,u_2\right)=\prod^2_{j=1}\Gamma\left(u_j+\tfrac{i\nu_j}{2}\right)\Gamma\left(u_j-\tfrac{i\nu_j}{2}\right).
\end{equation}

The boundary (i.e. late-time) limit is straightforward to take using the Mellin-Barnes representation. Considering the late-time limit of a single leg, say $\eta_2 \to 0$, the power series expansion of the Wightman function \eqref{MBwightman} in $\eta_2$ is given by the residues of the poles in the corresponding Mellin variable $u_2$, which are at
\begin{equation}
    u_2=-n\mp\tfrac{i\nu}{2}, \qquad n = 0, 1, 2, 3, ...\,.
\end{equation}
The leading contributions as $\eta_2 \to 0$ are given by the residues of the poles with $n=0$, so that\footnote{Note that for the Principal Series $\nu \in \mathbb{R}$ both terms are leading in the limit $z \rightarrow 0$, while for representations with $\Delta = \tfrac{d}{2}+i\nu \in \mathbb{R}$ one of them dominates.}
\begin{equation}\label{etlimwightan}
   \lim_{\eta_2 \to 0} G_{\pm \mp, \vec{k}}\left(\eta_1,\eta_2\right)=F^{(\nu)}_{\pm,\vec{k}}(\eta_1;\eta_2)+F^{(-\nu)}_{\pm,\vec{k}}(\eta_1;\eta_2),
\end{equation}
which defines de Sitter bulk-to-boundary propagators:\footnote{For convenience we defined the normalisation:
\begin{equation}
    {\cal N}_{\nu}\left(\eta_2\right)=\left(-\eta_2\right)^{\tfrac{d}{2}+i\nu} \frac{\Gamma\left(-i\nu\right)}{4\pi L^{d-1}}.
\end{equation}}
\begin{shaded}
\noindent \emph{Mellin-Barnes representation for de Sitter bulk-to-boundary propagators}
\begin{multline}\label{pmF}
   F^{(\nu)}_{\pm,\vec{k}}(\eta;\eta_0)=\left(-\eta\right)^{\tfrac{d}{2}-i\nu} {\cal N}_{\nu}\left(\eta_0\right)
   \int^{i\infty}_{-i\infty} \frac{ds}{2\pi i}\,e^{\delta^\pm_\nu\left(s\right)} \Gamma\left(s+\tfrac{i\nu}{2}\right)\Gamma\left(s-\tfrac{i\nu}{2}\right)\\ \times \left(-\frac{\eta k}{2}\right)^{-2s+i\nu},
\end{multline}
\end{shaded}
\noindent where $\eta_0 \sim 0$, with complex conjugate phases: 
\begin{equation}\label{bubophase}
    \delta^\pm_\nu\left(s\right)=\mp i\pi\left(s+\tfrac{i\nu}{2}\right).
\end{equation}
In \eqref{pmF} we adopted the Mellin variable $s$ in place of $u_1$, which we use henceforth for propagators associated to external legs connected to the boundary. The integrals in \eqref{pmF} are in fact the Mellin-Barnes representations for Hankel functions of the first and second kind \cite{MellinBook},
\begin{subequations}
  \begin{align}\label{hankel}
    F^{(\nu)}_{+,\vec{k}}(\eta;\eta_0)&=-i\pi \left(\frac{k}{2}\right)^{+i\nu} {\cal N}_{\nu}\left(\eta_0\right)\left(-\eta\right)^{\tfrac{d}{2}}e^{+ \nu \pi}  H^{\left(2\right)}_{i\nu}\left(-\eta k\right),\\
    F^{(\nu)}_{-,\vec{k}}(\eta;\eta_0)&=i\pi   \left(\frac{k}{2}\right)^{+i\nu} {\cal N}_{\nu}\left(\eta_0\right)\left(-\eta\right)^{\tfrac{d}{2}}e^{- \nu \pi}H^{\left(1\right)}_{i\nu}\left(-\eta k\right),
\end{align}  
\end{subequations}
as consistent with the known expressions for mode functions of scalar fields in de Sitter. The expression \eqref{pmF} makes manifest that bulk-to-boundary propagators in dS$_{d+1}$ can be obtained as analytic continuations of EAdS$_{d+1}$ bulk-to-boundary propagators \eqref{mellinK}.\footnote{In particular, we have the following relation between the Hankel function and the modified Bessel function of the second kind \cite{watson1944treatise}:
\begin{equation}
    K_{i\nu}\left(z\right)=\frac{\pi i}{2} e^{-\frac{ \pi \nu}{2}}H^{\left(1\right)}_{i\nu}\left(iz\right).
\end{equation}}

The Mellin-Barnes representations \eqref{MBwightman} and \eqref{pmF} for propagators in de Sitter space are the primary tool with which we obtain the late-time correlators in this work. The integrals over conformal time reduce to simple integrals of the power-law type, giving expressions for the late-time correlators as Mellin-Barnes integrals in the momenta which, as we shall see, provide a useful framework with which their properties can be studied, and a natural language in which conformal correlators can be described in momentum space.

\paragraph{Late-time two-point function.} In a similar way we can obtain the late-time two-point function by also sending $\eta_1 \to 0$ in \eqref{etlimwightan}. Focusing on a single bulk-to-boundary propagator \eqref{pmF}, the non-analytic terms in $k$ are generated by the residues of the poles:
\begin{equation}
    s=-\tfrac{i\nu}{2}+n, \qquad n=0,1,2,3,...\,.
\end{equation}
The second set of poles in \eqref{pmF} at $s=\tfrac{i\nu}{2}+n$ generate only analytic terms which in position space don't give rise to long-distance correlations. The leading term is generated by the leading $\Gamma$-function pole (with $n=0$), which gives:
\begin{multline}\label{2ptlongdist}
 \hspace*{-0.7cm}  \lim_{\eta_1,\eta_2 \to 0} \langle 0 | \phi_{\vec{k}}\left(\eta_1\right)\phi_{-\vec{k}}\left(\eta_2\right) |0 \rangle^\prime=\frac{1}{4\pi L^{d-1}}\left[\Gamma\left(-i\nu\right)^2 \left(\frac{k^2\eta_1\eta_2}{4}\right)^{\frac{d}{2}+i\nu}+ \nu \to -\nu \right]+\text{local},
\end{multline}
where
\begin{equation}
    \langle 0 | \phi_{\vec{k}}\left(\eta_1\right)\phi_{\vec{k}^\prime}\left(\eta_2\right) |0 \rangle = \left(2\pi\right)^d \delta^{d}(\vec{k}+\vec{k}^\prime\,) \langle 0 | \phi_{\vec{k}}\left(\eta_1\right)\phi_{-\vec{k}}\left(\eta_2\right) |0 \rangle^\prime.
\end{equation}

The ``+local" in \eqref{2ptlongdist} denotes analytic terms in $k_I$ which do not encode long-range correlations. For light particles \eqref{CS}, where $\nu \to i\mu$ with $|\mu| \in \left(0,\frac{d}{2}\right)$, the expectation value decreases exponentially with time $\eta^{\frac{d}{2}\pm i\nu}\sim e^{-\left(\frac{d}{2}\mp \mu\right)t}$. For massive particles \eqref{PS}, where $\nu \in \mathbb{R}$, we see an oscillatory behaviour $\eta^{\frac{d}{2}\pm i\nu}\sim e^{-\left(\frac{d}{2}\pm i\nu\right)t}$ due to particle creation in the expanding universe, for which the expectation value is exponentially suppressed for large $\nu$, $|\Gamma\left(\pm i\nu\right)^2|\sim e^{-\pi \nu}$.

\section{Contact diagrams}
\label{sec::contactdiagrams}

The most basic correlators are those generated by local contact interactions. We start off with the simplest contact diagrams, which are those generated by three scalars. In section \ref{subsec::nptcontact} we show that the Mellin-Barnes representation of late-time 3-point contact diagrams trivially extends to $n$-point contact diagrams. In all cases the late-time correlators are given by generalised Hypergeometric functions. In sections \ref{subsec::3pttwocc} and \ref{subsecc::mlesssc} we discuss the simplifications which occur when one or more of the scalars is conformally coupled or massless. In section \ref{subsec::3ptlimits} we demonstrate the utility of the Mellin-Barnes representation in the study of kinematic limits in the phase space of momenta.

\begin{figure}[h]
    \centering
    \captionsetup{width=0.9\textwidth}
    \includegraphics[scale=0.45]{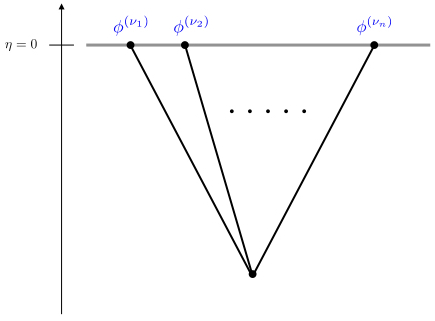}
    \caption{Contact diagram contributing to the correlator $ \langle \phi^{(\nu_1)}\phi^{(\nu_2)}\,...\,\phi^{(\nu_n)}\rangle$ at late times $\eta_0 \sim 0$.}
    \label{fig:ncontact}
\end{figure}

\subsection{Three general scalars}
\label{subsec::3ptscalargen}

Consider the cubic interaction $\phi_1\phi_2\phi_3$ for general scalar fields $\phi_i$ of scaling dimension $\Delta_k=\tfrac{d}{2}+i\nu_k$. This interaction is unique on-shell. Strictly speaking, in the following we shall assume that the $\nu_k$ belong to the Principal series, i.e. $\nu_k \in \mathbb{R}$, though the results extend beyond the Principal Series with due care about the analytic continuation in $\nu_k$ (which we discuss further below).

The 3pt correlator is given by
\begin{subequations}\label{3pt000}
\begin{align}
     \langle \phi^{(\nu_1)}_{\vec{k}_1}\phi^{(\nu_2)}_{\vec{k}_2}\phi^{(\nu_3)}_{\vec{k}_3}\rangle &= \left(2\pi\right)^d \delta^{(d)}\left(\vec{k}_1+\vec{k}_2+\vec{k}_3\right)\langle \phi^{(\nu_1)}_{\vec{k}_1}\phi^{(\nu_2)}_{\vec{k}_2}\phi^{(\nu_3)}_{\vec{k}_3}\rangle^\prime,\\ \langle \phi^{(\nu_1)}_{\vec{k}_1}\phi^{(\nu_2)}_{\vec{k}_2}\phi^{(\nu_3)}_{\vec{k}_3}\rangle^\prime&=\langle \phi^{(\nu_1)}_{\vec{k}_1}\phi^{(\nu_2)}_{\vec{k}_2}\phi^{(\nu_3)}_{\vec{k}_3}\rangle^\prime_{+}+\langle \phi^{(\nu_1)}_{\vec{k}_1}\phi^{(\nu_2)}_{\vec{k}_2}\phi^{(\nu_3)}_{\vec{k}_3}\rangle^\prime_{-},
\end{align}
\end{subequations}
where the $+(-)$ sub-indices indicate the contributions from the (anti)-time-ordered branches of the in-in contour, which at late-times $\eta_0 \sim 0$ is given by:
\begin{align}
    \langle \phi^{(\nu_1)}_{\vec{k}_1}\phi^{(\nu_2)}_{\vec{k}_2}\phi^{(\nu_3)}_{\vec{k}_3}\rangle^\prime_{\pm}&=\pm i \int^{\eta_0}_{-\infty}\frac{d\eta}{\left(-\eta/L\right)^{d+1}} F^{(\nu_1)}_{\vec{k}_1,\pm}(\eta;\eta_0)F^{(\nu_2)}_{\vec{k}_2,\pm}(\eta;\eta_0)F^{(\nu_3)}_{\vec{k}_3,\pm}(\eta;\eta_0),
\end{align}
in terms of the bulk-to-boundary propagators \eqref{etlimwightan}. 

The Mellin-Barnes representation \eqref{pmF} of the bulk-to-boundary propagators renders the integral over conformal time to a simple power-law integral:\footnote{For ease of presentation we introduced the following notation for late-time $n$-point correlators:
\begin{subequations}
\begin{align}
\label{nptnorm}
    {\cal N}_n\left(\eta_0,k_i\right)&=\prod^n_{j=1}\left(\tfrac{ k_j}{2}\right)^{i\nu_j}{\cal N}_{\nu_j}\left(\eta_0\right)\\
    \int \left[ds\right]_n&= \int^{i\infty}_{-i\infty}\frac{ds_1}{2\pi i}...\frac{ds_n}{2\pi i}.
\end{align}
\end{subequations}
}
\begin{multline}
  \langle \phi^{(\nu_1)}_{\vec{k}_1}\phi^{(\nu_2)}_{\vec{k}_2}\phi^{(\nu_3)}_{\vec{k}_3}\rangle^\prime_{\pm}\\ \hspace*{1cm} = \pm i L^{d+1} \underbrace{\left(\prod^3_{j=1}\left(\tfrac{ k_j}{2}\right)^{i\nu_j}{\cal N}_{\nu_j}\left(\eta_0\right)\right)}_{{\cal N}_3\left(\eta_0,k_i\right)}\,  \int \left[ds\right]_3\,\rho_{\nu_1,\nu_2,\nu_3}\left(s_1,s_2,s_3\right)\prod^3_{j=1}e^{\delta^\pm_{\nu_j}\left(s_j\right)}\left(\tfrac{ k_j}{2}\right)^{-2s_j} 
  \\ \times \int^{\eta_0}_{-\infty}d\eta \left(-\eta\right)^{\frac{d}{2}-1-2\left(s_1+s_2+s_3\right)},
\end{multline}
where the overall constant ${\cal N}_3\left(\eta_0,k_i\right)$ arises from the bulk-to-boundary propagators and we combined the $\Gamma$-function factors from each leg into the function:
\begin{equation}\label{rhomeasure3pt}
  \rho_{\nu_1,\nu_2,\nu_3}\left(s_1,s_2,s_3\right) = \left(\prod^3_{j=1} \Gamma\left(s_j+\tfrac{i\nu_j}{2}\right)\Gamma\left(s_j-\tfrac{i\nu_j}{2}\right)\right).
\end{equation}
The requirement that the $\eta$-integral converges restricts the integration contour for Mellin-Barnes integrals, in particular:
\begin{subequations}
\begin{align}
     \int^{\eta_0}_{-\infty}d\eta \left(-\eta\right)^{\frac{d}{2}-1-2\left(s_1+s_2+s_3\right)} &= -\frac{\left(-\eta_0\right)^{\frac{d}{2}-2\left(s_1+s_2+s_3\right)}}{\tfrac{d}{2}-2\left(s_1+s_2+s_3\right)}, \\\mathfrak{Re}\left[\tfrac{d}{2}-2\left(s_1+s_2+s_3\right)\right]&<0. \label{s3restr}
\end{align}
\end{subequations}
The leading contribution in the late-time limit $\eta_0 \rightarrow 0$ is therefore encoded in the residue of the single pole at
\begin{equation}
    \tfrac{d}{4}-\left(s_1+s_2+s_3\right)\sim0,
\end{equation}
so that in the late-time limit the integral over conformal time is encoded in a Dirac delta function,
\begin{subequations}
\begin{align}
    i \pi \delta \left(\tfrac{d}{4}-s_1-s_2-s_3\right) &= \lim_{\eta_0 \to 0}\left[\int^{\eta_0}_{-\infty}d\eta \left(-\eta\right)^{\frac{d}{2}-1-2\left(s_1+s_2+s_3\right)}\right]\\
    &=-\frac{\left(-\eta_0\right)^{\frac{d}{2}-2\left(s_1+s_2+s_3\right)}}{\tfrac{d}{2}-2\left(s_1+s_2+s_3\right)},
\end{align}
\end{subequations}
which gives
\begin{subequations}\label{pm3ptcontrib}
\begin{align}
 \hspace*{-0.425cm} \langle \phi^{(\nu_1)}_{\vec{k}_1}\phi^{(\nu_2)}_{\vec{k}_2}\phi^{(\nu_3)}_{\vec{k}_3}\rangle^\prime_{\pm} &= \pm i \frac{L^{d+1}}{2}\, e^{\mp \left(\frac{d}{2}+i\left(\nu_1+\nu_2+\nu_3\right)\right) \frac{\pi i}{2}} {\cal N}_3\left(\eta_0,k_i\right) I^{\left(\nu_1,\nu_2,\nu_3\right)}_{\vec{k}_1,\vec{k}_2,\vec{k}_3}, \\ \label{3ptconfstr}
  \hspace*{-0.425cm} I^{\left(\nu_1,\nu_2,\nu_3\right)}_{\vec{k}_1,\vec{k}_2,\vec{k}_3} &= \int \left[ds\right]_3\,2\pi i\, \delta\left(\tfrac{d}{4}-s_1-s_2-s_3\right)I^{\left(\nu_1,\nu_2,\nu_3\right)}_{\vec{k}_1,\vec{k}_2,\vec{k}_3}\left(s_1,s_2,s_3\right), \\
  I^{\left(\nu_1,\nu_2,\nu_3\right)}_{\vec{k}_1,\vec{k}_2,\vec{k}_3}\left(s_1,s_2,s_3\right)&=\rho_{\nu_1,\nu_2,\nu_3}\left(s_1,s_2,s_3\right)\prod^3_{j=1}\left(\frac{ k_j}{2}\right)^{-2s_j},
\end{align}
\end{subequations}
where we used the Dirac delta distribution to translate the sum of the phase factors \eqref{bubophase} from each bulk-to-boundary propagator \eqref{pmF} into an overall phase for each $\pm$ contribution. The $+$ and $-$ contributions thus differ only by a phase. Eliminating one of the Mellin variables, say $s_3$, gives the correlation function as a double Mellin-Barnes integral,
\begin{multline}\label{2varrep3pt}
    I^{\left(\nu_1,\nu_2,\nu_3\right)}_{\vec{k}_1,\vec{k}_2,\vec{k}_3}=\left(\frac{k_3}{2}\right)^{-\frac{d}{2}}\int \left[ds\right]_2\,\prod^2_{j=1} \Gamma\left(s_j+\tfrac{i\nu_j}{2}\right)\Gamma\left(s_j-\tfrac{i\nu_j}{2}\right)\\
    \times \Gamma\left(\tfrac{d}{4}-s_1-s_2+\tfrac{i\nu_3}{2}\right)\Gamma\left(\tfrac{d}{4}-s_1-s_2-\tfrac{i\nu_3}{2}\right)\left(\frac{k_1}{k_3}\right)^{-2s_1}\left(\frac{k_2}{k_3}\right)^{-2s_2},
\end{multline}
which is a function of the two ratios $k_1/k_{3}$ and $k_2/k_{3}$.

\begin{figure}[t]
    \centering
    \captionsetup{width=0.95\textwidth}
    \includegraphics[scale=0.45]{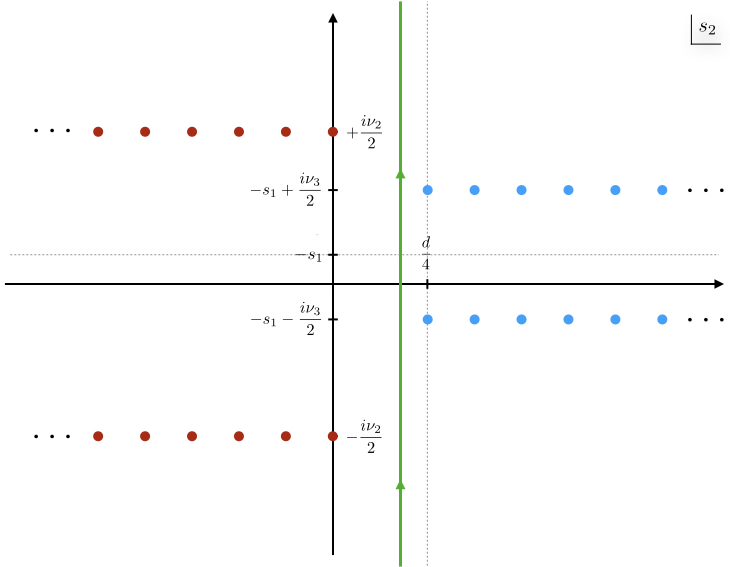}
    \caption{Integration contour (Green) for the Mellin-Barnes representation \eqref{2varrep3pt} of the three-point conformal structure. W.l.o.g. we focus on the integral in the Mellin variable $s_2$ and take $\mathfrak{Re}\left(s_1\right)=0$. In this figure all scaling dimensions are taken to lie on the Principal Series \eqref{PS}, $\nu_j \in \mathbb{R}$. Poles of the $\Gamma$-functions $\Gamma\left(s_2\pm \tfrac{i\nu_2}{2}\right)$ are displayed in red and the poles of $\Gamma\left(\tfrac{d}{4}-s_1-s_2\pm\tfrac{i\nu_3}{2}\right)$ in blue, which the integration contour is prescribed to separate.}
    \label{fig:3pts2}
\end{figure}

Combining the contributions from the $+$ and $-$ branches of the in-in contour, the resulting expression for the late-time three-point function is: 
\begin{shaded}
\noindent\emph{Mellin-Barnes representation of late-time scalar 3pt contact diagram}
\begin{equation}\label{total3ptdoublemellin}
     \langle \phi^{(\nu_1)}_{\vec{k}_1}\phi^{(\nu_2)}_{\vec{k}_2}\phi^{(\nu_3)}_{\vec{k}_3}\rangle^\prime = L^{d+1} {\cal N}_3\left(\eta_0,k_i\right) \sin\left(\left(\tfrac{d}{2}+i\left(\nu_1+\nu_2+\nu_3\right)\right) \tfrac{\pi}{2}\right)I^{\left(\nu_1,\nu_2,\nu_3\right)}_{\vec{k}_1,\vec{k}_2,\vec{k}_3}.
\end{equation}
\end{shaded}

\noindent A few comments are in order:

\noindent $\bullet$ As is standard treatment of Mellin-Barnes integrals (see e.g. \cite{MellinBook}), the  integration contour runs parallel to the imaginary axis and is suitably indented so it separates the sequences of poles encoded in $\Gamma$-functions of the type $\Gamma\left(s_j+a_j\right)$ from those of the type $\Gamma\left(-s_j+b_j\right)$, see figure \ref{fig:3pts2}. See also appendix \ref{app:mellinbarnesints} where various pertinent properties of Mellin-Barnes integrals are reviewed. This contour prescription is well-defined for parameters $a_j$ and $b_j$, where the sequences of poles from $\Gamma$-functions $\Gamma\left(s_j+a_j\right)$ do not collide with those from a $\Gamma$-function of the type $\Gamma\left(-s_j+b_j\right)$. This is always the case for Principal Series representations \eqref{PS}, where such sequences of poles are strictly separated and they can only move along the imaginary axis as the $\nu_j$ vary. Away from the Principal Series however, the sequences of poles can move along the real axis. In this case, for a given boundary dimension $d$, certain special values of $\nu_j$ lead to the collision of such sequences of poles and the prescribed integration contour becomes ``pinched". See e.g. figure \ref{fig:pinching}. This generates singularities which require careful regularisation to define the result for those values of $\nu_j$ and $d$ (either by analytic continuation or by the addition of appropriate counter-terms).\footnote{See also \cite{Bzowski:2013sza,Bzowski:2015pba,Bzowski:2015yxv,Bzowski:2017poo,Bzowski:2018fql} for extensive studies of the regularisation and renormalisation of conformal structures in momentum space.} We shall consider some explicit examples of this type in section \ref{subsecc::mlesssc} and at the end of sections \ref{subsec::3pttwocc} and \ref{subsec::seriesexpfrommellin}.\\

\noindent $\bullet$ The Mellin-Barnes integral \eqref{2varrep3pt} is nothing but Appell's $F_4$ function \cite{appell1880series,AppelletKampe}, which is a generalised Hypergeometric function of two-variables (see appendix \ref{subsec::appgenhypf}). In particular, 
\begin{align}
    & I^{\left(\nu_1,\nu_2,\nu_3\right)}_{\vec{k}_1,\vec{k}_2,\vec{k}_3} =    \left(\frac{k_3}{2}\right)^{-\frac{d}{2}} \Gamma\left(-i\nu_1\right) \Gamma\left(-i\nu_2\right) \Gamma\left(\tfrac{d}{4}+\tfrac{i\left(\nu_1+\nu_2+\nu_3\right)}{2}\right) \Gamma\left(\tfrac{d}{4}+\tfrac{i\left(\nu_1+\nu_2-\nu_3\right)}{2}\right)\\ & \times 
     \left(\frac{k_1}{k_3}\right)^{i\nu_1}\left(\frac{k_2}{k_3}\right)^{i\nu_2} F_4\left(\tfrac{d}{4}+\tfrac{i}{2}\left(\nu_1+\nu_2+\nu_3\right), \tfrac{d}{4}+\tfrac{i}{2}\left(\nu_1+\nu_2-\nu_3\right);i\nu_1+1,i\nu_2+1;\frac{k_1^2}{k_3^2},\frac{k_2^2}{k_3^2}\right). \nonumber
\end{align}
The appearance of the function $F_4$ is expected since special conformal invariance on scalar 3pt correlators in momentum space translates into Appell's system of partial differential equations for $F_4$ \cite{Coriano:2013jba,Bzowski:2013sza} (see also \cite{Davydychev:1992xr,Antoniadis:2011ib}), where a unique physical solution is selected by the absence of divergences in co-linear momentum configurations (also known as ``collapsed triangle limits"). In \S \ref{subsec::3ptlimits} we will show how limits in the phase space of momenta, including the ``collapsed triangle limits", can be studied systematically using the Mellin-Barnes representation.\\

\noindent $\bullet$ The phases $e^{\mp \left(\frac{d}{2}+i\left(\nu_1+\nu_2+\nu_3\right)\right) \frac{\pi i}{2}}$, however, cannot be fixed by conformal symmetry alone. These are determined by the boundary condition at early times, which is encoded in the propagators \eqref{pmF}. When combining the contributions from the different branches of the in-in contour, these phases give the sine function in \eqref{total3ptdoublemellin} which encodes the interference pattern between the different processes.\\

\noindent $\bullet$ Either of the $s_1$ and $s_2$ Mellin-Barnes integrals in \eqref{2varrep3pt} can be expressed in terms of a Gauss Hypergeometric function ${}_2F_1\left(a_1,a_2;a_3;z\right)$ in the variable $z=1-\left(k_i/k_3\right)^2$ (see appendix \ref{subsec::app2f1}), with parameters $a_i$ depending on the remaining Mellin variable.
Note that the resulting expression for the correlator would anyway still be a double Mellin-Barnes integral, since the Gauss Hypergeometric function itself is defined by a single Mellin-Barnes integral -- see equation \eqref{app2f1mellin}. This representation will come in use in section \ref{subsec::opelimit}, when we derive the OPE expansion of the corresponding exchange four-point function. The representation \eqref{2varrep3pt} on the other hand is more suitable for exploring other limits in the phase space of momenta, such as the soft limit of external legs -- which we consider in \S \ref{subsubsec::soft3pt}.\\

\noindent $\bullet$ While for a general triplet of scalars the 3pt correlator \eqref{3pt000} is given in terms of the Appell function $F_4$, in some special cases away from the Principal Series this simplifies to more familiar functions. A known example is when two of the three scalars are conformally coupled, where the function $F_4$ reduces to a single Gauss Hypergeometric function ${}_2F_1$ \cite{Arkani-Hamed:2015bza}. We shall work out this example in detail in section \S \ref{subsec::3pttwocc}, along with its extension to $n$-point diagrams. From the perspective of this work, however, it is worth keeping in mind that the Gauss Hypergeometric function, like the Appell function, is ultimately defined by a Mellin-Barnes integral (see appendix \ref{app2f1mellin}). The latter provides the analytic continuation of the Hypergeometric series beyond its radius of convergence, and played a central role in deriving many of its properties. This includes: their analytic structure, transformation formulae and asymptotic expansions -- examples of which we shall see in this work. The same is true for Appell's functions and other generalised Hypergeometric functions, as a consequence of their Mellin-Barnes representation.  \\

\noindent $\bullet$ While in the above we only considered the simplest cubic interaction of general scalars $\phi_1 \phi_2 \phi_3$, the Mellin-Barnes approach can also be used to compute late-time correlators generated by derivative interactions. Cubic interactions of scalars fields are unique on-shell, so the result is represented by the same Mellin-Barnes integral except for an overall polynomial factor in the $\nu_j$ which encodes derivative structure of the interaction. \\

In the following section we show how the formalism presented here extends without difficulty to $n$-point contact diagrams.

\subsection{$n$ general scalars}
\label{subsec::nptcontact}

The approach introduced in the previous section extends immediately to an arbitrary number of external legs. Consider the contact interaction of $n$ general scalar fields $\phi_1\phi_2...\,\phi_n$. Proceeding as before, with the Mellin-Barnes representation \eqref{pmF} of the bulk-to-boundary propagators the $+$ and $-$ contributions to the late-time $n$-point function take the following form 
\begin{multline}
 \hspace*{-0.5cm} \langle \phi^{(\nu_1)}_{\vec{k}_1}\phi^{(\nu_2)}_{\vec{k}_2}...\,\phi^{(\nu_n)}_{\vec{k}_n}\rangle^\prime_{\pm} = \pm iL^{d+1}\, {\cal N}_n\left(\eta_0,k_i\right)  \int \left[ds\right]_n\,\rho_{\nu_1,\nu_2,...,\nu_n}\left(s_1,s_2,...,s_n\right)\prod^n_{j=1}e^{\delta^\pm_{\nu_j}\left(s_j\right)}\left(\tfrac{ k_j}{2}\right)^{-2s_j}\\ \times \int^{\eta_0}_{-\infty} \frac{d\eta}{\left(-\eta\right)^{d+1}} \prod^n_{i=1} \left(-\eta\right)^{\tfrac{d}{2}-2s_i},
\end{multline}
where the overall constant is defined in \eqref{nptnorm} and the extension of the function \eqref{rhomeasure3pt} to $n$-external legs is
\begin{equation}\label{rhomeasurenpt}
  \rho_{\nu_1,\nu_2,...\,,\nu_n}\left(s_1,s_2,...\,,s_n\right) = \left(\prod^n_{j=1} \Gamma\left(s_j+\tfrac{i\nu_j}{2}\right)\Gamma\left(s_j-\tfrac{i\nu_j}{2}\right)\right) .
\end{equation}

As before, the integral over conformal time has been reduced to a power law:
\begin{subequations}
\begin{align}
    &\int^{\eta_0}_{-\infty} \frac{d\eta}{\left(-\eta\right)^{d+1}} \prod^n_{i=1} \left(-\eta\right)^{\tfrac{d}{2}-2s_i}=-\frac{\left(-\eta_0\right)^{\tfrac{d(n-2)}{2}-2\left(s_1+s_2+...+s_n\right)}}{\tfrac{d(n-2)}{2}-2\left(s_1+s_2+...\,+s_n\right)}, \\ &\mathfrak{Re}\left[\tfrac{d(n-2)}{2}-2\left(s_1+s_2+...\,+s_n\right)\right]<0,
\end{align}
\end{subequations}
where the leading contribution in the late-time limit $\eta_0 \rightarrow 0$ is encoded in the single pole at
\begin{equation}
    \frac{d(n-2)}{4}-\left(s_1+s_2+...\,+s_n\right)\sim0,
\end{equation}
so that the integration over conformal time is encoded in a Dirac delta function:
\begin{subequations}\label{connptpm}
\begin{align}
   \langle \phi^{(\nu_1)}_{\vec{k}_1}\phi^{(\nu_2)}_{\vec{k}_2}...\,\phi^{(\nu_n)}_{\vec{k}_n}\rangle^\prime_{\pm}  &= \pm i \frac{L^{d+1}}{2} e^{\mp\left(\tfrac{d(n-2)}{2}+i\left(\nu_1+...+\nu_n\right)\right)\frac{\pi i}{2}} {\cal N}_n\left(\eta_0,k_i\right) I^{\left(\nu_1,...,\nu_n\right)}_{\vec{k}_1,...,\vec{k}_n},
\end{align}
\end{subequations}
where
\begin{align} \label{nptconfint}
   \hspace*{-0.5cm} I^{\left(\nu_1,...,\nu_n\right)}_{\vec{k}_1,...,\vec{k}_n} &= \int \left[ds\right]_n\, 2\pi i\, \delta\left(\tfrac{d(n-2)}{4}-s_1-...-s_{n}\right)I^{\left(\nu_1,...,\nu_n\right)}_{\vec{k}_1,...,\vec{k}_n}\left(s_1,...,s_n\right), \\
   I^{\left(\nu_1,...,\nu_n\right)}_{\vec{k}_1,...,\vec{k}_n}\left(s_1,...,s_n\right) &= \rho_{\nu_1,\nu_2,...,\nu_n}\left(s_1,...,s_{n}\right) \prod^{n}_{j=1} \left(\frac{k_j}{2}\right)^{-2s_j},
\end{align}
and as before we pulled out the total phase factor coming from the bulk-to-boundary propagators \eqref{pmF}, which makes manifest that the $+$ and $-$ contributions differ by a phase. By eliminating one of the Mellin variables, say $s_n$, this is a function of the $n-1$ ratios of the momenta by $k_n$:
\begin{multline} 
   I^{\left(\nu_1,...,\nu_n\right)}_{\vec{k}_1,...,\vec{k}_n} = \left(\frac{k_n}{2}\right)^{-\frac{d\left(n-2\right)}{2}}\int \left[ds\right]_{n-1}\,\prod^{n-1}_{j=1} \Gamma\left(s_j+\tfrac{i\nu_j}{2}\right)\Gamma\left(s_j-\tfrac{i\nu_j}{2}\right)\\
  \times \Gamma\left(\tfrac{d\left(n-2\right)}{4}-s_1-...-s_{n-1}+\tfrac{i\nu_n}{2}\right)\Gamma\left(\tfrac{d\left(n-2\right)}{4}-s_1-...-s_{n-1}-\tfrac{i\nu_n}{2}\right)\prod^{n-1}_{j=1} \left(\frac{k_j}{k_n}\right)^{-2s_j}.
\end{multline}

The full late-time $n$-point correlator is 
\begin{shaded}
\noindent\emph{Mellin-Barnes representation of the late-time scalar $n$-pt correlation function}
\begin{equation}\label{nptscalargen}
   \langle \phi^{(\nu_1)}_{\vec{k}_1}\phi^{(\nu_2)}_{\vec{k}_2}...\,\phi^{(\nu_n)}_{\vec{k}_n}\rangle^\prime  =  L^{d+1}   {\cal N}_n\left(\eta_0,k_i\right)\sin\left(\left(\tfrac{d(n-2)}{2}+i\left(\nu_1+...+\nu_n\right)\right)\tfrac{\pi }{2}\right)I^{\left(\nu_1,...,\nu_n\right)}_{\vec{k}_1,...,\vec{k}_n}.
\end{equation}
\end{shaded} 
\noindent This is a generalised Hypergeometric function of $n-1$ variables and extends the results of the previous section without any obstacle to $n$-point contact diagrams.

\subsection{$n$-point contact diagrams with conformally coupled scalars}
\label{subsec::3pttwocc}

For certain special values of the scaling dimensions away from the Principal Series \eqref{PS}, the correlators simplify. An interesting example is when an external scalar is conformally coupled, where $\nu=\frac{i}{2}$, which is in the complementary series \eqref{CS}. The reason for this is that the pair of Gamma functions in the Mellin-Barnes representation \eqref{pmF} of the bulk-to-boundary propagator reduce to a single Gamma function via the Legendre duplication formula:
\begin{equation}
    \Gamma\left(s+\tfrac{i\nu}{2}\right)\Gamma\left(s-\tfrac{i\nu}{2}\right) \quad \overset{\nu \; \rightarrow \; \frac{i}{2}}{\rightarrow} \quad 2^{\tfrac{3}{2}-2s} \sqrt{\pi} \Gamma\left(2s-\tfrac{1}{2}\right).
\end{equation}
 Within a correlator, the integral over the Mellin variable associated to a conformally coupled scalar therefore takes the form:
\begin{align}\label{cccontint}
    \int^{i\infty}_{-i\infty} \frac{ds}{2\pi i}\,\Gamma\left(2s-\tfrac{1}{2}\right)\Gamma\left(t-2s\right)z^{2s} = \frac{\sqrt{z+1}}{2} \, \Gamma \left(t-\tfrac{1}{2}\right)\left(\tfrac{1}{z}+1\right)^{-t} ,
\end{align}
where $z$ is a function of the momenta $k_i$. The derivation is given in appendix \ref{subsec::appccs}. The Mellin-Barnes integral associated to any conformally coupled scalar in a late-time correlator may therefore be lifted.

A simple illustrative example of this mechanism is when all but one scalar is conformally coupled. In this case, the $n$-point structure \eqref{nptconfint} can be represented by just a single Mellin-Barnes integral: 
\begin{subequations}\label{ccsclnptconfstr}
\begin{align}
    I^{\left(i/2,...,i/2,\nu_n\right)}_{\vec{k}_1,...,\vec{k}_n} &= \int^{i\infty}_{-i\infty} \frac{ds}{2\pi i}\,I^{\left(i/2,...,i/2,\nu_n\right)}_{\vec{k}_1,...,\vec{k}_n}\left(s\right),\\ \label{ccsclnptconfstr2}
  I^{\left(i/2,...,i/2,\nu_n\right)}_{\vec{k}_1,...,\vec{k}_n}\left(s\right)&= \pi^{\tfrac{n-1}{2}} \frac{2^{d(n-4)+\frac{n+1}{2}}}{\sqrt{k_1...\,k_{n-1}}}\left(\frac{k_n}{2}\right)^{-\frac{d(n-2)}{2}+\tfrac{n-1}{2}}  
    \left(2\frac{k_1+...+k_{n-1}}{k_n}\right)^{\tfrac{n-1}{2}}\\ \nonumber & \hspace*{-1.5cm}\times \Gamma\left(\tfrac{d(n-2)}{4}+\tfrac{i\nu_n}{2}-s\right)\Gamma\left(\tfrac{d(n-2)}{4}-\tfrac{i\nu_n}{2}-s\right)\Gamma\left(2s-\tfrac{n-1}{2}\right)
    \left(2\frac{k_1+...+k_{n-1}}{k_n}\right)^{-2s},
\end{align}
\end{subequations}
where we applied \eqref{cccontint} to lift $n-2$ Mellin integrals associated to the conformally coupled scalars in \eqref{connptpm}. The remaining Mellin-Barnes integral actually defines a Gauss Hypergeometric function of argument $\tfrac{k_n-k_1...-k_{n-1}}{2k_n}$ (see appendix \ref{subsec::app2f1}), so that we can equivalently write:
\begin{multline}\label{2f1ccscaln}
       I^{\left(i/2,...,i/2,\nu_n\right)}_{\vec{k}_1,...,\vec{k}_n}  =  \pi^{n/2}\left(\frac{k_n}{2}\right)^{-\frac{d(n-2)}{2}+\tfrac{n-1}{2}}  
    \frac{\Gamma \left(\tfrac{d(n-2)}{2}+i\nu_n-\tfrac{n-1}{2}\right)\Gamma\left(\tfrac{d(n-2)}{2}-i\nu_n-\tfrac{n-1}{2}\right)}{2^{2(d-n)+\tfrac{n-1}{2}}\sqrt{k_1...\,k_{n-1}}\Gamma \left(\tfrac{d(n-2)}{2}+1-\tfrac{n}{2}\right)}\\ \times  {}_2F_1\left(\tfrac{d(n-2)}{2}+i\nu_n-\tfrac{n-1}{2},\tfrac{d(n-2)}{2}-i\nu_n-\tfrac{n-1}{2};\tfrac{d(n-2)}{2}+1-\tfrac{n}{2};\tfrac{k_n-k_1...-k_{n-1}}{2k_n}\right).
\end{multline}
The late-time correlator \eqref{nptscalargen} in this case therefore reduces to:
\begin{shaded}
\noindent\emph{$n$-point contact diagram with $n-1$ conformally coupled scalars and a general scalar}
\begin{multline}\label{ccgennpt}
     \langle \phi^{(i/2)}_{\vec{k}_1}\phi^{(i/2)}_{\vec{k}_2}...\,\phi^{(i/2)}_{\vec{k}_{n-1}}\phi^{(\nu_n)}_{\vec{k}_n}\rangle^\prime = \pi^{n/2} L^{d+1}{\cal N}_n\left(\eta_0,k_i\right) \sin\left(\left(\tfrac{d(n-2)}{2}-\tfrac{n-1}{2}+i\nu_n\right)\tfrac{\pi}{2}\right) \\ 
   \times \frac{1}{\sqrt{k_1...\,k_{n-1}}}\left(\frac{k_n}{2}\right)^{-\frac{d(n-2)}{2}+\tfrac{n-1}{2}}  
    \frac{\Gamma \left(\tfrac{d(n-2)}{2}+i\nu_n-\tfrac{n-1}{2}\right)\Gamma\left(\tfrac{d(n-2)}{2}-i\nu_n-\tfrac{n-1}{2}\right)}{2^{2(d-n)+\tfrac{n-1}{2}}\Gamma \left(\tfrac{d(n-2)}{2}+1-\tfrac{n}{2}\right)}\\ \times  {}_2F_1\left(\tfrac{d(n-2)}{2}+i\nu_n-\tfrac{n-1}{2},\tfrac{d(n-2)}{2}-i\nu_n-\tfrac{n-1}{2};\tfrac{d(n-2)}{2}+1-\tfrac{n}{2};\tfrac{k_n-k_1...-k_{n-1}}{2k_n}\right),
\end{multline}
\end{shaded}
\noindent which extends to an arbitrary $d$ and $n$ the $d=3$ and $n=3$ result obtained in \cite{Arkani-Hamed:2015bza}.

The above expression further simplifies when $\nu_n$ also takes certain special values away from the Principal Series \eqref{PS}, in which case the Mellin-Barnes representation is no longer required. Naturally this is the case when all external scalars are conformally coupled, i.e. when also $\nu_n=\frac{i}{2}$. In this case, the Gauss Hypergeometric function in \eqref{ccgennpt} reduces to 
\begin{equation}
    {}_2F_1\left(a,b;c;z\right)\overset{b=c}{=}\left(1-z\right)^{-a},
\end{equation}
so that we have:
\begin{multline}\label{totalccs4pt}
      \langle \phi^{(i/2)}_{\vec{k}_1}...\,\phi^{(i/2)}_{\vec{k}_n}\rangle^\prime  = 2\pi^{n/2} L^{d+1}{\cal N}_n\left(\eta_0,k_i\right)\sin\left(\left(\tfrac{d(n-2)}{2}-\tfrac{n}{2}\right)\tfrac{\pi}{2}\right) \\
   \times \frac{2^{d(n-4)+\tfrac{n}{2}}}{\sqrt{k_1...\,k_{n}}} 
   \frac{\Gamma \left(\tfrac{d(n-2)}{2}-\tfrac{n}{2}\right)}{\underbrace{\left(k_1+k_2+...+k_n\right)}_{k_t}{}^{\tfrac{d(n-2)}{2}-\tfrac{n}{2}}}. 
\end{multline}
This extends to general $d$ and $n$ the result obtained for $d=3,\, n=4$ in \cite{Arkani-Hamed:2015bza}, and $d=3,\, n=3$ in \cite{Creminelli:2011mw} (see also \cite{Boyanovsky:2011xn,Anninos:2014lwa}). Notice that for $d>1$ and $n>2$ the argument of the Gamma function satisfies:
\begin{equation}
    \frac{d(n-2)}{2}-\frac{n}{2} > -1,
\end{equation}
so that the only pole of the Gamma function is when  $\frac{d(n-2)}{2}-\frac{n}{2} \sim 0$, for which there is also a zero of the sine function. The expression \eqref{totalccs4pt} is therefore non-singular for all $d>1$ and $n>2$.

The remaining zeros of the sine function gives the values of $d$ and $n$ for which the late-time $\phi^4$ contact diagram is vanishing, which is when:
\begin{equation}
    d\left(n-2\right)-n=4m+4, \qquad m \in \mathbb{N}_0.
\end{equation}
Note that when $n=4$ this occurs when $d$ is even.

\begin{figure}[t]
    \centering
    \captionsetup{width=0.95\textwidth}
    \includegraphics[scale=0.5]{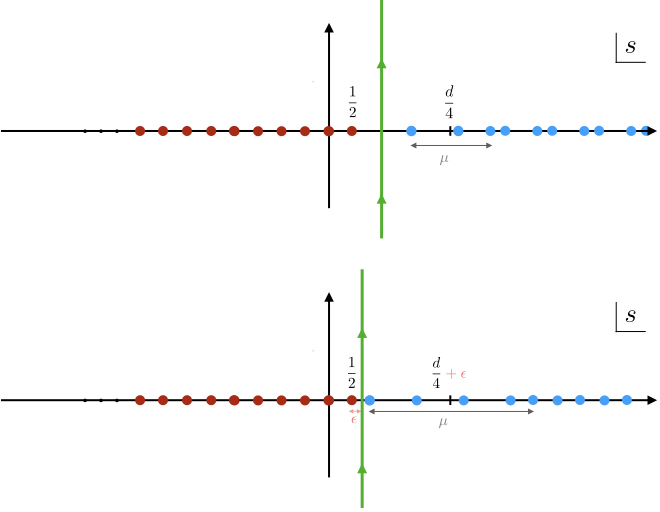}
    \caption{Poles of the Mellin representation for the conformal structure \eqref{ccsclnptconfstr} when the $n$-th scalar does not lie on the Principal Series, $\nu_n=i\mu$ with $\mu \in \mathbb{R}$. For ease of presentation, we display the $n=3$ case. As $\mu$ varies, the poles of $\Gamma\left(\frac{d}{4}\pm \frac{i\nu_n}{2}-s\right)$ (in blue) move along the real axis while the poles of $\Gamma\left(2s-1\right)$ (in red) remain fixed. When $\mu=\frac{d}{2}-1$ these poles collide and the contour separating them (in Green) becomes pinched. The pinching is regulated by sending $d \to d+\epsilon$, with $\epsilon >0$.}
    \label{fig:pinching}
\end{figure}

\paragraph{Contour pinching.} The contact diagrams considered in this section provide an interesting and moreover simple example of contour pinching discussed at the end of section \ref{subsec::3ptscalargen}. When the general scalar $\phi^{\left(\nu_n\right)}$ lies on the Principal Series \eqref{PS}, the integration contour in the Mellin-Barnes integral \eqref{ccsclnptconfstr2} is un-pinched since, for $\nu_n \in \mathbb{R}$, the poles of the $\Gamma$-function $\Gamma\left(2s-\tfrac{n-1}{2}\right)$:
\begin{equation}\label{nonnunpoles}
    s=\frac{n-1}{4}-\frac{m}{2}, \qquad m \in \mathbb{N}_0,
\end{equation}
are sharply separated from the poles of the $\Gamma$-functions $\Gamma\left(\tfrac{d(n-2)}{4}\pm \tfrac{i\nu_n}{2}-s\right)$:
\begin{equation}\label{nunpoles}
    s=\frac{d}{4}\pm\frac{i\nu_n}{2}+m, \qquad m \in \mathbb{N}_0,
\end{equation}
like in figure \ref{fig:3pts2}. When $\nu \to i \mu$ with $\mu \in \mathbb{R}$ however, for certain values of $\mu$ these poles overlap (see figure \ref{fig:pinching}). This occurs when
\begin{equation}
    \mu = \frac{d\left(n-2\right)}{2}-\frac{n-1}{2}.
\end{equation}
The contour pinching can be regulated by sending $d \to d + \epsilon$, which for $n>2$ gives an infinitesimal right-ward shift of the poles \eqref{nunpoles} with respect to the poles \eqref{nonnunpoles}. 

A simple example is the conformally coupled scalar, $\mu=\frac{1}{2}$, which pinches the integration contour when $d\left(n-2\right)-n=0$.\footnote{For example, when $d=3$ this requires $n=3$. For $d>3$ there is no pole pinching when $n=3$} With the above regularisation, the integral \eqref{ccsclnptconfstr} becomes:
\begin{multline}
    I^{\left(i/2,...,i/2,i/2\right)}_{\vec{k}_1,...,\vec{k}_n} =  \pi^{\tfrac{n-1}{2}} \frac{2^{d(n-4)+\frac{n+1}{2}}}{\sqrt{k_1...\,k_{n-1}}}\left(\frac{k_n}{2}\right)^{-\frac{1}{2}}  
    \left(2\frac{k_1+...+k_{n-1}}{k_n}\right)^{\tfrac{n-1}{2}}\\ \times \int^{i\infty}_{-i\infty} \frac{ds}{2\pi i}\, \Gamma\left(\epsilon+\tfrac{n}{4}-\tfrac{1}{4}-s\right)\Gamma\left(\epsilon+\tfrac{n}{4}+\tfrac{1}{4}-s\right)\Gamma\left(2s-\tfrac{n-1}{2}\right)\\ \times 
    \left(2\frac{k_1+...+k_{n-1}}{k_n}\right)^{-2s},
\end{multline}
where the regulator $\epsilon>0$ ensures that the integration contour is un-pinched. The integral can then be evaluated by closing the contour to the right of the imaginary axis, which encloses the poles \eqref{nonnunpoles}. Re-summing the residues gives: 
\begin{multline}\label{epsccsc3pt}
    I^{\left(i/2,...,i/2,i/2\right)}_{\vec{k}_1,...,\vec{k}_n}=\pi^{\tfrac{n-1}{2}} \frac{2^{d(n-4)+\frac{n}{2}+1}}{\sqrt{k_1...\,k_{n}}} \left[ \frac{\sqrt{\pi}}{2\epsilon}-\sqrt{\pi}\left(\gamma+\log\left(2\frac{k_1+...+k_n}{k_n}\right)\right)+O\left(\epsilon\right)\right].
\end{multline}
The simple pole at $\epsilon=0$ is equivalent to the pole of the $\Gamma$-function in \eqref{totalccs4pt} at ${d\left(n-2\right)-n=0}$. This is cancelled by the sine-factor, which arises upon combining the contributions from the $+$ and $-$ branches of the in-in contour (as in \eqref{nptscalargen}):
\begin{subequations}
\begin{align}
    \langle \phi^{(i/2)}_{\vec{k}_1}\phi^{(i/2)}_{\vec{k}_2}...\,\phi^{(i/2)}_{\vec{k}_{n-1}}\phi^{(i/2)}_{\vec{k}_n}\rangle^\prime&=\lim_{\epsilon \to 0}\left[L^{d+1}{\cal N}_{n}\left(\eta_0,k_i\right)\sin\left(\tfrac{\epsilon \pi }{2}\right)I^{\left(i/2,...,i/2,i/2\right)}_{\vec{k}_1,...,\vec{k}_n}\right]\\
    &=\pi^{\frac{n}{2}+1}L^{d+1}{\cal N}_{n}\left(\eta_0,k_i\right)\frac{2^{d(n-4)+\tfrac{n}{2}}}{\sqrt{k_1...\,k_{n}}},
\end{align}
\end{subequations}
which matches \eqref{totalccs4pt} when $d\left(n-2\right)-n=0$.

\subsection{Massless scalars}
\label{subsecc::mlesssc}

Another interesting simple example of contour pinching is the three-point function of massless scalars for $d=3$, for which $\nu_j=\frac{3i}{2}$. Keeping $d$ general for the moment and fixing $\nu_j=\frac{3i}{2}$, the Mellin-Barnes representation \eqref{2varrep3pt} of the three-point structure reads
\begin{align}
  \hspace*{-0.5cm}  I^{\left(\frac{3i}{2},\frac{3i}{2},\frac{3i}{2}\right)}_{\vec{k}_1,\vec{k_2},\vec{k}_3}&=2\sqrt{2} \pi ^{\frac{3}{2}}k^{-\frac{d}{2}}_3\int \left[ds\right]_2\, (4 s_1-1) (4 s_2-1)  (d-1-4 s_1-4 s_2)  \\ \nonumber
    & \hspace*{2cm}\times \Gamma \left(2s_1-\tfrac{3}{2}\right) \Gamma \left(2 s_2-\tfrac{3}{2}\right)\Gamma \left(\tfrac{d-3}{2}-2 s_1-2 s_2\right) \left(\frac{k_1}{k_3}\right)^{-2s_1}\left(\frac{k_2}{k_3}\right)^{-2s_2}\\ \nonumber
    \hspace*{-0.5cm}&= \sqrt{2} \pi^{\frac{3}{2}}\frac{\left(k_2+k_3\right)^{1-\frac{d}{2}}}{\left(k_2k_3\right)^{\frac{3}{2}}}  \int^{i\infty}_{-i\infty} \frac{ds_1}{2\pi i}(4 s_1-1) (d-4-4 s_1)  \left(k_2 k_3 (d-4 s_1-2)+2k^2_2+2 k^2_3\right)\\ \nonumber
    & \hspace*{6cm} \times  \Gamma \left(2s_1-\tfrac{3}{2}\right) \Gamma \left(\tfrac{d-6}{2}-2s_1\right)\left(\frac{k_1}{k_2+k_3}\right)^{-2 s_1}, 
\end{align}
where in the second equality we lifted the integral in $s_2$, closing the integration contour to the left of the imaginary axis. The poles in the remaining Mellin-variable $s_1$ are
\begin{subequations}
\begin{align}
    s_1&=\frac{3}{4}-\frac{n}{2}, \qquad n \in \mathbb{N}_0,\\
    s_1&=\frac{d-6}{4}+\frac{m}{2}, \qquad m \in \mathbb{N}_0,\label{epspoles3mlss}
\end{align}
\end{subequations}
which overlap for $d \leq 9$. To consider the correlator for $d=3$, we set $d \to 3+\epsilon$ to regulate the contour pinching as before. Closing the integration contour to the right, which encloses the poles \eqref{epspoles3mlss}, one obtains
\begin{multline}\label{mlessd3eps}
 \sin\left(\left(\tfrac{\epsilon}{2}-3\right)\tfrac{\pi}{2}\right)  I^{\left(\frac{3i}{2},\frac{3i}{2},\frac{3i}{2}\right)}_{\vec{k}_1,\vec{k_2},\vec{k}_3} = \frac{1}{\epsilon} \frac{8 \sqrt{2} \pi^{\frac{3}{2}} }{(k_1k_2k_3)^{\frac{3}{2}}}\sum\limits^3_{i=1}k^3_i\\-\frac{4\sqrt{2} \pi^{\frac{3}{2}}}{9 (k_1k_2k_3)^{\frac{3}{2}}} \left[ 3k_1k_2k_3 -3\sum_{i\ne j}k^2_ik_j+ \left(3\log k_t+3 \gamma -4\right)\sum^3_{i=1}k^3_i\right]+O\left(\epsilon\right),
\end{multline}
where we included the interference factor \eqref{total3ptdoublemellin} which arises from combining the contributions from the $+$ and $-$ branches of the in-in contour, and expanded in $\epsilon$. We see that, in contrast to the three-point function of conformally coupled scalars considered in the previous section, the pole at $\epsilon=0$ does not cancel upon combining all contributions along the in-in contour. In this case the correlator cannot be defined by analytic continuation in $d$. The pole in $\epsilon$ is proportional to the local term
\begin{equation}\label{ct3mless}
    \frac{A}{(k_1k_2k_3)^{\frac{3}{2}}}\sum\limits^3_{i=1}k^3_i,
\end{equation}
and so can be compensated by a local counter-term of this type. This is in accordance with the analyses \cite{Falk:1992sf,Zaldarriaga:2003my,Seery:2008qj,Creminelli:2011mw}, which each have the same non-divergent piece given by the second line of \eqref{mlessd3eps} up to a finite local term proportional to \eqref{ct3mless}.

\subsection{Kinematic limits of the momenta \`a la Mellin-Barnes}
\label{subsec::3ptlimits}

The Mellin-Barnes representation is a useful tool for studying kinematic limits of correlators in the phase space of momenta, which are encoded in the leading poles of certain Gamma functions in the Mellin integrand. This will be demonstrated in the following sections for various limits of interest: the soft momentum limit (section \ref{subsubsec::soft3pt}), collapsed triangle limits (section \ref{subsubsec3pt::Collapsed triangle limit}) and the high energy limit (section \ref{subsubsec::helimit}). We focus for ease of illustration on three-point functions, though the methodology applies in general. We shall also consider interesting limits of four-point exchange diagrams in section \ref{sec::exch}.

\subsubsection{Soft momentum limit}
\label{subsubsec::soft3pt}

Suppose we want to take the soft momentum limit, say $k_1 \rightarrow 0$, of the general three-point function \eqref{total3ptdoublemellin}. With the Mellin formalism, this is simply given by the residue of the first pole in the series: $s_1 = -\tfrac{i\nu_1}{2} - n$, $n=0,1,2,...\,$, which is encoded by one of the two Gamma functions in the Mellin representation \eqref{pmF} of the propagator for the field $\phi^{(\nu_1)}_{\vec{k}_1}$:\footnote{The second Gamma function in the Mellin representation \eqref{pmF} encodes the series of poles $s_1 = -\tfrac{i\nu_1}{2} - n$, $n \in \mathbb{N}_0$, which give purely local contributions (i.e. contributions analytic in $k_1$) that do not give rise to long distance correlations in position space.}
\begin{multline}
     \langle \phi^{(\nu_1)}_{\vec{k}_1}\phi^{(\nu_2)}_{\vec{k}_2}\phi^{(\nu_3)}_{\vec{k}_3}\rangle^\prime_{\pm}\Big|_{k_1\rightarrow 0,\, k_2 \sim k_3} = \pm i \frac{L^{d+1}}{2} {\cal N}_3\left(\eta_0,k_i\right) e^{\mp \left(\frac{d}{2}+i\left(\nu_1+\nu_2+\nu_3\right)\right) \frac{\pi i}{2}}    \\ \times \Gamma\left(-i\nu_1\right)\left(\tfrac{k_1}{2}\right)^{2i\nu_1}\left(\tfrac{k_3}{2}\right)^{-\frac{d}{2}+i\left(\nu_3+\nu_2-\nu_1\right)}\int^{i\infty}_{-i\infty}\frac{ds_2}{2\pi i}\,
     \Gamma\left(s_2+\tfrac{i\nu_2}{2}\right)\Gamma\left(s_2-\tfrac{i\nu_2}{2}\right) \\ \times \Gamma\left(\tfrac{d}{4}+\tfrac{i\nu_3}{2}+\tfrac{i\nu_1}{2}-s_2\right)\Gamma\left(\tfrac{d}{4}-\tfrac{i\nu_3}{2}+\tfrac{i\nu_1}{2}-s_2\right),
\end{multline}
where we recall that momentum conservation implies that $\lim_{k_1\rightarrow 0}\left(\tfrac{k_2}{k_3}\right)=1$. The remaining $s_2$-integral can be evaluated using Barnes' first lemma, which gives:
\begin{multline}
    \langle \phi^{(\nu_1)}_{\vec{k}_1}\phi^{(\nu_2)}_{\vec{k}_2}\phi^{(\nu_3)}_{\vec{k}_3}\rangle^\prime_{\pm}\Big|_{k_1\rightarrow 0,\, k_2 \sim k_3} = \pm i \frac{L^{d+1}}{2}  {\cal N}_3\left(\eta_0,k_i\right) e^{\mp \left(\frac{d}{2}+i\left(\nu_1+\nu_2+\nu_3\right)\right) \frac{\pi i}{2}}   \\\times \frac{\Gamma\left(-i\nu_1\right)}{\Gamma\left(\tfrac{d}{2}+i\nu_1\right)}\left(\tfrac{k_1}{2}\right)^{2i\nu_1}\left(\tfrac{k_3}{2}\right)^{-\frac{d}{2}+i\left(\nu_2+\nu_3-\nu_1\right)}\Gamma\left(\tfrac{d}{4}+\tfrac{i\left(\nu_2+\nu_3+\nu_1\right)}{2}\right)\Gamma\left(\tfrac{d}{4}+\tfrac{i\left(\nu_2-\nu_3+\nu_1\right)}{2}\right)\\ \times \Gamma\left(\tfrac{d}{4}+\tfrac{i\left(-\nu_2+\nu_3+\nu_1\right)}{2}\right)\Gamma\left(\tfrac{d}{4}+\tfrac{i\left(-\nu_2-\nu_3+\nu_1\right)}{2}\right).
\end{multline}
Note that if the field $\phi^{\left(\nu_1\right)}$ is exactly massless, $\nu_1=\frac{di}{2}$, the three-point function vanishes in the soft limit.

For higher-point functions (of any type) one proceeds in exactly the same way to obtain the soft-limit $k_1 \to 0$, taking the residue of the leading pole $s_1=-\frac{i\nu_1}{2}$ in the Mellin-Barnes representation of the corresponding bulk-to-boundary propagator that generates non-analytic contributions in $k_1$.

\subsubsection{Collapsed triangle limit}
\label{subsubsec3pt::Collapsed triangle limit}

\begin{figure}[h]
    \centering
    \captionsetup{width=0.95\textwidth}
    \includegraphics[scale=0.35]{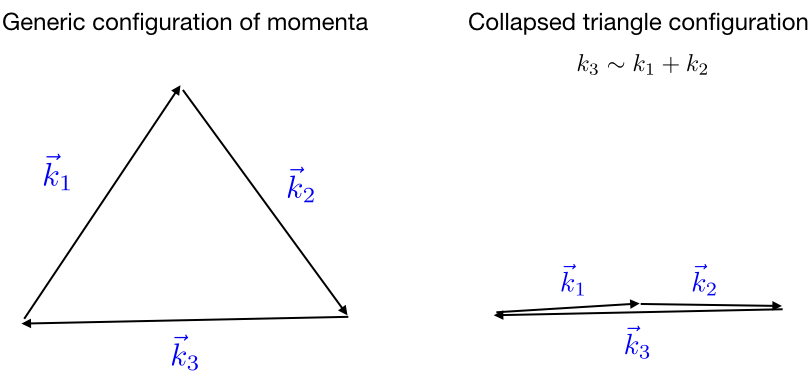}
    \caption{Generic configuration of momenta for a three-point function vs. the collinear configuration (collapsed triangle) $k_3 \sim k_1+k_2$, where the length of one side is the sum of the other two.}
    \label{fig:collaspedmom}
\end{figure}

Collapsed triangle configurations are those for which the momenta are collinear e.g. $k_3 = k_1+k_2$, see figure \ref{fig:collaspedmom}. The correlators should be non-singular in such configurations, which is related to the adiabatic vacuum condition \cite{Chen:2006nt,Holman:2007na,LopezNacir:2011kk,Flauger:2013hra,Aravind:2013lra}.

We can study the singularity structure of correlators at the level of the Mellin-Barnes integral \eqref{total3ptdoublemellin} by using the following trick. Parameterising the distance from the collapsed configuration by $\epsilon$, we replace $k_3=k_1+k_2+\epsilon$ and use the basic formula:
\begin{equation}\label{smirnform}
    \left(z+\epsilon\right)^{-\lambda} = \frac{1}{\Gamma\left(\lambda\right)} \int^{i\infty}_{-i\infty}\frac{du}{2\pi i} \Gamma\left(u+\lambda\right)\Gamma\left(-u\right)\frac{\epsilon^u}{z^{u+\lambda}},
\end{equation}
where in this example $z=k_1+k_2$. The series expansion of the correlator for $\epsilon<1$ is given by closing the integration contour to the right, where the leading contribution in the $\epsilon \rightarrow 0$ limit is given by the residue of the pole with smallest real part. This integration contour always includes the series of poles $u=0,1,2,...$ encoded the Gamma function $\Gamma\left(-u\right)$, which do not generate any singularities in $\epsilon$ since the exponent is non-negative on these poles. 

Further poles in $u$ are hidden in the integration over the Mellin variables in \eqref{total3ptdoublemellin}, the dependence on which enters \eqref{smirnform} through $\lambda = -2s_2$ where, for convenience, we shifted $s_2 \rightarrow s_2 - s_1$ so that the $s_1$ integral is disentangled from the $u$-integral. The location of these poles can actually be determined without having to evaluate any integral. They can be generated through either of the following two mechanisms (which are detailed in appendix \ref{appendix::polegen}):
\begin{enumerate}
    \item \emph{Poles at the values of $u$ for which series of poles in the other Mellin variables collide.}
    
    In the example under consideration, this occurs for:  
    \begin{align}\label{collisu}
        u = \pm i\nu_2 -n^\prime, \quad n^\prime \in \mathbb{N}_0,
    \end{align}
    i.e. those values for which the series of poles in the Gamma function $\Gamma\left(u+\lambda\right)$ overlap with the poles of either $\Gamma\left(s_2\pm\tfrac{i\nu_2}{2}\right)$ in the propagator \eqref{pmF}.
    
    The poles \eqref{collisu} are not enclosed by the $u$-integration contour in \eqref{smirnform} when we close it to the right (as required for $\epsilon <1$). This mechanism therefore does not generate any singularities in the correlator \eqref{total3ptdoublemellin} in the collapsed triangle limit $\epsilon \rightarrow 0$.
    
    \item \emph{Poles at the values of $u$ for which the other Mellin integrals diverge.}
    
    Using Stirling's formula, the asymptotic behaviour of the $s_2$ integral as $|\mathfrak{Im}\left[s_2\right]| \rightarrow \infty$ is
    \begin{equation}
       \sim e^{-2\pi |\mathfrak{Im}\left[s_2\right]|+\log (\left| \mathfrak{Im}\left[s_2\right]\right| ) \left(u-\mathfrak{Re}\left[s_1\right]+\tfrac{d}{2}-2\right)},
    \end{equation}
    which is exponentially suppressed for all values of $u$. This mechanism therefore also does not generate any singularities in the correlator \eqref{total3ptdoublemellin} in the collapsed triangle limit $\epsilon \rightarrow 0$, though it is responsible for singularities in the high energy limit -- as we shall see in the following section. 
\end{enumerate}

We have thus ruled out the existence of poles with $\mathfrak{Re}\left[u\right]<0$ when we close the $u$-integration contour in \eqref{smirnform} to the right, which confirms the regularity of the correlator \eqref{total3ptdoublemellin} in the collapsed triangle limit $\epsilon \rightarrow 0$.

While above we focused on the collapsed triangle configuration $k_3=k_1+k_2$, in exactly the same way we can verify there are no singularities in the other collinear momentum configurations, such as $k_1-k_2\mp k_3=0$.

\subsubsection{High energy limit $k_t = k_1+k_2+k_3\rightarrow 0$}
\label{subsubsec::helimit}

Using the same trick we can study the singularity of the correlator \eqref{total3ptdoublemellin} in the limit $k_t \equiv k_1+k_2+k_3 \rightarrow 0$. The coefficient of this singularity is related to the high-energy limit of the flat space amplitude \cite{Maldacena:2011nz,Raju:2012zr,Raju:2012zs,Arkani-Hamed:2015bza}.

As for the collapsed triangle limit in the previous section, we replace $k_3=z+k_t$ and use the formula \eqref{smirnform} to parameterise the exponent of $k_t$ with the Mellin variable $u$. The poles which lie to the right of the $u$-integration contour encode the singularity structure as $k_t \rightarrow 0$. The only difference with respect to the collapsed triangle limit of the previous section is that the parameter $z$ now takes negative values: $z=-\left(k_1+k_2\right)$. This does not change the series of poles \eqref{collisu} generated by the collision of poles in the Mellin variable $s_2$. It does, however, change the asymptotic behaviour of the $s_2$ integral as $|\mathfrak{Im}\left[s_2\right]| \rightarrow \infty$, which now goes as:
\begin{equation}
 \sim  e^{-2\pi\left(|\mathfrak{Im}\left[s_2\right]|+\mathfrak{Im}\left[s_2\right]\right)+\log (\left| \mathfrak{Im}\left[s_2\right]\right| ) \left(u-\mathfrak{Re}\left[s_1\right]+\tfrac{d}{2}-2\right)}.
\end{equation}
We see that there is a possible divergence when $\mathfrak{Im}\left[s_2\right]<0$, where $\left(|\mathfrak{Im}\left[s_2\right]|+\mathfrak{Im}\left[s_2\right]\right)=0$, for which the integrand is no longer exponentially suppressed. The leading term is now sensitive to the Mellin integral in $s_1$. For simplicity let's suppose the scalar associated to $s_1$ is conformally coupled. In this case the $s_1$ Mellin-Barnes integral can be lifted (see section \ref{subsec::3pttwocc}), and the asymptotic behaviour of the $s_2$ integrand for $\mathfrak{Im}\left[s_2\right] \rightarrow -\infty$ is:
\begin{equation}
    \sim  e^{\log (\left| \mathfrak{Im}\left[s_2\right]\right| ) \left(u+\tfrac{d-5}{2}\right)}.
\end{equation}
The integral thus diverges for $u+\tfrac{d-5}{2} \geq -1$, signalling the presence of a series of poles at (see appendix \ref{subsec::appgenhypf}):
\begin{equation}\label{asymu}
    u=\tfrac{3-d}{2}+n, \quad n \in \mathbb{N}_0.
\end{equation}
These poles lie to the right of the $u$-integration contour and so contribute to the series expansion of the correlator in $k_t <1$ together with the poles at
\begin{equation}\label{gammup}
    u=n^\prime, \quad n^\prime \in \mathbb{N}_0,
\end{equation}
which are encoded by the Gamma function $\Gamma\left(-u\right)$ in the formula \eqref{smirnform}.  The leading contribution in the limit $k_t \rightarrow 0$ is therefore given by the residue at:
\begin{subequations}
\begin{align}
    u &= \tfrac{3-d}{2}<0 \hspace*{0.5cm} \text{if} \hspace*{0.25cm} d>3, \\
    u&= 0 \hspace*{1.6cm} \text{if} \hspace*{0.25cm} d \leq 3.
\end{align}
\end{subequations}
For $d>3$ the correlator thus has a singularity of the form: 
\begin{equation}
    \langle \phi^{(i/2)}_{\vec{k}_1}\phi^{(\nu_2)}_{\vec{k}_2}\phi^{(\nu_3)}_{\vec{k}_3}\rangle^\prime_{\pm}\Big|_{k_t\rightarrow 0} \sim k^{\frac{3-d}{2}}_{t}.
\end{equation}
For $d=3$ the pole at $u=0$ is actually double pole because when $d=3$ the series \eqref{asymu} and \eqref{gammup} coincide, so in this case the correlator has a logarithmic singularity:
\begin{equation}
     \langle \phi^{(i/2)}_{\vec{k}_1}\phi^{(\nu_2)}_{\vec{k}_2}\phi^{(\nu_3)}_{\vec{k}_3}\rangle^\prime_{\pm}\Big|_{d=3,\, k_t\rightarrow 0} \sim \log k_{t}.
\end{equation}

As we saw in section \ref{subsec::3pttwocc}, when the external scalars are all conformally coupled, every Mellin-Barnes integral can be lifted from the correlator -- as in \eqref{totalccs4pt}. In this case the singular behaviour as $k_t \rightarrow 0$ is manifest and agrees with the above.

\section{Exchange diagrams}

\label{sec::exch}

Having considered contact diagrams in the previous section, in this section we apply the Mellin formalism to tree-level exchange four-point functions in dS$_{d+1}$. In section \ref{subsec::genresult4pt} we derive the Mellin-Barnes representation for an exchange diagram involving general internal and external scalars, which takes the form of an integrated product of the Mellin-Barnes representations for the corresponding three-point contact diagrams. In the subsequent sections we discuss various features of the Mellin-Barnes representations, including how it encodes the Operator Product Expansion in section \ref{subsec::opelimit} and the Effective Field Theory expansion in section \ref{subsec::seriesexpfrommellin}.

\subsection{General external and internal scalars}
\label{subsec::genresult4pt}

In the following we consider the tree-level exchange of a general scalar $\phi^{\left(\nu\right)}$ between two pairs of scalars $\phi^{\left(\nu_j\right)}$ which is generated by the zero-derivative cubic vertices considered in section \ref{subsec::3ptscalargen}. The contributions from the different branches of the in-in contour in this case are:
\begin{subequations}\label{4pt0000exch}
\begin{align}
\langle \phi^{(\nu_1)}_{\vec{k}_1}\phi^{(\nu_2)}_{\vec{k}_2}\phi^{(\nu_3)}_{\vec{k}_3}\phi^{(\nu_4)}_{\vec{k}_4} \rangle=&\left(2\pi\right)^d \delta^{(d)}\left(\vec{k}_1+\vec{k_2}+\vec{k}_3+\vec{k}_4\right)\langle\phi^{(\nu_1)}_{\vec{k}_1}\phi^{(\nu_2)}_{\vec{k}_2}\phi^{(\nu_3)}_{\vec{k}_3}\phi^{(\nu_4)}_{\vec{k}_4} \rangle^\prime,\\
     \langle \phi^{(\nu_1)}_{\vec{k}_1}\phi^{(\nu_2)}_{\vec{k}_2}\phi^{(\nu_3)}_{\vec{k}_3}\phi^{(\nu_4)}_{\vec{k}_4} \rangle^\prime =& \sum_{\pm {\hat \pm}}\langle \phi^{(\nu_1)}_{\vec{k}_1}\phi^{(\nu_2)}_{\vec{k}_2}\phi^{(\nu_3)}_{\vec{k}_3}\phi^{(\nu_4)}_{\vec{k}_4} \rangle^\prime_{\pm {\hat \pm}},
\end{align}
\end{subequations}
where each branch receives a contribution from the ${\sf s}$-, ${\sf t}$ and ${\sf u}$-exchange channels, which we denote by:\footnote{We employ the sans-serif notation ${\sf s}$, ${\sf t}$ and ${\sf u}$ to denote the three \emph{exchange channels}, which are not to be confused with the \emph{Mellin variables} $s$, $t$ and $u$.}
\begin{align}
    \langle \phi^{(\nu_1)}_{\vec{k}_1}\phi^{(\nu_2)}_{\vec{k}_2}\phi^{(\nu_3)}_{\vec{k}_3}\phi^{(\nu_4)}_{\vec{k}_4} \rangle^\prime_{\pm {\hat \pm}}= \sum_{\alpha={\sf s},{\sf t},{\sf u}}{}^{(\alpha)}\langle \phi^{(\nu_1)}_{\vec{k}_1}\phi^{(\nu_2)}_{\vec{k}_2}\phi^{(\nu_3)}_{\vec{k}_3}\phi^{(\nu_4)}_{\vec{k}_4} \rangle^\prime_{\pm {\hat \pm}}.
\end{align}

\begin{figure}[t]
    \centering
    \captionsetup{width=0.95\textwidth}
    \includegraphics[scale=0.4]{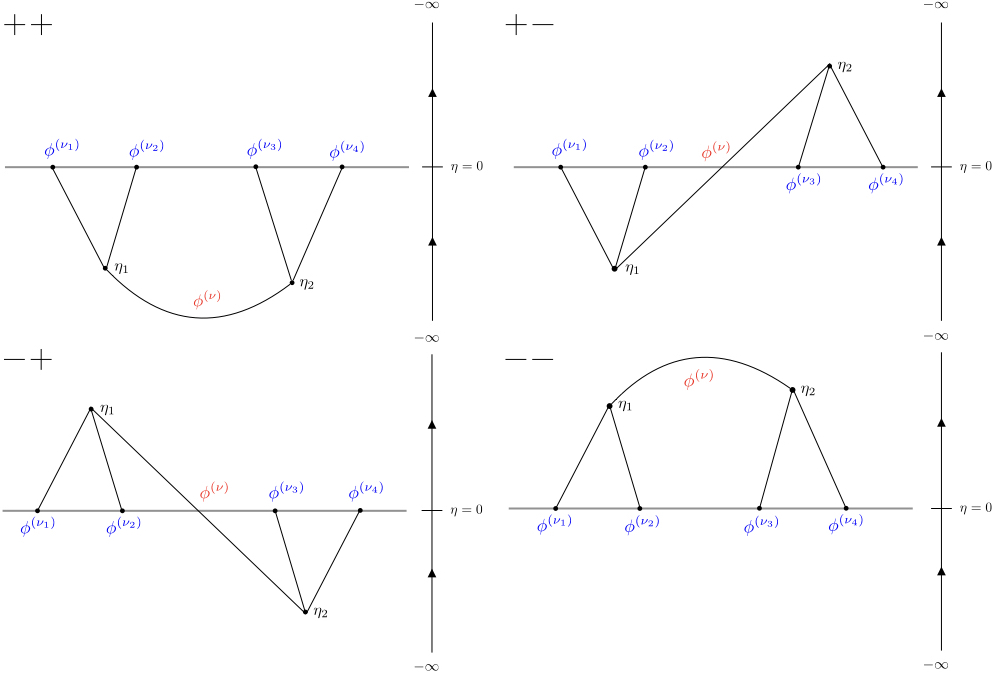}
    \caption{Tree exchange diagrams of a general scalar $\phi^{\left(\nu\right)}$ contributing to the correlation function $\langle \phi^{(\nu_1)}\phi^{(\nu_2)}\phi^{(\nu_3)}\phi^{(\nu_4)} \rangle$ at late times $\eta_0 \sim 0$. The $+-$ and $-+$ contributions (top right and bottom left) factorise into a product of two late-time three-point contact diagrams, one from the $+$ branch of the in-in contour and the other from the $-$ branch.}
    \label{fig:exchcontribs}
\end{figure}

In the following we shall focus without loss of generality on the ${\sf s}$-channel exchange, where (see figure \ref{fig:exchcontribs})
\begin{multline}
    {}^{({\sf s})}\langle \phi^{(\nu_1)}_{\vec{k}_1}\phi^{(\nu_2)}_{\vec{k}_2}\phi^{(\nu_3)}_{\vec{k}_3}\phi^{(\nu_4)}_{\vec{k}_4} \rangle^\prime_{\pm {\hat \pm}}=\left(\pm i\right)\left({\hat \pm} i\right)  \int^{\eta_0}_{-\infty} \frac{d\eta_1}{\left(-\eta_1/L\right)^{d+1}} 
   \int^{\eta_0}_{-\infty} \frac{d\eta_2}{\left(-\eta_2/L\right)^{d+1}}
   \\ \times   F^{\left(\nu_1\right)}_{\pm,\vec{k}_1}(\eta_1;\eta_0)F^{\left(\nu_2\right)}_{\pm,\vec{k}_2}(\eta_1;\eta_0)G_{\pm {\hat \pm},\vec{k}_I}\left(\eta_1;\eta_2\right) F^{\left(\nu_3\right)}_{{\hat \pm},\vec{k}_3}(\eta_2;\eta_0)F^{\left(\nu_4\right)}_{{\hat \pm},\vec{k}_4}(\eta_2;\eta_0),
\end{multline}
 and $\vec{k}_I=\vec{k}_1+\vec{k}_2=-\vec{k}_3-\vec{k}_4$ is the exchanged momentum.\footnote{Note that for the ${\sf t}$- and ${\sf u}$-channel exchanges we have instead $\vec{k}_I=\vec{k}_1+\vec{k}_4$ and $\vec{k}_I=\vec{k}_1+\vec{k}_3$, respectively.} The Keldysh bulk-to-bulk propagators $G_{\pm {\hat \pm},\vec{k}_I}$ were defined in equation \eqref{SKprop} and their Mellin-Barnes representation is given by \eqref{MBwightman}. In the following we discuss the contributions from the different branches of the in-in contour, before combining them together in equation \eqref{total4ptsch}.

\paragraph{$+-$ and $-+$ contributions.} These are the simplest contributions since the Keldysh bulk-to-bulk propagators factorise into a product of a time-ordered and anti-time-ordered bulk-to-boundary propagators (see section \ref{subsec::mellinfourierspace}) so that, in turn, the $+-$ and $-+$ contributions factorise into a product of $+$ and $-$ contributions to late-time three-point functions -- which we evaluated in section \ref{subsec::3ptscalargen}. See figure \ref{fig:exchcontribs} for a diagrammatic understanding of this property. In particular, the corresponding Keldysh propagator can be expressed as
\begin{equation}
    G_{\pm \mp,\vec{k}_I}\left(\eta_1,\eta_2\right)=\frac{1}{4\pi} \left(\frac{k}{2}\right)^{-2i\nu}\left({\cal N}_\nu\left(\eta_0\right)\right)^{-2}F^{(-\nu)}_{\pm,\vec{k}_I}\left(\eta_1;\eta_0\right)F^{(\nu)}_{\mp,-\vec{k}_I}\left(\eta_2;\eta_0\right),
\end{equation}
from which it immediately follows 
\begin{subequations}\label{pmmpfactor}
\begin{align}
 \hspace*{-0.75cm} {}^{({\sf s})}\langle \phi^{(\nu_1)}_{\vec{k}_1}\phi^{(\nu_2)}_{\vec{k}_2}\phi^{(\nu_3)}_{\vec{k}_3}\phi^{(\nu_4)}_{\vec{k}_4} \rangle^\prime_{\pm \mp} &=  \frac{1}{4\pi}\frac{{\cal N}_4\left(\eta_0,k_i\right)}{{\cal N}_\nu\left(\eta_0\right)^{2}}  \left(\frac{k}{2}\right)^{-2i\nu} \langle \phi^{(\nu_1)}_{\vec{k}_1}\phi^{(\nu_2)}_{\vec{k}_2}\phi^{(\nu)}_{\vec{k}_I}\rangle^\prime_{\pm} \langle \phi^{(-\nu)}_{-\vec{k}_I} \phi^{(\nu_3)}_{\vec{k}_3}\phi^{(\nu_4)}_{\vec{k}_4}\rangle^\prime_{\mp} \\
\hspace*{-0.41cm} &=\frac{L^{2(d+1)}}{16\pi}{\cal N}_4\left(\eta_0,k_i\right)e^{\pm \left(\nu_1+\nu_2\right)\frac{\pi}{2}} e^{\mp \left(\nu_3+\nu_4\right)\tfrac{\pi}{2}}  I^{\left(\nu_1,\nu_2,\nu\right)}_{\vec{k}_1,\vec{k}_2,\vec{k}_I}I^{\left(-\nu,\nu_3,\nu_4\right)}_{-\vec{k}_I,\vec{k}_3,\vec{k}_4}, 
\end{align}
\end{subequations}
where in the second equality we inserted the expression \eqref{pm3ptcontrib} for the three-point factors. Combining the two contributions gives:
\begin{multline}\label{+--+MBfin}
    {}^{({\sf s})}\langle \phi^{(\nu_1)}_{\vec{k}_1}\phi^{(\nu_2)}_{\vec{k}_2}\phi^{(\nu_3)}_{\vec{k}_3}\phi^{(\nu_4)}_{\vec{k}_4} \rangle^\prime_{+-}+{}^{({\sf s})}\langle \phi^{(\nu_1)}_{\vec{k}_1}\phi^{(\nu_2)}_{\vec{k}_2}\phi^{(\nu_3)}_{\vec{k}_3}\phi^{(\nu_4)}_{\vec{k}_4} \rangle^\prime_{-+}\\ = \frac{L^{2(d+1)}}{8\pi} {\cal N}_4\left(\eta_0,k_i\right)\cos\left(\left(\nu_1+\nu_2-\nu_3-\nu_4\right)\tfrac{\pi i}{2}\right) I^{\left(\nu_1,\nu_2,\nu\right)}_{\vec{k}_1,\vec{k}_2,\vec{k}_I}I^{\left(-\nu,\nu_3,\nu_4\right)}_{-\vec{k}_I,\vec{k}_3,\vec{k}_4},
\end{multline}
where the cosine function encodes the interference between the two processes.

\paragraph{$++$ and $--$ contributions.} In contrast, the $++$ and $--$ contributions are not factorised and contain contact terms arising from the collision of the two bulk points.\footnote{This is not possible for the $+-$ and $-+$ contributions, where the two bulk points lie on different branches of the in-in contour.} This is manifest from the $\theta$-function terms in the corresponding Keldysh propagators \eqref{SKprop}. Each contribution consists of two terms, one for $\eta_2 <\eta_1$ and the other for $\eta_2 > \eta_1$: 
\begin{multline}
\nonumber
     {}^{({\sf s})}\langle \phi^{(\nu_1)}_{\vec{k}_1}\phi^{(\nu_2)}_{\vec{k}_2}\phi^{(\nu_3)}_{\vec{k}_3}\phi^{(\nu_4)}_{\vec{k}_4} \rangle^\prime_{\pm \pm}={}^{({\sf s})}\langle \phi^{(\nu_1)}_{\vec{k}_1}\phi^{(\nu_2)}_{\vec{k}_2}\phi^{(\nu_3)}_{\vec{k}_3}\phi^{(\nu_4)}_{\vec{k}_4} \rangle^\prime_{\pm \pm,<}+{}^{({\sf s})}\langle \phi^{(\nu_1)}_{\vec{k}_1}\phi^{(\nu_2)}_{\vec{k}_2}\phi^{(\nu_3)}_{\vec{k}_3}\phi^{(\nu_4)}_{\vec{k}_4} \rangle^\prime_{\pm \pm,>},
\end{multline}
which read
\begin{subequations}\label{++<>}
\begin{align}
   \hspace*{-0.5cm}   {}^{({\sf s})}\langle \phi^{(\nu_1)}_{\vec{k}_1}\phi^{(\nu_2)}_{\vec{k}_2}\phi^{(\nu_3)}_{\vec{k}_3}\phi^{(\nu_4)}_{\vec{k}_4} \rangle^\prime_{++,<}&=\left(+ i\right)\left(+ i\right)  \int^{\eta_0}_{-\infty} \frac{d\eta_1d\eta_2}{\left(-\eta_1/L\right)^{d+1}\left(-\eta_2/L\right)^{d+1}} \theta\left(\eta_1-\eta_2\right)
  \nonumber \\  & \hspace*{-2cm} \times   F^{\left(\nu_1\right)}_{+,\vec{k}_1}(\eta_1;\eta_0)F^{\left(\nu_2\right)}_{+,\vec{k}_2}(\eta_1;\eta_0)G_{-+,\vec{k}_I}\left(\eta_1;\eta_2\right) F^{\left(\nu_3\right)}_{+,\vec{k}_3}(\eta_2;\eta_0)F^{\left(\nu_4\right)}_{+,\vec{k}_4}(\eta_2;\eta_0), \\
   \hspace*{-0.5cm}   {}^{({\sf s})}\langle \phi^{(\nu_1)}_{\vec{k}_1}\phi^{(\nu_2)}_{\vec{k}_2}\phi^{(\nu_3)}_{\vec{k}_3}\phi^{(\nu_4)}_{\vec{k}_4} \rangle^\prime_{++,>}&=\left(+ i\right)\left(+ i\right)  \int^{\eta_0}_{-\infty} \frac{d\eta_1d\eta_2}{\left(-\eta_1/L\right)^{d+1}\left(-\eta_2/L\right)^{d+1}} \theta\left(\eta_2-\eta_1\right)
  \nonumber \\  & \hspace*{-2cm} \times   F^{\left(\nu_1\right)}_{+,\vec{k}_1}(\eta_1;\eta_0)F^{\left(\nu_2\right)}_{+,\vec{k}_2}(\eta_1;\eta_0)G_{+-,\vec{k}_I}\left(\eta_2;\eta_1\right) F^{\left(\nu_3\right)}_{+,\vec{k}_3}(\eta_2;\eta_0)F^{\left(\nu_4\right)}_{+,\vec{k}_4}(\eta_2;\eta_0),
\end{align}
\end{subequations}
and
\begin{subequations}\label{--<>}
\begin{align}
  \hspace*{-0.5cm}  {}^{({\sf s})}\langle \phi^{(\nu_1)}_{\vec{k}_1}\phi^{(\nu_2)}_{\vec{k}_2}\phi^{(\nu_3)}_{\vec{k}_3}\phi^{(\nu_4)}_{\vec{k}_4} \rangle^\prime_{--,<}&=\left(- i\right)\left(- i\right)  \int^{\eta_0}_{-\infty} \frac{d\eta_1d\eta_2}{\left(-\eta_1/L\right)^{d+1}\left(-\eta_2/L\right)^{d+1}} \theta\left(\eta_1-\eta_2\right)
  \nonumber \\  & \hspace*{-2cm} \times   F^{\left(\nu_1\right)}_{-,\vec{k}_1}(\eta_1;\eta_0)F^{\left(\nu_2\right)}_{-,\vec{k}_2}(\eta_1;\eta_0)G_{+-,\vec{k}_I}\left(\eta_2;\eta_1\right) F^{\left(\nu_3\right)}_{-,\vec{k}_3}(\eta_2;\eta_0)F^{\left(\nu_4\right)}_{-,\vec{k}_4}(\eta_2;\eta_0), \\
   \hspace*{-0.5cm}   {}^{({\sf s})}\langle \phi^{(\nu_1)}_{\vec{k}_1}\phi^{(\nu_2)}_{\vec{k}_2}\phi^{(\nu_3)}_{\vec{k}_3}\phi^{(\nu_4)}_{\vec{k}_4} \rangle^\prime_{--,>}&=\left(- i\right)\left(- i\right)  \int^{\eta_0}_{-\infty} \frac{d\eta_1d\eta_2}{\left(-\eta_1/L\right)^{d+1}\left(-\eta_2/L\right)^{d+1}} \theta\left(\eta_2-\eta_1\right)
  \nonumber \\  & \hspace*{-2cm} \times   F^{\left(\nu_1\right)}_{-,\vec{k}_1}(\eta_1;\eta_0)F^{\left(\nu_2\right)}_{-,\vec{k}_2}(\eta_1;\eta_0)G_{-+,\vec{k}_I}\left(\eta_1;\eta_2\right) F^{\left(\nu_3\right)}_{-,\vec{k}_3}(\eta_2;\eta_0)F^{\left(\nu_4\right)}_{-,\vec{k}_4}(\eta_2;\eta_0).
\end{align}
\end{subequations}
In this case, the Mellin-Barnes representation reduces the conformal time integrals to the basic form:
\begin{subequations}\label{++--CTints}
\begin{align}
    \int^{\eta_0}_{-\infty} d\eta_2 d\eta_1\, \theta\left(\eta_1-\eta_2\right) \left(-\eta_1\right)^{\alpha}\left(-\eta_2\right)^{\beta} &= \frac{\left(-\eta_0\right)^{\alpha+\beta+2}}{\left(\beta+1\right)\left(\alpha+\beta+2\right)},\\
    \int^{\eta_0}_{-\infty} d\eta_2 d\eta_1\, \theta\left(\eta_2-\eta_1\right) \left(-\eta_1\right)^{\alpha}\left(-\eta_2\right)^{\beta} &= \frac{\left(-\eta_0\right)^{\alpha+\beta+2}}{\left(\alpha+1\right)\left(\alpha+\beta+2\right)},
\end{align}
\end{subequations}
where, for both contributions \eqref{++<>} and \eqref{--<>},
\begin{align}
    \alpha=\tfrac{d}{2}-1-2\left(s_1+s_2+u_1\right), \qquad    \beta=\tfrac{d}{2}-1-2\left(s_3+s_4+u_2\right),
\end{align}
where the Mellin variables $s_j$ are assigned to bulk-to-boundary propagators \eqref{pmF} for the external fields $\phi^{\left(\nu_j\right)}_{\vec{k}_j}$ and the $u_j$ to the Wightman function \eqref{MBwightman} for the internal field. For convergence of the integrals \eqref{++--CTints} we impose the following constraints on the Mellin-Barnes contours: 
\begin{equation}\label{--++contourconds}
    \mathfrak{Re}\left[\tfrac{d}{2}-2\left(s_1+s_2+u_1\right)\right]<0, \qquad \mathfrak{Re}\left[\tfrac{d}{2}-2\left(s_3+s_4+u_2\right)\right]<0. 
\end{equation}
From \eqref{++--CTints} we see that the late-time limit $\eta_0 \to 0$ is encoded in the residue of the pole at:
\begin{equation}
    \alpha+\beta+2=d-2\left(s_1+s_2+s_3+s_4+u_1+u_2\right) \sim 0,
\end{equation}
where the restrictions \eqref{--++contourconds} require the contour to intersect the real axis to the right of this pole, meaning that we must close the contour to the left of the imaginary axis. At late times this gives the contributions \eqref{++<>} and \eqref{--<>} as an integrated product of three-point structures:
\begin{subequations}\label{uintppmm<>}
\begin{multline}
   {}^{({\sf s})}\langle \phi^{(\nu_1)}_{\vec{k}_1}\phi^{(\nu_2)}_{\vec{k}_2}\phi^{(\nu_3)}_{\vec{k}_3}\phi^{(\nu_4)}_{\vec{k}_4} \rangle^\prime_{\pm \pm,<}= -e^{\pm \left(\nu_1+\nu_2+\nu_3+\nu_4\right)\frac{\pi}{2}}\frac{L^{2\left(d+1\right)} }{16\pi} {\cal N}_4\left(\eta_0,k_i\right) \\ \times \int^{i\infty}_{-i\infty} \frac{du}{2\pi i} \frac{e^{\mp 2u\pi i}}{u+\epsilon} I^{\left(\nu_1,\nu_2,\nu\right)}_{\vec{k}_1e^{\pm \pi i},\vec{k}_2e^{\pm \pi i},\vec{k}_I}\left(\tcr{+}u\right) I^{\left(\nu_3,\nu_4,\nu\right)}_{\vec{k}_3,\vec{k}_4,\vec{k}_I}\left(\tcb{-}u\right),
\end{multline}
\begin{multline}
   {}^{({\sf s})}\langle \phi^{(\nu_1)}_{\vec{k}_1}\phi^{(\nu_2)}_{\vec{k}_2}\phi^{(\nu_3)}_{\vec{k}_3}\phi^{(\nu_4)}_{\vec{k}_4} \rangle^\prime_{\pm \pm,>} = -e^{\pm \left(\nu_1+\nu_2+\nu_3+\nu_4\right)\frac{\pi}{2}}\frac{L^{2\left(d+1\right)} }{16\pi} {\cal N}_4\left(\eta_0,k_i\right)\\
    \times \int^{i\infty}_{-i\infty} \frac{du}{2\pi i} \frac{e^{\mp 2u\pi i}}{u+\epsilon} I^{\left(\nu_1,\nu_2,\nu\right)}_{\vec{k}_1,\vec{k}_2,\vec{k}_I}\left(\tcb{-}u\right)I^{\left(\nu_3,\nu_4,\nu\right)}_{\vec{k}_3e^{\pm \pi i},\vec{k}_4e^{\pm \pi i},\vec{k}_I}\left(\tcr{+}u\right),
\end{multline}
\end{subequations}
where the restrictions \eqref{--++contourconds} translate into $\mathfrak{Re}\left[u\right]>0$, which we reflected in the $\epsilon$ prescription. The three-point structures are given by
\begin{multline}\label{3ptu}
   I^{\left(\nu_1,\nu_2,\nu_3\right)}_{\vec{k}_1,\vec{k}_2,\vec{k}_3}\left(u\right)= \int \left[ds\right]_2 \,\rho_{\nu_1,\nu_2,\nu_3}\left(s_1,s_2,w-u\right) \\ \times \left(\frac{k_3}{2}\right)^{-2w}\prod^2_{j=1}\left(\frac{k_j}{2}\right)^{-2s_j}\Big|_{w=\tfrac{d}{4}-s_1-s_2},
\end{multline}
which for $u=0$ coincide with \eqref{pm3ptcontrib}:
\begin{equation}
    I^{\left(\nu_1,\nu_2,\nu_3\right)}_{\vec{k}_1,\vec{k}_2,\vec{k}_3}\left(u=0\right)=I^{\left(\nu_1,\nu_2,\nu_3\right)}_{\vec{k}_1,\vec{k}_2,\vec{k}_3}.
\end{equation}
The $u=0$ poles displayed in \eqref{uintppmm<>} thus generate factorised contributions to the exchange diagram.\footnote{This makes manifest that the factorised contributions can be obtained from the $+-$ and $-+$ contributions \eqref{pmmpfactor} to the exchange four-point function through the analytic continuation: 
\begin{subequations}
\begin{align}
 {}^{({\sf s})}\langle \phi^{(\nu_1)}_{\vec{k}_1}\phi^{(\nu_2)}_{\vec{k}_2}\phi^{(\nu_3)}_{\vec{k}_3}\phi^{(\nu_4)}_{\vec{k}_4} \rangle^\prime_{\pm \pm,<}\Big|_{\text{factorised}}&=-e^{\mp\left(\nu_1+\nu_2+\nu_3+\nu_4\right)}{}^{({\sf s})}\langle \phi^{(\nu_1)}_{\vec{k}_1e^{\pm \pi i}}\phi^{(\nu_2)}_{\vec{k}_2e^{\pm \pi i}}\phi^{(\nu_3)}_{\vec{k}_3}\phi^{(\nu_4)}_{\vec{k}_4} \rangle^\prime_{\pm \mp},\\
   {}^{({\sf s})}\langle \phi^{(\nu_1)}_{\vec{k}_1}\phi^{(\nu_2)}_{\vec{k}_2}\phi^{(\nu_3)}_{\vec{k}_3}\phi^{(\nu_4)}_{\vec{k}_4} \rangle^\prime_{\pm \pm,>}\Big|_{\text{factorised}}&=-e^{\mp\left(\nu_1+\nu_2+\nu_3+\nu_4\right)}{}^{({\sf s})}\langle \phi^{(\nu_1)}_{\vec{k}_1}\phi^{(\nu_2)}_{\vec{k}_2}\phi^{(\nu_3)}_{\vec{k}_3e^{\pm \pi i}}\phi^{(\nu_4)}_{\vec{k}_4e^{\pm \pi i}} \rangle^\prime_{\mp \pm},
\end{align}
\end{subequations}
where we recall that the constant ${\cal N}_4\left(\eta_0,k_i\right)$ depends on the external momenta. This feature was exhibited in \cite{Arkani-Hamed:2015bza} for external conformally coupled scalars in $d=3$.} The organisation of the different contributions to the exchange diagram become more transparent upon lifting the $u$-integral, the details for which we give in appendix \ref{appsubsec::exchdia}. The resulting expression for the total $++$ and $--$ contributions to the ${\sf s}$-channel exchange is: 
\begin{multline}\label{++--MBfin}
    {}^{({\sf s})}\langle \phi^{(\nu_1)}_{\vec{k}_1}\phi^{(\nu_2)}_{\vec{k}_2}\phi^{(\nu_3)}_{\vec{k}_3}\phi^{(\nu_4)}_{\vec{k}_4} \rangle^\prime_{++}+{}^{({\sf s})}\langle \phi^{(\nu_1)}_{\vec{k}_1}\phi^{(\nu_2)}_{\vec{k}_2}\phi^{(\nu_3)}_{\vec{k}_3}\phi^{(\nu_4)}_{\vec{k}_4} \rangle^\prime_{--}\\ = \frac{L^{2\left(d+1\right)}}{8\pi} {\cal N}_4\left(\eta_0,k_i\right) \int \left[ds\right]_4 \,
     \text{cosec} \left(\pi(w+{\bar w})\right) \delta_{++\,--}\left(w,{\bar w}\right)\\ \times I^{\left(\nu_1,\nu_2,\nu\right)}_{\vec{k}_1,\vec{k}_2,\vec{k}_I}\left(s_1,s_2,w\right)I^{\left(-\nu,\nu_3,\nu_4\right)}_{-\vec{k}_I,\vec{k}_3,\vec{k}_4}\left(s_3,s_4,{\bar w}\right)\Big|_{{}^{w=\frac{d}{4}-s_1-s_2}_{{\bar w}=\frac{d}{4}-s_3-s_4}},
\end{multline}
in terms of the Mellin-Barnes representation for the three-point conformal structure \eqref{3ptconfstr}, and the function 
\begin{multline}
     \delta_{++\,--}\left(w,{\bar w}\right)=\cosh\left(\pi \nu\right)\sin \left(\pi\left(\tfrac{i\left(\nu_1+\nu_2+\nu_3+\nu_4\right)}{2}+\tfrac{d}{2}-w-{\bar w}\right)\right)\\
     -\frac{1}{2} \left[ \sin \left(\pi  (\tfrac{d}{2}+\tfrac{i\left(\nu_1+\nu_2+\nu_3+\nu_4\right)}{2}+w-{\bar w})\right)+ w \leftrightarrow {\bar w}\right],
\end{multline}
gives the zeros of the Mellin integrand, encoding the interference between the different processes. Comparing with the expression \eqref{pmmpfactor} for the  $+-$ and $-+$ contributions, we see that the cosecant factor is responsible for the non-factorisation of the $++$ and $--$ contributions.

We shall discuss the properties of the expression \eqref{+--+MBfin} below, after combining it with the $+-$ and $-+$ contributions to obtain the final result for the exchange diagram.

\paragraph{Combined contributions.} The expression \eqref{+--+MBfin} for the $+-$ and $-+$ contributions can be re-expressed in the same form as \eqref{++--MBfin} simply using equation \eqref{3ptconfstr}. The interference factor in this case reads:
\begin{align}\label{IFpmmp}
     \delta_{+-\,-+}\left(w,{\bar w}\right)&= \cos\left(\left(\nu_1+\nu_2-\nu_3-\nu_4\right)\tfrac{\pi i}{2}\right)\text{sin} \left(\pi(w+{\bar w})\right) 
     \\&=\tfrac{1}{2}\sin \left( \pi  (\tfrac{i\left(\nu_3+\nu_4-\nu_1-\nu_2\right)}{2}-w-{\bar w})\right)+\tfrac{1}{2}\sin \left( \pi  (\tfrac{i\left(\nu_1+\nu_2-\nu_3-\nu_4\right)}{2}-w-{\bar w})\right).\nonumber
\end{align}

Combined with \eqref{++--MBfin}, this gives the following Mellin-Barnes representation of the ${\sf s}$-channel exchange diagram: 
\begin{shaded}\noindent\emph{General scalar 4pt exchange in the ${\sf s}$-channel}
\begin{multline}\label{total4ptsch}
  {}^{\left({\sf s}\right)}\langle \phi^{(\nu_1)}_{\vec{k}_1}\phi^{(\nu_2)}_{\vec{k}_2}\phi^{(\nu_3)}_{\vec{k}_3}\phi^{(\nu_4)}_{\vec{k}_4} \rangle^\prime
         = \frac{L^{2\left(d+1\right)}}{8\pi}  {\cal N}_4\left(\eta_0,k_i\right)\int \left[ds\right]_4 \,
     \text{cosec} \left(\pi(w+{\bar w})\right) \delta\left(w,{\bar w}\right)\\ \times I^{\left(\nu_1,\nu_2,\nu\right)}_{\vec{k}_1,\vec{k}_2,\vec{k}_I}\left(s_1,s_2,w\right)I^{\left(-\nu,\nu_3,\nu_4\right)}_{-\vec{k}_I,\vec{k}_3,\vec{k}_4}\left(s_3,s_4,{\bar w}\right)\Big|_{{}^{w=\frac{d}{4}-s_1-s_2}_{{\bar w}=\frac{d}{4}-s_3-s_4}},
\end{multline}
\end{shaded}
\noindent with total interference factor:
\begin{align}\label{interferencefactor_total}
    \delta\left(w,{\bar w}\right)=&\delta_{++\,--}\left(w,{\bar w}\right)+\delta_{+-\,-+}\left(w,{\bar w}\right)\\ \nonumber
   =&\cosh\left(\pi \nu\right)\sin \left(\pi\left(\tfrac{i\left(\nu_1+\nu_2+\nu_3+\nu_4\right)}{2}+\tfrac{d}{2}-w-{\bar w}\right)\right)\\ \nonumber &+\tfrac{1}{2}\sin \left( \pi  (\tfrac{i\left(\nu_3+\nu_4-\nu_1-\nu_2\right)}{2}-w-{\bar w})\right)+\tfrac{1}{2}\sin \left( \pi  (\tfrac{i\left(\nu_1+\nu_2-\nu_3-\nu_4\right)}{2}-w-{\bar w})\right)\\ \nonumber &-\frac{1}{2} \left[ \sin \left(\pi  (\tfrac{d}{2}+\tfrac{i\left(\nu_1+\nu_2+\nu_3+\nu_4\right)}{2}+w-{\bar w})\right)+ w \leftrightarrow {\bar w}\right].
\end{align}

We note that Barnes integrals of the type \eqref{total4ptsch} are known in the Mathematics literature as (four-variable) Meijer-G functions \cite{meijer1941multiplikationstheoreme}.\footnote{Technically it is a sum of Meijer-G functions owing to the interference factor \eqref{interferencefactor_total}.} 

In the following we discuss various properties of the expression \eqref{total4ptsch}.

\noindent $\bullet$ The expression \eqref{total4ptsch} for the exchange diagram is an integrated product of three-point conformal structures\footnote{The product of three-point structures appearing in the expression \eqref{total4ptsch} is precisely the Mellin-Barnes representation for the boundary conformal partial wave which is dual to the exchanged field in the bulk. This connection is made more concrete in \cite{ToAppear}, which provides a different approach to obtain the expression \eqref{total4ptsch} purely from the knowledge of the boundary conformal partial wave.} sewn together by the factor:
\begin{equation}\label{interfercosec}
   \underbrace{\text{cosec} \left(\pi(w+{\bar w})\right)}_{\text{EFT}} \delta\left(w,{\bar w}\right).
\end{equation}
The poles of cosecant function are responsible for the non-factorised contributions to the exchange four-point function, the residues of which generate an infinite sum of contact terms which constitute the Effective Field Theory (EFT) expansion of the exchange four-point function. In particular, the $\csc$-function has two sequences of poles:
\begin{subequations}\label{cosecpoles}
\begin{align}\label{cosecpoles1}
    w+{\bar w}&=-n, \qquad n=0,1,2,...\,, \\
     w+{\bar w}&=1+m \qquad m=0,1,2,...\,,
\end{align}
\end{subequations}
where the expansion of the correlator in $k_I$ is obtained by closing the Mellin contour on the first series \eqref{cosecpoles1}, whose residues generate only non-negative integer powers of $k^2_I$:
\begin{equation}\label{cosecanalpoles}
    \left(\frac{k_I}{2}\right)^{-2(w+{\bar w})} \; \overset{\eqref{cosecpoles1}}{\to} \; \left(\frac{k^2_I}{4}\right)^n.
\end{equation}
In section \ref{subsec::seriesexpfrommellin} we detail how to extract the coefficients of this expansion from the Mellin-Barnes representation in the case of external conformally coupled scalars.\\

\noindent $\bullet$ The factorised contributions to the exchange diagram are instead encoded in the sequences of $\Gamma$-function poles in $w,\,{\bar w}$ associated to the exchanged particle $\phi^{\left(\nu\right)}$. These in particular originate from the Mellin-Barnes representation \eqref{MBwightman} for its Wightman function, which are the following four sequences:
\begin{subequations}\label{factpoles}
\begin{align}\label{factpoles1}
    w&=\pm\frac{i\nu}{2}-n, \qquad n \in \mathbb{N}_0,\\
    {\bar w}&={\hat \pm}\frac{i\nu}{2}-m, \qquad m \in \mathbb{N}_0.\label{factpoles2}
\end{align}
\end{subequations}
That these correspond to factorised contributions in the exchange diagram can be straightforwardly observed from the fact that the interference factor \eqref{interfercosec} is constant on each series:
\begin{multline}\label{phase4opelim}
    \text{cosec} \left(\pi(w+{\bar w})\right) \delta\left(w,{\bar w}\right)\Big|_{w=\pm\frac{i\nu}{2}-n \text{ or } {\bar w}=\pm\frac{i\nu}{2}-m}\\=2\sin \pi \left(\tfrac{d}{4}+\tfrac{i\left(\nu_1+\nu_2\mp \nu\right)}{2}\right)\sin \pi \left(\tfrac{d}{4}+\tfrac{i\left(\nu_3+\nu_4\mp \nu\right)}{2}\right), \quad \forall\, n, m.
\end{multline}
Notice that the constant is the product of the interference factors for the corresponding three-point functions \eqref{total3ptdoublemellin}. Non-analytic terms in the exchanged momentum are generated when the series of poles in $w$ and ${\bar w}$ have correlated signs in front of $\nu$, while analytic terms are generated when the signs are anti-correlated:
\begin{subequations}
\begin{align}\label{0nonanalkI}
    &\left(\frac{k_I}{2}\right)^{-2(w+{\bar w})} \; \overset{\pm \pm}{\to} \; \left(\frac{k^2_I}{4}\right)^{\mp i\nu+(m+n)}, \\
    &\left(\frac{k_I}{2}\right)^{-2(w+{\bar w})} \; \overset{\pm \mp}{\to} \; \left(\frac{k^2_I}{4}\right)^{m+n}, \label{0analkI}
\end{align}
\end{subequations}
where the $\pm$ above the arrows denote the signs of the poles in \eqref{factpoles1} and \eqref{factpoles2} respectively. The non-analytic terms \eqref{0nonanalkI} are characteristic of a particle exchange and thus serve as a signal for particle production \cite{Arkani-Hamed:2015bza}. The tail of analytic terms \eqref{0analkI} accompanying \eqref{0nonanalkI} are required for the absence of singularities in the collapsed triangle configurations $k_I \sim k_1+k_2$ and $k_I \sim k_3+k_4$ (see \cite{Arkani-Hamed:2015bza,Arkani-Hamed:2018kmz}), and should not be confused with the contact contributions \eqref{cosecanalpoles}.\footnote{A simple way to understand this point is that the factorised contributions satisfy the homogeneous conformal invariance condition on four-point correlators (see e.g. section 5.2.1 of  \cite{Arkani-Hamed:2015bza} or section 3.3 of \cite{Arkani-Hamed:2018kmz} for a more recent treatment) and so by definition do not contain contact terms.} In section \ref{subsec::opelimit} we extract the coefficients of the terms \eqref{0nonanalkI} and \eqref{0analkI} from the Mellin-Barnes representation \eqref{total4ptsch}.  \\

\noindent $\bullet$ It is interesting to note that the form \eqref{total4ptsch} of the exchange four-point function can be fixed by conformal symmetry, except the precise expression for the interference factor $\delta\left(w,{\bar w}\right)$, which encodes the early time boundary condition (Bunch Davies). In the spirit of the ``Cosmological Bootstrap" \cite{Arkani-Hamed:2015bza,Arkani-Hamed:2018kmz}, the interference factor may be fixed by demanding the absence of singularities in the collapsed triangle configurations (i.e. the adiabatic vacuum condition) mentioned in the above bullet point. One might envisage using this as a guiding principle to ``Bootstrap" the Mellin-Barnes representation for more general exchange four-point functions for spinning fields.
\\

\noindent $\bullet$ Notice that there is an ambiguity in the splitting of the cosecant into Gamma functions:
\begin{multline}\label{splitcosec}
    \text{cosec}\left(\pi z+\pi n\right)=\left(-1\right)^n \text{cosec}\left(\pi z\right) \\ \rightarrow \quad \text{cosec}\left(\pi z\right) = \left(-1\right)^n\Gamma\left(z+n\right)\Gamma\left(1-z-n\right),
\end{multline}
owing to the periodicity of the sine function. As we have seen, the Mellin-Barnes integration contour is sensitive to the above splitting, since it prescribed to the separate poles of all Gamma functions of the type $\Gamma\left(a+s\right)$ from those of the type $\Gamma\left(b-s\right)$. The different Mellin-Barnes contours generated by this freedom give exchange diagrams which differ by contact terms (analytic in $k_I$), which correspond to different choices of higher-derivative improvement terms in the cubic vertices (which vanish on-shell). For the exchange diagram \eqref{total4ptsch} we used the non-derivative cubic interaction of the type $\phi_1 \phi_2 \phi_3 $, which corresponds to the splitting \eqref{splitcosec} with $n=0$ and $z=w+{\bar w}$, which we arrive to by keeping track of the Gamma functions before we combine them into the cosecant function (the details of which are given in section \ref{appsubsec::exchdia}, where we evaluate the $u$-integral).  \\

\noindent $\bullet$ The representation of the exchange diagram as a quadruple Mellin-Barnes integral is the best one can do for generic scalar fields, which is a function of four variables: $k_I/k_i$, $i=1,2,3,4$. As we saw in section \ref{subsec::3pttwocc}, for certain special values of the scaling dimensions away from the Principal Series, the number of Mellin-Barnes integrals required is reduced. A simple example which we shall see in section \ref{subsec::seriesexpfrommellin} is when all external scalars are conformally coupled, which the expression \eqref{total4ptsch} simplifies to a double-Mellin-Barnes integral for a general exchanged field and is a function of two variables $k_I/k_{12}$ and $k_I/k_{34}$. The simplest expression for general $d$ is when the exchanged scalar is also conformally coupled, which can be given in terms of Gauss Hypergeometric functions
\begin{align} 
  &  {}^{({\sf s})}\langle \phi^{(i/2)}_{\vec{k}_1}\phi^{(i/2)}_{\vec{k}_2}\phi^{(i/2)}_{\vec{k}_3}\phi^{(i/2)}_{\vec{k}_4} \rangle^\prime = {\cal N}_4\left(\eta_0,k_i\right)  \frac{4\pi^2L^{2\left(d+1\right)}}{k_I\sqrt{k_1k_2k_3k_4}}\\ & \hspace*{1.75cm}\nonumber \times \left[\Gamma\left(\frac{d-3}{2}\right)^2\left(k_1+k_2+k_I\right)^{3-d}\left(k_3+k_4+k_I\right)^{3-d} \right. \\ \nonumber
   & \hspace*{1.75cm} \left. + 2 \sin\left(\tfrac{\pi d}{2}\right)\left(k_1+k_2+k_I\right)^{3-d}
   \frac{\Gamma (d-3)}{d-3}{}_2F_1\left(\frac{d-3}{2},d-3;\frac{d-1}{2};\frac{k_I-k_3-k_4}{k_1+k_2+k_I}\right)\right. \\ \nonumber
  &  \hspace*{1.75cm} \left.  + 2 \sin\left(\tfrac{\pi d}{2}\right)\left(k_3+k_4+k_I\right)^{3-d}
   \frac{\Gamma (d-3)}{d-3}{}_2F_1\left(\frac{d-3}{2},d-3;\frac{d-1}{2};\frac{k_I-k_1-k_2}{k_3+k_4+k_I}\right)\right]. \nonumber
\end{align}
 Notice that this expression is singular for $d=3$, which arises from pinching of the Mellin integration contour. At the end of section \ref{subsec::seriesexpfrommellin} we show how the Mellin-Barnes representation defines this correlator for $d=3$ by analytic continuation.
\\

\noindent $\bullet$ Another way to obtain the exchange four-point function is by solving the conformal invariance conditions as an EFT expansion \cite{Arkani-Hamed:2015bza}. This idea was implemented in \cite{Arkani-Hamed:2018kmz} for external conformally coupled scalars when $d=3$. One can then apply the weight-shifting operator provided in \cite{Arkani-Hamed:2015bza} to obtain the exchange with external massless-scalars interacting in a shift-symmetric fashion from the result for external conformally coupled scalars.\footnote{More generally, weight-shifting operators which relate spinning correlators to scalar correlators are available in position space \cite{Sleight:2016dba,Sleight:2016hyl,Sleight:2017krf,Castro:2017hpx,Sleight:2017fpc,Chen:2017yia,Costa:2018mcg} and more recently in momentum space \cite{Isono:2018rrb,Arkani-Hamed:2018kmz,Chu:2018kec,Isono:2019ihz}} The Mellin-Barnes representation \eqref{total4ptsch} instead gives an expression for the exchange four-point function in general $d$ and for generic internal and external scalars, from which the EFT expansion and non-perturbative corrections can be extracted by evaluating the appropriate residues -- as discussed in the bullet points above. We shall make contact with the result of \cite{Arkani-Hamed:2018kmz} in section \ref{subsec::seriesexpfrommellin}.\\

The above features of the Mellin-Barnes representation will be considered in further detail in the following sections.

\subsection{OPE Expansion}
\label{subsec::opelimit}

\begin{figure}[h]
    \centering
    \captionsetup{width=0.95\textwidth}
    \includegraphics[scale=0.5]{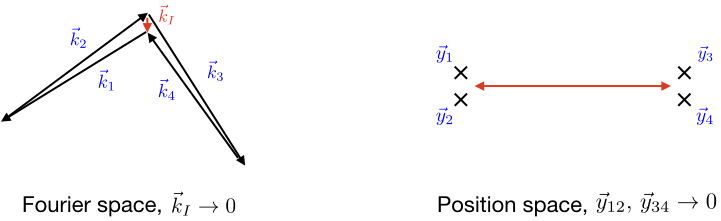}
    \caption{Left: Configuration of momenta in the OPE (collapsed) limit $k_I \to 0$ for exchange four-point functions, where $\vec{k}_I=\vec{k}_1+\vec{k}_2$. Right: In position space this corresponds to the points forming two pairs which are separated far from each other.}
    \label{fig:OPE_limit}
\end{figure}

A limit of particular interest is the collapsed limit $k_I \rightarrow 0$, in which the non-analytic terms in $k_I$ signal the exchange of the physical (on-shell) single-particle state \cite{Arkani-Hamed:2015bza} -- see equation \eqref{2ptlongdist}. In position space, this is the Operator Product Expansion (OPE) limit. See figure \ref{fig:OPE_limit}.\footnote{Note that here we are referring to the Operator Product Expansion on the $d$-dimensional boundary at $\eta_0=0$, not in the $(d+1)$-dimensional bulk.} 

As we saw at the end of the previous section, these terms are encoded in the factorised contributions to the exchange diagram. In the following we shall detail how to use the Mellin formalism to derive the expansion of these contributions in $k_I$, from which we can read off the OPE limit of the exchange four-point function. To this end it is useful to make the change of variables $s_1 \to s_1-s_2$ and $s_3 \to s_3-s_4$, so that the exchange four-point function \eqref{total4ptsch} takes the form:
\begin{multline}
  \hspace*{-0.5cm} \langle \phi^{(\nu_1)}_{\vec{k}_1}\phi^{(\nu_2)}_{\vec{k}_2}\phi^{(\nu_3)}_{\vec{k}_3}\phi^{(\nu_4)}_{\vec{k}_4} \rangle^\prime = {\cal N}_4\left(\eta_0,k_i\right)\frac{L^{2\left(d+1\right)}}{8\pi}  \left(\frac{k_1k_3}{2}\right)^{-\tfrac{d}{2}} \int^{i\infty}_{-i\infty} \frac{dw}{2\pi i}\frac{d{\bar w}}{2\pi i}\,
     \text{cosec} \left(\pi(w+{\bar w})\right) \delta\left(w,{\bar w}\right)\\ \hspace*{0.5cm} \times \Gamma\left(w+\tfrac{i\nu}{2}\right)\Gamma\left(w-\tfrac{i\nu}{2}\right) \left(\frac{k_I}{k_1}\right)^{-2w}
     I\left(\tfrac{i\nu_2}{2},-\tfrac{i\nu_2}{2};\tfrac{d}{4}+\tfrac{i\nu_1}{2}-w,\tfrac{d}{4}-\tfrac{i\nu_1}{2}-w;\tfrac{k_2}{k_1}\right)
      \\ \hspace*{0.5cm} \times
    \Gamma\left({\bar w}+\tfrac{i\nu}{2}\right)\Gamma\left({\bar w}-\tfrac{i\nu}{2}\right) \left(\frac{k_I}{k_3}\right)^{-2{\bar w}} I\left(\tfrac{i\nu_4}{2},-\tfrac{i\nu_4}{2};\tfrac{d}{4}+\tfrac{i\nu_3}{2}-{\bar w},\tfrac{d}{4}-\tfrac{i\nu_3}{2}-{\bar w};\tfrac{k_4}{k_3}\right),
\end{multline}
where $w=\tfrac{d}{4}-s_1$, ${\bar w}=\tfrac{d}{4}-s_3$, and integrals in $s_2$ and $s_4$ have factorised into the two Mellin integrals $I\left(a,b;c,d;z\right)$, which were defined in \eqref{3ptccorigs1} and are each given by a Gauss Hypergeometric function:
\begin{multline}
   I\left(\tfrac{i\nu_2}{2},-\tfrac{i\nu_2}{2};\tfrac{d}{4}+\tfrac{i\nu_1}{2}-w,\tfrac{d}{4}-\tfrac{i\nu_1}{2}-w;\tfrac{k_2}{k_1}\right) = \frac{1}{\Gamma\left(\tfrac{d}{2}-2w\right)}\prod_{\pm, {\hat \pm}} \Gamma \left(\tfrac{d}{4}-w
   \pm\tfrac{i\nu_1}{2}{\hat \pm}\tfrac{i\nu_2}{2}\right) \\  \times \left(\tfrac{k_2}{k_1}\right)^{i\nu_2} {}_2F_{1}\left(\tfrac{d}{4}-w
   +\tfrac{i(\nu_1+\nu_2)}{2},\tfrac{d}{4}-w
   -\tfrac{i(\nu_1-\nu_2)}{2};\tfrac{d}{2}-2w;1-\left(\tfrac{k_2}{k_1}\right)^2\right).
\end{multline}
See equation \eqref{1mz2gh}.

There are four factorised contributions to the exchange diagram, which are encoded in the four series of poles \eqref{factpoles} -- two for each of the Mellin variables $w$ and ${\bar w}$. As discussed around \eqref{0nonanalkI}, the following combinations generate non-analytic terms in $k_I$:
\begin{subequations}
\begin{align}
    w&=\tfrac{i\nu}{2}-n, && \hspace*{-3cm} {\bar w}=\tfrac{i\nu}{2}-m, \\
    w&=-\tfrac{i\nu}{2}-n,  && \hspace*{-3cm} {\bar w}=-\tfrac{i\nu}{2}-m,
\end{align}
\end{subequations}
while the combinations
\begin{subequations}
\begin{align}
    w&=\tfrac{i\nu}{2}-n, && \hspace*{-3cm} {\bar w}=-\tfrac{i\nu}{2}-m, \\
    w&=-\tfrac{i\nu}{2}-n,  && \hspace*{-3cm} {\bar w}=\tfrac{i\nu}{2}-m,
\end{align}
\end{subequations}
generate the accompanying analytic terms \eqref{0analkI} required for the absence of unphysical singularities. Evaluating the residues gives the expansion 
{\footnotesize
\begin{multline}\label{factexp}
   \hspace*{-0.75cm}   {}^{\left({\sf s}\right)}\langle \phi^{(\nu_1)}_{\vec{k}_1}\phi^{(\nu_2)}_{\vec{k}_2}\phi^{(\nu_3)}_{\vec{k}_3}\phi^{(\nu_4)}_{\vec{k}_4} \rangle^\prime\Big|_{\text{factorised}}= {\cal N}_4\left(\eta_0,k_i\right)\frac{L^{2\left(d+1\right)}}{4\pi}  \left(\frac{k_1k_3}{4}\right)^{-\tfrac{d}{2}} \sin \pi \left(\tfrac{d}{4}+\tfrac{i\left(\nu_1+\nu_2+\nu\right)}{2}\right)\sin \pi \left(\tfrac{d}{4}+\tfrac{i\left(\nu_3+\nu_4+\nu\right)}{2}\right)\\ \nonumber  \times \left[\left(\frac{k^2_I}{k_1k_3}\right)^{i\nu} \sum^\infty_{n,m=0} c^{\left(\nu_1,\nu_2,\nu\right)}_{n}c^{\left(\nu_3,\nu_4,\nu\right)}_{m}\left(\frac{k^2_I}{k^2_1}\right)^n\left(\frac{k^2_I}{k^2_3}\right)^m + \sum^\infty_{n,m=0} c^{\left(\nu_1,\nu_2,\nu\right)}_{n}c^{\left(\nu_3,\nu_4,-\nu\right)}_{m}\left(\frac{k^2_I}{k^2_1}\right)^n\left(\frac{k^2_I}{k^2_3}\right)^m\right]\\
   + \nu \to -\nu,
\end{multline}}
\noindent with coefficients:
\begin{align}
   & c^{\left(\nu_1,\nu_2,\nu\right)}_{n}=\frac{\left(-1\right)^n}{n!}\frac{\Gamma\left(-i\nu-n\right)}{\Gamma\left(\tfrac{d}{2}+i\nu+2n\right)}  \prod_{\pm, {\hat \pm}} \Gamma \left(\tfrac{d}{4}+\tfrac{i\nu}{2}+n
   \pm\tfrac{i\nu_1}{2}{\hat \pm}\tfrac{i\nu_2}{2}\right) \\ \nonumber & \hspace*{0.75cm} \times \left(\frac{k_2}{k_1}\right)^{i\nu_2}{}_2F_{1}\left(\tfrac{d}{4}+\tfrac{i\nu}{2}+n
   +\tfrac{i(\nu_1+\nu_2)}{2},\tfrac{d}{4}+\tfrac{i\nu}{2}+n
   -\tfrac{i(\nu_1-\nu_2)}{2};\tfrac{d}{2}+i\nu+2n;1-\left(\tfrac{k_2}{k_1}\right)^2\right).
\end{align}

The leading terms in OPE limit $k_{I} \to 0$ are given by the $n=m=0$ contributions and focusing on the non-analytic terms:
\begin{shaded}
\noindent\emph{OPE limit of a general scalar exchange}
\begin{multline}
    {}^{\left({\sf s}\right)}\langle \phi^{(\nu_1)}_{\vec{k}_1}\phi^{(\nu_2)}_{\vec{k}_2}\phi^{(\nu_3)}_{\vec{k}_3}\phi^{(\nu_4)}_{\vec{k}_4} \rangle^\prime \sim \frac{L^{2(d+1)}}{4\pi} \left(\frac{k_{12}k_{34}}{16}\right)^{-\frac{d}{2}}  {\cal N}_4\left(\eta_0,k_i\right) \\
 \times  \left[\left(\frac{4k^2_I}{k_{12}k_{34}}\right)^{i\nu}\frac{\Gamma (-i \nu )^2}{\Gamma \left(\frac{d}{2}+i\nu\right)^2}\sin \left(\pi  \left(\tfrac{d}{4}+\tfrac{(\nu +\nu_1+\nu_2)i}{2}\right)\right)\sin \left(\pi  \left(\tfrac{d}{4}+\tfrac{(\nu +\nu_3+\nu_4)i}{2}\right)\right)
   \right. \\ \left.
  \times \prod_{\pm, {\hat \pm}} \Gamma \left(\tfrac{d}{4}+\tfrac{i\nu}{2}\pm\tfrac{i\nu_1}{2}{\hat \pm}\tfrac{i\nu_2}{2}\right)\Gamma \left(\tfrac{d}{4}+\tfrac{i\nu}{2}   \pm\tfrac{i\nu_3}{2}{\hat \pm}\tfrac{i\nu_4}{2}\right) \right. \\ \left.
  + \nu \rightarrow -\nu \right],
\end{multline}
\end{shaded}
\noindent where we used that $k_1 \sim k_2$ and $k_3 \sim k_4$ as $k_I \to 0$. This expression agrees with, and generalises to general external scalars and general boundary dimension $d$, the existing results \cite{Arkani-Hamed:2015bza} available in $d=3$ for the OPE limit of exchange four-point correlators with external conformally coupled scalars $\nu_j=\frac{i}{2}$ and external massless scalars $\nu_j=\frac{3i}{2}$. 

Note that the above expression contains oscillatory terms in $\log\left(k^2_I/\left(k_{12}k_{34}\right)\right)$ for a massive exchanged particle on the Principal Series, $\nu \in \mathbb{R}$, where the phase of the oscillation depends on the scaling dimensions of the fields participating in the interaction and in particular on the interference factor \eqref{phase4opelim} (which in turn is fixed by the interference factors of the corresponding three-point correlators \eqref{total3ptdoublemellin}). This property was observed in \cite{Arkani-Hamed:2015bza} (see also \cite{Noumi:2012vr}) in $d=3$ for equal external conformally coupled or massless scalars. 

\subsection{EFT expansion from Mellin} 
\label{subsec::seriesexpfrommellin}

Similarly we can extract the EFT expansion of the exchange four-point from the Mellin-Barnes representation \eqref{total4ptsch}, which is encoded in the poles \eqref{cosecpoles1} of the cosecant function. For ease of presentation we shall consider the case where all external scalars are conformally coupled where, by virtue of the mechanism detailed at the beginning of section \ref{subsec::3pttwocc}, two of the four Mellin-Barnes integrals can be lifted:
\begin{multline}\label{extccscexch}
   {}^{\left({\sf s}\right)}\langle \phi^{(i/2)}_{\vec{k}_1}\phi^{(i/2)}_{\vec{k}_2}\phi^{(i/2)}_{\vec{k}_3}\phi^{(i/2)}_{\vec{k}_4} \rangle^\prime
         = {\cal N}_4\left(\eta_0,k_i\right)\frac{L^{2\left(d+1\right)}}{8\pi}\int^{i\infty}_{-i\infty} \frac{dw}{2\pi i}\frac{d{\bar w}}{2\pi i} \, \,
     \text{cosec} \left(\pi(w+{\bar w})\right) \delta\left(w,{\bar w}\right)\\ \times I^{\left(i/2,i/2,\nu\right)}_{\vec{k}_1,\vec{k}_2,\vec{k}_I}\left(\tfrac{d}{4}-w\right)I^{\left(-\nu,i/2,i/2\right)}_{-\vec{k}_I,\vec{k}_3,\vec{k}_4}\left(\tfrac{d}{4}-{\bar w}\right),
\end{multline}
recalling the Mellin representation \eqref{ccsclnptconfstr2} for the three-point conformal structure involving two conformally coupled scalars and a general scalar. This expression is a function of two variables:
\begin{equation}
    u=\frac{k_I}{k_{12}}, \qquad v=\frac{k_I}{k_{34}}.
\end{equation}

It is useful to make the change of variables $w \to w-{\bar w}$, for which the poles \eqref{cosecpoles} of the cosecant function are in the variable $w$ only and the two Mellin-Barnes integrals in \eqref{extccscexch} become functions of $u$ and $u/v$ respectively. Equivalently we could have sent ${\bar w} \to {\bar w}-w$, which would instead give an expansion in $v$ and $v/u$. The expansion in $u$ is obtained by closing the integration contour for $w$ to the left of the imaginary axis, which selects the residues of the poles \eqref{cosecpoles1}. This gives:
\begin{multline}
   {}^{\left({\sf s}\right)}\langle \phi^{(i/2)}_{\vec{k}_1}\phi^{(i/2)}_{\vec{k}_2}\phi^{(i/2)}_{\vec{k}_3}\phi^{(i/2)}_{\vec{k}_4} \rangle^\prime\Big|_{\text{EFT}} = {\cal N}_4\left(\eta_0,k_i\right) L^{2\left(d+1\right)}\sin \left(\frac{\pi  d}{2}\right) \frac{u^{\frac{d}{2}-1}v^{\frac{d}{2}-1}}{2^{d-3}\sqrt{k_1k_2k_3k_4}} \left(\frac{k_I}{2}\right)^{2-d}\\ \times \int^{i\infty}_{-i\infty} \frac{d{\bar w}}{2\pi i} \,(\cos (2 \pi  {\bar w})-\cosh (\pi  \nu ))\Gamma \left({\bar w}-\tfrac{i \nu }{2}\right) \Gamma \left({\bar w}+\tfrac{i \nu }{2}\right)\Gamma \left(\tfrac{d}{2}-1-2 {\bar w}\right) \left(\frac{u}{v}\right)^{2 {\bar w}}\\ \times \sum^\infty_{m=0} \Gamma \left(\tfrac{d}{2}-1+2 m+2 {\bar w}\right) \Gamma \left(-m-{\bar w}-\tfrac{i \nu }{2}\right) \Gamma \left(-m-{\bar w}+\tfrac{i \nu }{2}\right) \left(\frac{u}{2}\right)^{2m} .
\end{multline}
Similarly, we can evaluate the remaining ${\bar w}$-integral as an expansion in $u/v$. To obtain an expansion which is valid, say, for $u<v$, we must close the Mellin-Barnes contour to the right of the imaginary axis. This encloses the following series of poles:\footnote{Naively one would also expect the following sequences of poles: 
\begin{equation}
    {\bar w}=-m\pm \frac{i\nu}{2}+n^\prime, \qquad n^\prime \in \mathbb{N}_{\geq 0}
\end{equation}
to contribute, which would in addition generate non-analytic terms in the exchanged momentum $k_I$. These are however cancelled by the zeros of the factor $(\cos (2 \pi  {\bar w})-\cosh (\pi  \nu ))$ which originates from the interference factor \eqref{interferencefactor_total}. This is consistent with the observation that the cosecant factor generates only contact contributions to the exchange diagram, as discussed below equation \eqref{interfercosec}.}
\begin{align}
{\bar w}=\frac{d-2}{4}+m+\frac{n}{2}, \qquad n \in \mathbb{N}_{\geq 0},
\end{align}
which gives: 
\begin{shaded}
\noindent \emph{EFT expansion of the exchange 4pt function with external conformally coupled scalars}
\begin{multline}\label{contactseries}
   {}^{\left({\sf s}\right)}\langle \phi^{(i/2)}_{\vec{k}_1}\phi^{(i/2)}_{\vec{k}_2}\phi^{(i/2)}_{\vec{k}_3}\phi^{(i/2)}_{\vec{k}_4}\rangle^\prime\Big|_{\text{EFT}}\\ = -4  \pi^2  L^{2\left(d+1\right)} \sin \left(\frac{\pi  d}{2}\right) \frac{ {\cal N}_4\left(\eta_0,k_i\right)}{\sqrt{k_1k_2k_3k_4}}\left(\frac{k_I}{2}\right)^{2-d} \sum^\infty_{m,n=0} c_{mn} u^{2m+d-2}\left(\frac{u}{v}\right)^n,
\end{multline}
\end{shaded}
\noindent where the series coefficients are given by
\begin{equation}
   c_{mn} = \frac{\left(-1\right)^n}{2^{2m+d-1}n!} \frac{(d+n+2 m-3)!}{\left(\frac{d}{4}+\frac{n+i\nu-1}{2}\right)_{m+1}\left(\frac{d}{4}+\frac{n-i\nu-1}{2}\right)_{m+1}}.
\end{equation}
Setting $d=3$ these reduce to
\begin{equation}
   c_{mn} = \frac{\left(-1\right)^n\left(n+1\right)\left(n+2\right)...\left(n+2m\right)}{\left[\left(n+\tfrac{1}{2}\right)^2+\nu^2\right]\left[\left(n+\tfrac{5}{2}\right)^2+\nu^2\right]...\left[\left(n+\tfrac{1}{2}+2m\right)^2+\nu^2\right]},
\end{equation}
which recovers equation (3.24) in \cite{Arkani-Hamed:2018kmz}. In the above we derived the series expansion in the variables $u$ and $u/v$, but we could have also obtained an expansion in $v$ and $v/u$ by instead making the change of variables ${\bar w} \to {\bar w}-w$ at the level of the Mellin-Barnes representation \eqref{extccscexch}. The resulting expression would be the same as \eqref{contactseries} but with $u \leftrightarrow v$, as required by conformal invariance of four-point interactions.

Notice that there is a curious overall factor of $\sin \left(\frac{\pi  d}{2}\right)$ in \eqref{contactseries}, which is vanishing in even dimensions. This is consistent with the expression \eqref{totalccs4pt} for the four-point function generated by the $\phi^4$ contact interaction of conformally coupled scalars which, upon combining the contributions from the $+$ and $-$ branches of the in-in contour, is vanishing in even dimensions by virtue of the same sinusoidal factor.

The non-perturbative corrections to the EFT expansion are given by the factorised contributions \eqref{factexp}. For external conformally coupled scalars, as we saw in section \ref{subsec::3pttwocc} the three-point factors in each contribution are given by Gauss Hypergeometric functions \eqref{2f1ccscaln}. In particular, in this case we have:
\begin{equation}
 \sum^\infty_{n=0}c^{\left(
 \frac{i}{2},\frac{i}{2},\pm \nu\right)}_{n}\left(\frac{k^2_I}{k^2_1}\right)^{n}=\pi^{3/2}\left(\frac{k_I}{k_1}\right)^{1-\frac{d}{2}\mp i\nu}\frac{\beta_0}{\alpha_{\pm}} F_{\pm}\left(u\right),
\end{equation}
where (extending the $d=3$ notation \cite{Arkani-Hamed:2018kmz} to general $d$):
\begin{subequations}
\begin{align}
  F_{\pm}\left(u\right) &= \left(\frac{i u}{2 \nu}\right)^{\frac{d}{2}-1\pm i\nu} {}_2F_1\left(\tfrac{d}{4}\pm \tfrac{i\nu}{2},\tfrac{d}{4}\pm\tfrac{i\nu}{2}-\tfrac{1}{2};1\pm i\nu;u^2\right),\\
  \alpha_\pm&=-
 \left(\frac{i}{2\nu}\right)^{\frac{d}{2}-1\pm i\nu} \frac{\Gamma\left(1\pm i\nu\right)}{\Gamma\left(\tfrac{d}{4}\pm \tfrac{i\nu}{2}\right)\Gamma\left(\tfrac{d}{4}\pm\tfrac{i\nu}{2}-\tfrac{1}{2}\right)},\\
 \beta_0&=\frac{1}{i\sinh{\pi \nu}},
\end{align}
\end{subequations}
so that 
\begin{shaded}
\noindent \emph{Non-perturbative correction to the EFT expansion}
\begin{multline}\label{nonpertEFT}
    {}^{({\sf s})}\langle \phi^{(i/2)}_{\vec{k}_1}\phi^{(i/2)}_{\vec{k}_2}\phi^{(i/2)}_{\vec{k}_3}\phi^{(i/2)}_{\vec{k}_4} \rangle^\prime\Big|_{\text{factorised}} = \pi^2   {\cal N}_4\left(\eta_0,k_i\right) \left(\frac{k_I}{2}\right)^{2-d} \frac{L^{2\left(d+1\right)}}{2\sqrt{k_1k_2k_3k_4}} \\\times \frac{\beta^2_0}{\alpha_+\alpha_-}\left[\left(\cos{\pi\left(\tfrac{d}{2}- i\nu\right)}+1\right)\frac{\beta_0\alpha_+}{\alpha_-}F_{-}\left(u\right)F_{-}\left(v\right)+\left(\cos{\pi\left(\tfrac{d}{2}+i\nu\right)}+1\right)\frac{\beta_0\alpha_-}{\alpha_+}F_{+}\left(u\right)F_{+}\left(v\right)\right.\\-\left. \left(\cos{\pi\left(\tfrac{d}{2}+i\nu\right)}+1\right)\beta_0F_{+}\left(u\right)F_{-}\left(v\right)-\left(\cos{\pi\left(\tfrac{d}{2}-i\nu\right)}+1\right)\beta_0F_{-}\left(u\right)F_{+}\left(v\right)\right].
\end{multline}
\end{shaded}
\noindent Setting $d=3$ this recovers equation (3.37) of \cite{Arkani-Hamed:2018kmz}.

The EFT expansion \eqref{contactseries} and non-perturbative corrections \eqref{nonpertEFT} are valid for generic values of $\nu$ and $d$ where the integration contour in the Mellin-Barnes representation \eqref{extccscexch} for the exchange four-point function is un-pinched. This is always the case when $\nu$ is on the Principal Series \eqref{PS}, while away from the Principal Series extra care needs to be taken for the specific values of $\nu$ and $d$ for which pinching occurs. This is discussed in more detail in the following. 

\paragraph{Contour Pinching.} The case of external conformally coupled scalars considered above provides another interesting and moreover simple example of contour pinching discussed at the end of section \ref{subsec::3ptscalargen}. 

Recall that the exchange four-point function \eqref{extccscexch} is an integrated product of the three-point structures \eqref{ccsclnptconfstr2} for two conformally coupled scalars and a general scalar, which are weighted by the cosecant factor \eqref{interfercosec}. Therefore, the integration contour for the exchange \eqref{extccscexch} becomes pinched for the same values of $\nu$ as for the Mellin-Barnes integral for the three-point structures \eqref{ccsclnptconfstr}. The simplest example is the conformally coupled scalar in $d=3$, which for the three-point structures \eqref{extccscexch} was discussed at the end of section \ref{subsec::3pttwocc}.

To study the pinching for the exchange four-point function \eqref{extccscexch} it is useful to return to the representation \eqref{uintppmm<>} of the $++$ and $--$ contributions along the in-in contour.\footnote{Since the $+-$ and $-+$ contributions are factorised \eqref{pmmpfactor} they are given by \eqref{epsccsc3pt}.} For the conformally coupled scalar, following the same steps as in section \ref{subsec::3pttwocc} we have
\begin{multline}\label{cccuint}
    I^{\left(i/2,i/2,i/2\right)}_{\vec{k}_1,\vec{k}_2,\vec{k}_3}\left(u\right)=4\frac{\pi^{\frac{3}{2}}}{\sqrt{k_1k_2}}\left(\frac{k_3}{2}\right)^{1-\frac{d}{2}}\left(2\frac{k_1+k_2+k_3}{k_3}\right)^{\frac{3-d}{2}} \Gamma \left(\frac{d-4 u-3}{2}\right) \\ \times \left(2\frac{k_1+k_2+k_3}{k_3}\right)^{2 u},
\end{multline}
so that
\begin{subequations}\label{uintccsc}
\begin{multline}\label{uintccsc1}
   {}^{({\sf s})}\langle \phi^{(i/2)}_{\vec{k}_1}\phi^{(i/2)}_{\vec{k}_2}\phi^{(i/2)}_{\vec{k}_3}\phi^{(i/2)}_{\vec{k}_4} \rangle^\prime_{\pm \pm,<}= -{\cal N}_4\left(\eta_0,k_i\right)L^{2\left(d+1\right)}  \frac{\pi^2}{\sqrt{k_1k_2k_3k_4}}\left(\frac{k_I}{2}\right)^{2-d}\\\times \left(2\frac{k_I-k_1-k_2}{k_I}\right)^{\frac{3-d}{2}}\left(2\frac{k_3+k_4+k_I}{k_I}\right)^{\frac{3-d}{2}}
   \\ \times \int^{i\infty}_{-i\infty} \frac{du}{2\pi i} \frac{e^{\mp 2u\pi i}}{u+{\bar \epsilon}}\Gamma \left(\frac{d-3-4 u}{2}\right)\Gamma \left(\frac{d-3+4 u}{2}\right)  \left(\frac{k_I-k_1-k_2}{k_3+k_4+k_I}\right)^{2u},
\end{multline}
and
\begin{multline}\label{uintccsc2}
   {}^{({\sf s})}\langle \phi^{(i/2)}_{\vec{k}_1}\phi^{(i/2)}_{\vec{k}_2}\phi^{(i/2)}_{\vec{k}_3}\phi^{(i/2)}_{\vec{k}_4} \rangle^\prime_{\pm \pm,>} = -{\cal N}_4\left(\eta_0,k_i\right)L^{2\left(d+1\right)}  \frac{\pi^2}{\sqrt{k_1k_2k_3k_4}}\left(\frac{k_I}{2}\right)^{2-d}
   \\\times \left(2\frac{k_1+k_2+k_I}{k_I}\right)^{\frac{3-d}{2}}\left(2\frac{k_I-k_3-k_4}{k_I}\right)^{\frac{3-d}{2}}
   \\
    \times \int^{i\infty}_{-i\infty} \frac{du}{2\pi i} \frac{e^{\mp 2u\pi i}}{u+{\bar \epsilon}}\Gamma \left(\frac{d-3-4 u}{2}\right)\Gamma \left(\frac{d-3+4 u}{2}\right)  \left(\frac{k_I-k_3-k_4}{k_1+k_2+k_I}\right)^{2u}.
\end{multline}
\end{subequations}
Note that above we used ${\bar \epsilon}>0$ for the prescription of the $u$-integration contour, to distinguish it from the regulator $\epsilon$ for the contour pinching that we use in the following. For both integrals, the sequences $\Gamma$-function poles are
\begin{subequations}
\begin{align}\label{rightccschreg}
    u&=\frac{d-3}{4}+\frac{n}{2}, \qquad n \in \mathbb{N}_0,\\
    u&=-\frac{d-3}{4}-\frac{m}{2}, \qquad m \in \mathbb{N}_0,
\end{align}
\end{subequations}
which, as anticipated, overlap \emph{only} when $d=3$. As in section \ref{subsec::3pttwocc}, the contour pinching can be regulated by setting $d \to 3+ \epsilon$, with $ \epsilon>0$. We then evaluate the $u$-integrals by closing the integration contour to the right of the imaginary axis, which is more convenient as it avoids the single pole at $u=-{\bar \epsilon}$ and just enclosed the poles \eqref{rightccschreg}. Re-summing the residues, this gives
\begin{subequations}\label{ppmmcontccscd3}
\begin{multline}
   {}^{({\sf s})}\langle \phi^{(i/2)}_{\vec{k}_1}\phi^{(i/2)}_{\vec{k}_2}\phi^{(i/2)}_{\vec{k}_3}\phi^{(i/2)}_{\vec{k}_4} \rangle^\prime_{\pm \pm,<}= -{\cal N}_4\left(\eta_0,k_i\right)  \frac{\pi^2L^{8}}{\sqrt{k_1k_2k_3k_4}}\left(\frac{k_I}{2}\right)^{-1}\\
   \times \left[\frac{2}{\epsilon ^2}+\frac{1}{\epsilon}\left(2\log \left( \frac{k_I}{k_I+k_3+k_4}\right)-2 \gamma-\log\left(4\right)\pm i \pi \right)+\text{Li}_2\left(\frac{k_I-k_1-k_2}{k_3+k_4+k_I}\right)\right.\\\left.+\log \left(\frac{k_I}{k_I+k_3+k_4}\right) \left(-\log \left(4\right)+\log\left(\frac{k_I}{k_I+k_3+k_4}\right)-2 \gamma \pm i \pi \right)\right.\\ \left.+\gamma ^2-\frac{\pi ^2}{12}\mp i \pi \left(\gamma + \log 2\right)+\left(\log2\right)^2+\gamma\log 4
   +O\left(\epsilon\right)\right],
\end{multline}
and
\begin{multline}
   {}^{({\sf s})}\langle \phi^{(i/2)}_{\vec{k}_1}\phi^{(i/2)}_{\vec{k}_2}\phi^{(i/2)}_{\vec{k}_3}\phi^{(i/2)}_{\vec{k}_4} \rangle^\prime_{\pm \pm,>} = -{\cal N}_4\left(\eta_0,k_i\right)  \frac{\pi^2L^{8}}{\sqrt{k_1k_2k_3k_4}}\left(\frac{k_I}{2}\right)^{-1}
\\   \times \left[\frac{2}{\epsilon ^2}+\frac{1}{\epsilon}\left(2\log \left( \frac{k_I}{k_I+k_1+k_2}\right)-2 \gamma-\log\left(4\right)\pm i \pi \right)+\text{Li}_2\left(\frac{k_I-k_3-k_4}{k_1+k_2+k_I}\right)\right.\\\left.+\log \left(\frac{k_I}{k_I+k_1+k_2}\right) \left(-\log \left(4\right)+\log\left(\frac{k_I}{k_I+k_1+k_2}\right)-2 \gamma \pm i \pi \right)\right.\\ \left.+\gamma ^2-\frac{\pi ^2}{12}\mp i \pi \left(\gamma + \log 2\right)+\left(\log2\right)^2+\gamma\log 4
   +O\left(\epsilon\right)\right].
\end{multline}
\end{subequations}
The poles in $\epsilon$ cancel upon including the contributions \eqref{pmmpfactor} from the $+-$ and $-+$ branches of the in-in contour, which are:
\begin{multline}
    {}^{({\sf s})}\langle \phi^{(i/2)}_{\vec{k}_1}\phi^{(i/2)}_{\vec{k}_2}\phi^{(i/2)}_{\vec{k}_3}\phi^{(i/2)}_{\vec{k}_4} \rangle^\prime_{+-}+{}^{({\sf s})}\langle \phi^{(\nu_1)}_{\vec{k}_1}\phi^{(\nu_2)}_{\vec{k}_2}\phi^{(\nu_3)}_{\vec{k}_3}\phi^{(\nu_4)}_{\vec{k}_4} \rangle^\prime_{-+}
    =2{\cal N}_4\left(\eta_0,k_i\right)  \frac{\pi^2L^{8}}{\sqrt{k_1k_2k_3k_4}}\left(\frac{k_I}{2}\right)^{-1}\\ \times 
    \left[\frac{4}{\epsilon ^2}-\frac{1}{\epsilon}\left(2\log \left(\frac{4 (k_1+k_2+k_I) (k_3+k_4+k_I)}{k^2_I}\right)+4 \gamma \right)\right. \\ \left.-2 \log \left(k_I\right) \left(\log (4 (k_1+k_2+k_I) (k_3+k_4+k_I))+2 \gamma \right)+2 \log ^2\left(k_I\right)\right.  \\  \left.+\frac{1}{2} \log ((k_1+k_2+k_I) (k_3+k_4+k_I)) (\log (16 (k_1+k_2+k_I) (k_3+k_4+k_I))+2\log\left(4\right)+4 \gamma )\right.  \\ \left.+\frac{\pi ^2}{6}+2 \log ^2(2)+2 \gamma^2  +2\gamma\log (4)+O\left(\epsilon\right)\right],
\end{multline}
where we used \eqref{cccuint} with $u=0$ and sent $d \to 3+\epsilon$. Combined with \eqref{ppmmcontccscd3}, this gives the following expression for the exchange four-point function of conformally coupled scalars in $d=3$:
\begin{multline}\label{ccscd3end}
    {}^{\left({\sf s}\right)}\langle \phi^{\left(i/2\right)}_{\vec{k}_1}\phi^{\left(i/2\right)}_{\vec{k}_2}\phi^{\left(i/2\right)}_{\vec{k}_3}\phi^{\left(i/2\right)}_{\vec{k}_4} \rangle^\prime   ={\cal N}_4\left(\eta_0,k_i\right)\frac{4\pi^2L^{8}}{k_I \sqrt{k_1k_2k_3k_4}}\,\\ \times \left[\frac{ \pi ^2}3- \text{Li}_2\left(\frac{k_I-k_3-k_4}{k_I+k_1+k_2}\right)- \text{Li}_2\left(\frac{k_I-k_1-k_2}{k_I+k_3+k_4}\right)-\frac12 \log ^2\left(\frac{k_I+k_3+k_4}{k_I+k_1+k_2}\right)\right],
\end{multline}
which recovers equation (5.74) of \cite{Arkani-Hamed:2015bza}. 

It is interesting to note that for general $d>3$ the integrals \eqref{uintccsc} give Gauss Hypergeometric functions:
\begin{subequations}
\begin{multline}
   {}^{({\sf s})}\langle \phi^{(i/2)}_{\vec{k}_1}\phi^{(i/2)}_{\vec{k}_2}\phi^{(i/2)}_{\vec{k}_3}\phi^{(i/2)}_{\vec{k}_4} \rangle^\prime_{\pm \pm,<}= \pm i\,e^{\mp \frac{1}{2} i \pi  d}  {\cal N}_4\left(\eta_0,k_i\right)  \frac{4\pi^2L^{2\left(d+1\right)}}{k_I\sqrt{k_1k_2k_3k_4}}\\\times \left(k_3+k_4+k_I\right)^{3-d}
   \frac{\Gamma (d-3)}{d-3}{}_2F_1\left(\frac{d-3}{2},d-3;\frac{d-1}{2};\frac{k_I-k_1-k_2}{k_3+k_4+k_I}\right),
\end{multline}
and
\begin{multline}
   {}^{({\sf s})}\langle \phi^{(i/2)}_{\vec{k}_1}\phi^{(i/2)}_{\vec{k}_2}\phi^{(i/2)}_{\vec{k}_3}\phi^{(i/2)}_{\vec{k}_4} \rangle^\prime_{\pm \pm,>} = \pm i\,e^{\mp \frac{1}{2} i \pi  d}  {\cal N}_4\left(\eta_0,k_i\right)  \frac{4\pi^2L^{2\left(d+1\right)}}{k_I\sqrt{k_1k_2k_3k_4}}
\\\times \left(k_1+k_2+k_I\right)^{3-d}
   \frac{\Gamma (d-3)}{d-3}{}_2F_1\left(\frac{d-3}{2},d-3;\frac{d-1}{2};\frac{k_I-k_3-k_4}{k_1+k_2+k_I}\right),
\end{multline}
\end{subequations}
which can be obtained simply closing the integration contour on the poles \eqref{rightccschreg}. This gives the following expression for the exchange of a conformally coupled scalar for general $d>3$:
\begin{shaded}
\noindent \emph{Exchange four-point function for internal and external conformally coupled scalars}
\begin{multline}
    {}^{({\sf s})}\langle \phi^{(i/2)}_{\vec{k}_1}\phi^{(i/2)}_{\vec{k}_2}\phi^{(i/2)}_{\vec{k}_3}\phi^{(i/2)}_{\vec{k}_4} \rangle^\prime = {\cal N}_4\left(\eta_0,k_i\right)  \frac{4\pi^2L^{2\left(d+1\right)}}{k_I\sqrt{k_1k_2k_3k_4}}\\ \times \left[\Gamma\left(\frac{d-3}{2}\right)^2\left(k_1+k_2+k_I\right)^{3-d}\left(k_3+k_4+k_I\right)^{3-d} \right. \\
    \left. + 2 \sin\left(\tfrac{\pi d}{2}\right)\left(k_1+k_2+k_I\right)^{3-d}
   \frac{\Gamma (d-3)}{d-3}{}_2F_1\left(\frac{d-3}{2},d-3;\frac{d-1}{2};\frac{k_I-k_3-k_4}{k_1+k_2+k_I}\right)\right. \\
    \left. + 2 \sin\left(\tfrac{\pi d}{2}\right)\left(k_3+k_4+k_I\right)^{3-d}
   \frac{\Gamma (d-3)}{d-3}{}_2F_1\left(\frac{d-3}{2},d-3;\frac{d-1}{2};\frac{k_I-k_1-k_2}{k_3+k_4+k_I}\right)\right].
\end{multline}
\end{shaded}
\noindent 
As anticipated, this expression is singular when $d=3$. A finite expression is obtained by setting $d \to 3+\epsilon$ and expanding in $\epsilon$. As we know from the above analysis, the poles in $\epsilon$ arising from each term cancel to give the expression \eqref{ccscd3end}.

\section*{Acknowledgments}

We thank Dionysios Anninos and Ruben Monten for useful correspondence, and Ruben Monten for making me aware of the publication \cite{Falk:1992sf}. I am grateful to Massimo Taronna for useful discussions and collaboration on related work. Part of this work was carried out at the Universit\'e Libre de Bruxelles and Princeton University during 2018, which we gratefully acknowledge for support and hospitality. We also thank the University of Naples Federico II and the Erwin
Schr\"odinger International Institute for Mathematics and Physics for hospitality during the final stages. The author is supported by the European Union’s Horizon 2020 research and innovation programme under the Marie Sk\l odowska-Curie grant agreement No 793661. Earlier stages of this work were supported until Oct 2018 by a Marina Solvay Fellowship.

\appendix

\section{Mellin-Barnes Integrals}
\label{app:mellinbarnesints}

\begin{figure}[h]
    \centering
    \captionsetup{width=0.95\textwidth}
    \includegraphics[scale=0.45]{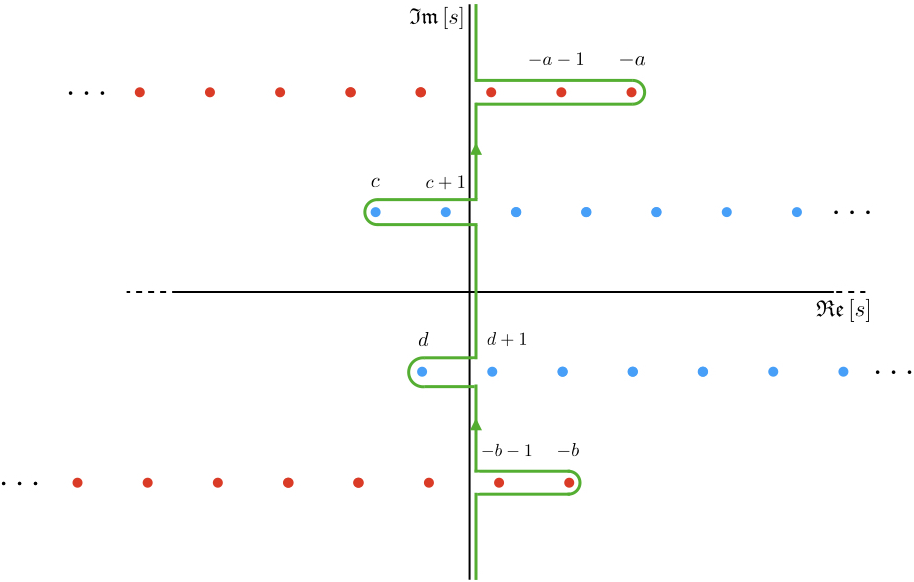}
    \caption{Integration contour (Green) for the Mellin-Barnes integral \eqref{blemma1}. This is chosen to separate the poles of the $\Gamma$ functions $\Gamma\left(a+s\right)$ and $\Gamma\left(b+s\right)$ (the red dots) which extend along the negative real axis from the sequences of poles of $\Gamma\left(c-s\right)$ and $\Gamma\left(d-s\right)$ (the blue dots) which extend along the positive real axis.}
    \label{fig:MellinBarnesint}
\end{figure}

Mellin-Barnes integrals \cite{barnes1,barnes2} are contour integrals involving products and ratios of Gamma functions in the integrand, which have the typical form
\begin{equation}\label{genmellinbarnsint}
    I\left(z\right)= \int^{\gamma+i\infty}_{\gamma-i\infty}\frac{ds}{2\pi i}\,\frac{\Gamma\left(a_1+A_1 s\right)...\Gamma\left(a_{n}+A_{n} s\right)\Gamma\left(b_1-B_1 s\right)...\Gamma\left(b_{m}-B_{m} s\right)}{\Gamma\left(c_1+C_1 s\right)...\Gamma\left(c_{p}+C_{p} s\right)\Gamma\left(d_1-D_1 s\right)...\Gamma\left(d_{q}-D_{q} s\right)}\,z^s,
\end{equation}
where $\gamma \in \mathbb{R}$ and $A_i$, $B_i$, $C_i$, $D_i >0 $.\footnote{When $A_i=B_i=C_i=D_i=1$ Barnes integrals of the form \eqref{genmellinbarnsint} are referred to in the Mathematics literature as Meijer-G functions \cite{meijer1941multiplikationstheoreme}.} The integration contour, which intersects the real axis at $\gamma$, runs parallel to the imaginary axis except when it is indented to separate the poles of the Gamma functions $\Gamma\left(a_i+A_i s\right)$ from the Gamma functions $\Gamma\left(b_i-B_i s\right)$. See e.g. figure \ref{fig:MellinBarnesint}.

A well known and simple example is given by Barnes' first lemma:
\begin{equation}\label{blemma1}
    \int^{\gamma+i\infty}_{\gamma-i\infty}\frac{ds}{2\pi i}\,\Gamma\left(a+s\right)\Gamma\left(b+s\right)\Gamma\left(c-s\right)\Gamma\left(d-s\right) = \frac{\Gamma\left(a+c\right)\Gamma\left(a+d\right)\Gamma\left(b+c\right)\Gamma\left(b+d\right)}{\Gamma\left(a+b+c+d\right)},
\end{equation}
which we employ often in this work. The integration contour is displayed in figure \ref{fig:MellinBarnesint}. To obtain the r.h.s. of the above equality we complete the integration contour with a circular arc of radius $R \rightarrow \infty$ and apply Cauchy's residue theorem. The arc at infinity does not contribute regardless of the side of the imaginary axis we close the contour, since the modulus of the integrand decays exponentially as: 
\begin{equation}\label{approxstirling}
    |\Gamma\left(a+s\right)\Gamma\left(b+s\right)\Gamma\left(c-s\right)\Gamma\left(d-s\right)| = O\left(e^{-2 \pi  R \left| \sin (\theta )\right|} e^{-\theta  \mathfrak{Im}\left[a+b+c+d\right]+\log (R) \mathfrak{Re}\left[a+b+c+d-2\right]}\right),
\end{equation}
where $s=R e^{i\theta}$ and we used Stirling's formula:
\begin{equation}
    \Gamma\left(z\right) = \sqrt{2\pi}e^{-z}z^{z-\tfrac{1}{2}}\left(1+{\cal O}\left(1/z\right)\right), \quad |z| \rightarrow 0, \quad |\arg z| < \pi.
\end{equation}
Note that technically we require $\mathfrak{Re}\left[a+b+c+d\right]> 1$ in order to neglect the arc at infinity, but this condition can be lifted afterwards by analytic continuation. The Gamma function poles in the r.h.s. of \eqref{blemma1} occur for the values of $a, b, c,$ and $d$ where poles of $\Gamma\left(a+s\right)\Gamma\left(b+s\right)$ overlap with poles of $\Gamma\left(c-s\right)\Gamma\left(d-s\right)$, which pinches the integration contour.

The above outlines the general approach for evaluating Mellin-Barnes integrals. In the following we give some explicit examples which cover the integrals we encounter in this work.

\subsection{Correlators with a conformally coupled scalar}
\label{subsec::appccs}

As explained in section \ref{subsec::3pttwocc}, when a correlator involves a conformally coupled scalar, the corresponding Mellin integral can be lifted using the formula:
\begin{align}\label{appccmbint}
    \int^{i\infty}_{-i\infty} \frac{ds}{2\pi i}\,\Gamma\left(2s-\tfrac{1}{2}\right)\Gamma\left(t-2s\right)z^{2s} = \frac{\sqrt{z+1}}{2} \, \Gamma \left(t-\tfrac{1}{2}\right)\left(\tfrac{1}{z}+1\right)^{-t}.
\end{align}
To prove this formula, as before we completing the integration contour with a circular arc of radius $R$. As $R \rightarrow \infty$, the modulus of the integrand decays exponentially:
\begin{equation}\label{stirlingccgen}
   \left|\Gamma\left(t-2s\right)\Gamma\left(2s-\tfrac{1}{2}\right)z^{2s}\right| = O\left(e^{-2 \pi  R \left| \sin (\theta )\right|}e^{2 R \cos (\theta ) \log (z)}e^{-\theta \mathfrak{Im}[u]+\log (R)\left[\mathfrak{Re}[u]-\frac{3}{2}\right]}\right),
\end{equation}
provided that we take $z > 1$ if we close the contour to the left, and $0<z<1$ if we close to the right. In closing the contour, say, to the left, Cauchy's residue theorem evaluates the integral as a series expansion in $1/z$, which can be re-summed to obtain the r.h.s of the formula \eqref{appccmbint}:
\footnote{Conversely, closing to the right gives the result as a series expansion in $z$, which re-sums to the same expression -- as expected by analytic continuation.} 
\begin{align}
    \int^{i\infty}_{-\infty} \frac{ds_2}{2\pi i}\,\Gamma\left(t-2s\right)\Gamma\left(2s-\tfrac{1}{2}\right)z^{2s} &= \frac{\sqrt{z}}{2}\sum_{n=0}\frac{1}{n!}\left(-\tfrac{1}{z}\right)^{n}\Gamma\left(t-\tfrac{1}{2}+n\right), \nonumber \\
    &= \frac{\sqrt{z+1}}{2} \, \Gamma \left(t-\tfrac{1}{2}\right)\left(\tfrac{1}{z}+1\right)^{-t}.
\end{align}

\subsection{Exchange diagrams}
\label{appsubsec::exchdia}

In this appendix we show how to lift the integral in the Mellin variable associated to the bulk-to-bulk propagator in the $++$ and $--$ contributions to the exchange four-point function. In equation \eqref{uintppmm<>} this is the $u$-integral. 

To this end it is useful to combine the various contributions. As we shall see in the following, the interference between the different processes manifest in simplifications to the Mellin integrand. We first combine the contributions for the different orderings of $\eta_1$ and $\eta_2$ along the same branch of the in-in contour:
\begin{align}\nonumber
     {}^{({\sf s})}\langle \phi^{(\nu_1)}_{\vec{k}_1}\phi^{(\nu_2)}_{\vec{k}_2}\phi^{(\nu_3)}_{\vec{k}_3}\phi^{(\nu_4)}_{\vec{k}_4} \rangle^\prime_{\pm \pm}&={}^{({\sf s})}\langle \phi^{(\nu_1)}_{\vec{k}_1}\phi^{(\nu_2)}_{\vec{k}_2}\phi^{(\nu_3)}_{\vec{k}_3}\phi^{(\nu_4)}_{\vec{k}_4} \rangle^\prime_{\pm \pm,<}+{}^{({\sf s})}\langle \phi^{(\nu_1)}_{\vec{k}_1}\phi^{(\nu_2)}_{\vec{k}_2}\phi^{(\nu_3)}_{\vec{k}_3}\phi^{(\nu_4)}_{\vec{k}_4} \rangle^\prime_{\pm \pm,>}, 
\end{align}
\begin{multline}\label{uintinitial}
   {}^{({\sf s})}\langle \phi^{(\nu_1)}_{\vec{k}_1}\phi^{(\nu_2)}_{\vec{k}_2}\phi^{(\nu_3)}_{\vec{k}_3}\phi^{(\nu_4)}_{\vec{k}_4} \rangle^\prime_{\pm \pm,<}+{}^{({\sf s})}\langle \phi^{(\nu_1)}_{\vec{k}_1}\phi^{(\nu_2)}_{\vec{k}_2}\phi^{(\nu_3)}_{\vec{k}_3}\phi^{(\nu_4)}_{\vec{k}_4} \rangle^\prime_{\pm \pm,>} \\
     = - e^{\pm \left(\nu_1+\nu_2+\nu_3+\nu_4-di\right)\frac{\pi}{2}}  \frac{L^{2\left(d+1\right)}}{16\pi}  {\cal N}_4\left(\eta_0,k_i\right)\\
    \times \int^{i\infty}_{-i\infty} \frac{du}{2\pi i} \int \left[ds\right]_4\,
    I^{\left(\nu_1,\nu_2,\nu\right)}_{\vec{k}_1,\vec{k}_2,\vec{k}_I}\left(s_1,s_2,w-u\right)I^{\left(\nu_3,\nu_4,\nu\right)}_{\vec{k}_3,\vec{k}_4,\vec{k}_I}\left(s_3,s_4,{\bar w}+u\right)
    \\
   \times  \left[ \frac{e^{\mp 2\left(u-w\right)\pi i}}{u+\epsilon} + \frac{e^{\pm 2\left(u+{\bar w}\right)\pi i}}{-u+\epsilon} \right]\Big|_{{}^{w=\frac{d}{4}-s_1-s_2}_{{\bar w}=\frac{d}{4}-s_3-s_4}},
\end{multline}
which we gave in terms of the Mellin-Barnes representation \eqref{3ptconfstr} of the general three-point conformal structure. To obtain this expression we made the change of variables $u \to -u$ in the contribution from the ordering $\eta_2 > \eta_1$, so the $u$-integration contour for the combined contributions is restricted by $-\epsilon < \mathfrak{Re}\left[u\right] < \epsilon$.\footnote{Equivalently we could have also made the change of variables $u \to -u$ in the contribution from the ordering $\eta_2 < \eta_1$.} Either of these bounds can be lifted by shifting the integration contour past either of poles at $u=\pm \epsilon$. Shifting the contour past the pole at $u=\epsilon$ gives\footnote{Likewise, if we could instead lift the restriction $-\epsilon < \mathfrak{Re}\left[u\right]$ by shifting the contour past the pole at $u=-\epsilon$.} 
\begin{multline}
       {}^{({\sf s})}\langle \phi^{(\nu_1)}_{\vec{k}_1}\phi^{(\nu_2)}_{\vec{k}_2}\phi^{(\nu_3)}_{\vec{k}_3}\phi^{(\nu_4)}_{\vec{k}_4} \rangle^\prime_{\pm \pm} +  {}^{({\sf s})}\langle \phi^{(\nu_1)}_{\vec{k}_1}\phi^{(\nu_2)}_{\vec{k}_2}\phi^{(\nu_3)}_{\vec{k}_3}\phi^{(\nu_4)}_{\vec{k}_4} \rangle^\prime_{\pm \pm,>\odot} \\= - e^{\pm \left(\nu_1+\nu_2+\nu_3+\nu_4-di\right)\frac{\pi}{2}} \frac{L^{2\left(d+1\right)}}{16\pi}  {\cal N}_4\left(\eta_0,k_i\right)\\
    \times \int^{i\infty}_{-i\infty} \frac{du}{2\pi i} \int \left[ds\right]_4\,
    I^{\left(\nu_1,\nu_2,\nu\right)}_{\vec{k}_1,\vec{k}_2,\vec{k}_I}\left(s_1,s_2,w-u\right)I^{\left(\nu_3,\nu_4,\nu\right)}_{\vec{k}_3,\vec{k}_4,\vec{k}_I}\left(s_3,s_4,{\bar w}+u\right)\\
    \times \frac{e^{\mp 2\left(u-w\right)\pi i}-e^{\pm 2\left(u+{\bar w}\right)\pi i}}{u+\epsilon}\Big|_{{}^{w=\frac{d}{4}-s_1-s_2}_{{\bar w}=\frac{d}{4}-s_3-s_4}},
\end{multline}
where now $-\epsilon < \mathfrak{Re}\left[u\right]$ and we subtracted the residue of the integrand in \eqref{uintinitial} at $u=\epsilon$:
\begin{multline}
    {}^{({\sf s})}\langle \phi^{(\nu_1)}_{\vec{k}_1}\phi^{(\nu_2)}_{\vec{k}_2}\phi^{(\nu_3)}_{\vec{k}_3}\phi^{(\nu_4)}_{\vec{k}_4} \rangle^\prime_{\pm \pm,>\odot}\\=e^{\pm (\nu_1+\nu_2+\nu_3+\nu_4) \frac{\pi}{2}} \frac{L^{2\left(d+1\right)}}{16\pi} {\cal N}_4\left(\eta_0,k_i\right)    I^{\left(\nu_1,\nu_2,\nu\right)}_{\vec{k}_1,\vec{k}_2,\vec{k}_I}I^{\left(\nu_3,\nu_4,\nu\right)}_{\vec{k}_3e^{\pm \pi i},\vec{k}_4e^{\pm \pi i},\vec{k}_I}.
\end{multline}
The contributions from the $++$ and $--$ contours then neatly combine as\footnote{To simplify this expression we used that: 
\begin{equation}\label{trigid}
    \cos \left(\theta_1\right)-\cos \left(\theta_2\right)=2\sin\left(\frac{\theta_1+\theta_2}{2}\right)\sin\left(\frac{\theta_2-\theta_1}{2}\right).
\end{equation}}
\begin{multline}\label{uintfinal}
      \sum_{\pm \pm}\left[ {}^{({\sf s})}\langle \phi^{(\nu_1)}_{\vec{k}_1}\phi^{(\nu_2)}_{\vec{k}_2}\phi^{(\nu_3)}_{\vec{k}_3}\phi^{(\nu_4)}_{\vec{k}_4} \rangle^\prime_{\pm \pm} +  {}^{({\sf s})}\langle \phi^{(\nu_1)}_{\vec{k}_1}\phi^{(\nu_2)}_{\vec{k}_2}\phi^{(\nu_3)}_{\vec{k}_3}\phi^{(\nu_4)}_{\vec{k}_4} \rangle^\prime_{\pm \pm,>\odot}\right]
       = \frac{L^{2\left(d+1\right)}}{4\pi} {\cal N}_4\left(\eta_0,k_i\right)\\ \times \int^{i\infty}_{-i\infty} \frac{du}{2\pi i} \int \left[ds\right]_4\, \,\frac{1}{u+\epsilon}\sin (\pi  ({\bar w}-w+2 u)) \sin \left(\pi\left(\tfrac{d+i(\nu_1+\nu_2+\nu_3+\nu_4)}{2}-w-{\bar w}\right)\right)\\\times I^{\left(\nu_1,\nu_2,\nu\right)}_{\vec{k}_1,\vec{k}_2,\vec{k}_I}\left(s_1,s_2,w-u\right)I^{\left(\nu_3,\nu_4,\nu\right)}_{\vec{k}_3,\vec{k}_4,\vec{k}_I}\left(s_3,s_4,{\bar w}+u\right)\Big|_{{}^{w=\frac{d}{4}-s_1-s_2}_{{\bar w}=\frac{d}{4}-s_3-s_4}},
 \end{multline}
where
\begin{multline}
    \sum_{\pm \pm} {}^{({\sf s})}\langle \phi^{(\nu_1)}_{\vec{k}_1}\phi^{(\nu_2)}_{\vec{k}_2}\phi^{(\nu_3)}_{\vec{k}_3}\phi^{(\nu_4)}_{\vec{k}_4} \rangle^\prime_{\pm \pm,>\odot}\\= \frac{1}{8\pi} L^{2\left(d+1\right)}  {\cal N}_4\left(\eta_0,k_i\right)\int \left[ds\right]_4 \, \cos\left(\left(\tfrac{d+(\nu_1+\nu_2+\nu_3+\nu_4)i}{2}-2{\bar w}\right)\pi\right)\\
  \times I^{\left(\nu_1,\nu_2,\nu\right)}_{\vec{k}_1,\vec{k}_2,\vec{k}_I}\left(s_1,s_2,w\right)I^{\left(\nu_3,\nu_4,\nu\right)}_{\vec{k}_3,\vec{k}_4,\vec{k}_I}\left(s_3,s_4,{\bar w}\right)\Big|_{{}^{w=\frac{d}{4}-s_1-s_2}_{{\bar w}=\frac{d}{4}-s_3-s_4}}.
\end{multline}
The sinusoidal factor in the $u$-integrand of \eqref{uintfinal} provides additional zeros where in the individual $++$ and $--$ contributions there would be poles. Physically, this represents the interference between processes on different branches of the in-in contour. On a practical level, this makes evaluating the $u$-integral after combining the contributions along the in-in contour simpler than evaluating the $u$-integral for each individual contribution first.\footnote{Evaluating the $u$-integral for each individual contribution first is still straightforward, just the resulting expressions generated by each contribution are more involved -- making it harder to combine them and to study how they interfere among each other.}

The $u$-integral we have left to evaluate is:
\begin{multline}\label{appuint}
     I_{\text{exch}}  =  \int^{i\infty}_{-i\infty} \frac{du}{2\pi i} \,\frac{1}{u+\epsilon}\sin (\pi  ({\bar w}-w+2 u))\\ \times \Gamma\left(w+\tfrac{i\nu}{2}-u\right)\Gamma\left(w-\tfrac{i\nu}{2}-u\right)\Gamma\left({\bar w}+\tfrac{i\nu}{2}+u\right)\Gamma\left({\bar w}-\tfrac{i\nu}{2}+u\right).
 \end{multline}
To evaluate the integral, it is simplest to close the integration contour to the right, so that it avoids the pole at $u=\epsilon$ and just encloses the following two series of Gamma function poles:
\begin{equation}
   u = w\pm\tfrac{i\nu}{2}+n, \quad n \in \mathbb{N}.
\end{equation}
Re-summing the contributions from the residues in each series gives the result for the integral as a sum of two generalised Hypergeometric functions ${}_3F_2$:
\begin{multline}
    I_{\text{exch}} = \Gamma \left(w+{\bar w}\right)\left[ \frac{\Gamma (-i \nu ) \sin \left(\pi \left(w+{\bar w}+i\nu\right)\right)\Gamma \left(w+{\bar w}+i\nu\right)}{w+\tfrac{i\nu}{2}} \right. \\
     \, \times {}_3F_2\left(w+{\bar w},w+\tfrac{i \nu }{2},w+{\bar w}+i\nu;w+\tfrac{i \nu }{2}+1,i \nu +1;1\right)\\
     + \left.\nu \rightarrow -\nu\right].
\end{multline}
Since generalised Hypergeometric functions (see \eqref{app3f2}) are defined by Mellin-Barnes integrals which, when inserted into the above equation, gives a Mellin-Barnes integral which is more cumbersome than the original integral \eqref{appuint}, naively it seems that we did not get any further than from where we started. Quite remarkably, there is an identity which relates precisely the above combination of ${}_3F_2$ to a single term involving only Gamma functions. This originates from the three-term relations between ${}_3F_2$ series at argument $z=1$ \cite{Baileybook}. It is: 
\begin{multline}
  \frac{\pi^3\Gamma \left(1-w-{\bar w}\right)}{\Gamma \left(1-w+\tfrac{i \nu }{2}\right)\Gamma \left(1-w-\tfrac{i \nu }{2}\right)\Gamma \left(1-{\bar w}+\tfrac{i \nu }{2}\right)\Gamma \left(1-{\bar w}-\tfrac{i \nu }{2}\right)}\\ = \left[\sin \left(\pi  (w-\tfrac{i \nu}{2})\right)\sin \left(\pi  (w+\tfrac{i \nu}{2})\right)\frac{\Gamma (-i \nu)\Gamma \left(w+{\bar w}+i\nu\right)\sin \left(\pi \left(w+{\bar w}+i\nu\right)\right)}{w+\tfrac{i \nu}{2}}
    \right.\\\left.
    \, \times {}_3F_2\left(w+{\bar w},w+\tfrac{i \nu }{2},w+{\bar w}+i\nu;w+\tfrac{i \nu }{2}+1,i \nu +1;1\right)\right.\\\left.
    + \nu \rightarrow -\nu \right].
\end{multline}
This gives us:
\begin{multline}
    I_{\text{exch}} = \Gamma \left(w+\tfrac{i \nu }{2}\right)\Gamma \left(w-\tfrac{i \nu }{2}\right)\Gamma \left({\bar w}+\tfrac{i \nu }{2}\right)\Gamma \left({\bar w}-\tfrac{i \nu }{2}\right)
    \\ \times \text{cosec}\left(\left(w+{\bar w}\right)\pi \right)\text{cosec}\left(\pi  ({\bar w}- \tfrac{i \nu}{2})\right)\text{cosec}\left(\pi  ({\bar w}+\tfrac{i \nu}{2})\right),
\end{multline}
whose definition does not involve any Mellin-Barnes integral.

\section{Mellin-Barnes representation of Hypergeometric functions}

\subsection{Gauss Hypergeometric function ${}_2F_1$}
\label{subsec::app2f1}

The Mellin-Barnes representation of the Gauss Hypergeometric function
\begin{equation}\label{app2f1mellin}
    {}_2F_1\left(a,b;c;z\right)= \frac{\Gamma\left(c\right)}{\Gamma\left(a\right)\Gamma\left(b\right)}\int^{i\infty}_{-\infty} \frac{ds}{2\pi i}\, \frac{\Gamma\left(a+s\right)\Gamma\left(b+s\right)\Gamma\left(-s\right)}{\Gamma\left(c+s\right)}\left(-z\right)^s,
\end{equation}
is the analytic continuation of the Gauss Hypergeometric series
\begin{equation}
 {}_2F_1\left(a,b;c;z\right)=\sum^{\infty}_{n=0}\frac{\left(a\right)_n\left(b\right)_n}{\left(c\right)_n} \frac{z^n}{n!}, \quad |z|<1,
\end{equation}
to any closed domain of the entire $z$-plane, which is cut along the real axis from $0$ to $\infty$.

There are linear relationships between Hypergeometric series with different domains of validity, which are straightforwardly proved using Mellin-Barnes integrals. Consider the integral
\begin{equation}\label{3ptccorigs1}
    I\left(a,b;c,d;z\right) = \int^{i\infty}_{-i\infty} \frac{ds}{2\pi i}\,\Gamma\left(s+a\right)\Gamma\left(s+b\right) \Gamma\left(c-s\right)\Gamma\left(d-s\right)
    z^{-2s},
\end{equation}
which is a generalisation of Barnes' first lemma \eqref{blemma1} to include a variable $z$. This type of integral (and minimal variations thereof) appears often in this work. The integral can be evaluated in the usual way by closing the integration contour either side of the imaginary axis and applying Cauchy's residue theorem, since on a circular arc of radius $R \rightarrow 0$ we have
\begin{multline}
    |\Gamma\left(s+a\right)\Gamma\left(s+b\right)\Gamma\left(c-s\right)\Gamma\left(d-s\right)
    z^{-2s}| \\= O\left(e^{-2 \pi  R \left| \sin (\theta )\right|-R \cos (\theta ) \log \left(z^2\right)}e^{-\theta  \mathfrak{Im}[a+b+c+d]+\log (R) \mathfrak{Re}\left[a+b+c+d-2\right]}\right),
\end{multline}
which requires $z^2>1$ if we close to the right and $z^2<1$ if we close to the left. Let us choose to close the contour to the left. This encloses the poles of the Gamma functions $\Gamma\left(s+a\right)$ and $\Gamma\left(s+b\right)$. Summing over the residues of each series of poles gives a sum of two Gauss Hypergeometric functions:
\begin{align}\label{sum2gausshyp}
     I\left(a,b;c,d;z\right) &= z^{2 a} \Gamma (a+c) \Gamma (d+a) \Gamma (b-a) \, _2F_1\left(a+c,d+a;a-b+1;z^2\right)\nonumber \\&+z^{2 b} \Gamma (c+b) \Gamma (b+d) \Gamma (a-b) \, _2F_1\left(c+b,b+d;-a+b+1;z^2\right),
\end{align}
one for each series of Gamma function poles. This combination can be identified with a single Gauss Hypergeometric function of argument $1-z^2$. To obtain such a transformation of the variable $z$, we consider the expansion of \eqref{3ptccorigs1} as a power series in $z^2-z_0$ for some $z_0\ne0$:\footnote{Here it was convenient to re-define $s\rightarrow s+a$ before expanding, so that each term in the expansion involves a product of four Gamma functions in $s$ from the resultant cancellation of the $\Gamma\left(s\right)$.}
\begin{multline}
 z^{-2a} I\left(a,b;c,d;z\right) =  \int^{i\infty}_{-\infty} \frac{ds}{2\pi i}\, \Gamma\left(a+c-s\right)\Gamma\left(d+a-s\right)\Gamma\left(s\right)\Gamma\left(s+b-a\right)\\
 \times \sum^\infty_{n=0}\frac{\left(-1\right)^n}{n!}\frac{\Gamma\left(s+n\right)}{\Gamma\left(s\right)}\left(z^2-z_0\right)^n z_0^{-s-n}.
\end{multline}
By choosing $z_0=1$ and inverting the order of integration and summation, we can evaluate the $s$-integral by applying Barnes' first lemma \eqref{blemma1} to each term in the sum:\footnote{For other choices of $z_0$, each term is an $s$-integral of the same form as \eqref{3ptccorigs1} with $z \rightarrow z_0$.}
\begin{align}\nonumber
   z^{-2a}   I\left(a,b;c,d;z\right) &= \Gamma\left(c+b\right)\Gamma\left(b+d\right) \sum^\infty_{n=0}\frac{\left(-1\right)^n}{n!}
     \frac{\Gamma\left(a+c+n\right)\Gamma\left(d+a+n\right)}{\Gamma\left(a+b+c+d+n\right)}  \left(z^2-1\right)^n \nonumber \\&= \frac{\Gamma\left(c+b\right)\Gamma\left(b+d\right)\Gamma\left(a+c\right)\Gamma\left(d+a\right)}{\Gamma\left(a+b+c+d\right)} \nonumber  \\ & \hspace*{3cm} \times {}_2F_{1}\left(a+c,d+a;a+b+c+d;1-z^2\right), \label{1mz2gh}
\end{align}
which is a single Gauss Hypergeometric function as advertised. 

A special case of the integral \eqref{3ptccorigs1} occurs when two of the four Gamma functions collapse into a single Gamma function via the Legendre duplication formula (when e.g. $a=b-\tfrac{1}{2}$), so that the integral reduces to the form:
\begin{equation}
    I\left(a,b;c,d;z\right) = 2^{1-2a} \sqrt{\pi} \int^{i\infty}_{-i\infty} \frac{ds}{2\pi i}\, \Gamma\left(c-s\right)\Gamma\left(d-s\right)\Gamma\left(2s+2a\right)
    \left(2z\right)^{-2s},
\end{equation}
which we encounter in contact diagrams \eqref{ccsclnptconfstr} in which all but one scalar is conformally coupled. Redefining $s\rightarrow s+a$ and expanding the integral around $z=2$, we can identify, in the same way as above, the sum of Gauss Hypergeometric functions \eqref{sum2gausshyp} (and hence also \eqref{1mz2gh}) with a single Gauss Hypergeometric function of argument $\tfrac{1-z}{2}$:
\begin{multline}\label{I1hyp2f1}
   I\left(a,b;c,d;z\right) =  z^{2a} 2^{-2(2a+c+d-1)} \frac{\pi \Gamma\left(2c+2a\right)\Gamma\left(2d+2a\right)}{\Gamma\left(c+d+2a+\tfrac{1}{2}\right)}\\ \times {}_2F_{1}\left(2c+2a,2d+2a;c+d+2a+\tfrac{1}{2};\tfrac{1-z}{2}\right).
\end{multline}

\subsection{Generalised Hypergeometric Functions}
\label{subsec::appgenhypf}

Mellin-Barnes integrals define generalised Hypergeometric functions \cite{gelfand1986general}. For example, the generalised Hypergeometric function ${}_3F_2$ is given by
\begin{equation}\label{app3f2}
    {}_3F_2\left(a,b;c,d;z\right)= \frac{\Gamma\left(c\right)\Gamma\left(d\right)}{\Gamma\left(a\right)\Gamma\left(b\right)}\int^{i\infty}_{-\infty} \frac{ds}{2\pi i}\, \frac{\Gamma\left(a+s\right)\Gamma\left(b+s\right)\Gamma\left(-s\right)}{\Gamma\left(c+s\right)\Gamma\left(d+s\right)}\left(-z\right)^s.
\end{equation}

As well as generalising Hypergeometric functions by increasing the number of parameters as above, we can also increase the number of variables. An example is the Appell function \cite{appell1880series,AppelletKampe}, which is a generalised Hypergeometric function of two variables. In this work we often encounter the Appell function $F_4$:
\begin{multline}\label{appellmellin}
     F_4\left(a_1,a_2,b_2,b_2;x,y\right) = \frac{\Gamma\left(b_1\right)\Gamma\left(b_2\right)}{\Gamma\left(a_1\right)\Gamma\left(a_2\right)} \int^{i\infty}_{-\infty} \frac{ds}{2\pi i}\frac{dt}{2\pi i}\,\frac{\Gamma\left(s+t+a_1\right)\Gamma\left(s+t+a_2\right)}{\Gamma\left(s+b_1\right)\Gamma\left(t+b_2\right)}\\\times \Gamma\left(-s\right)\Gamma\left(-t\right) \left(-x\right)^{s}\left(-y\right)^{t}.
\end{multline}

The late time three-point correlation function \eqref{3pt000} for generic external scalars is given by the above Appell function. This can be seen by noting that
\begin{subequations}
\begin{align}
 \Gamma\left(s_1+\tfrac{i\nu_1}{2}\right)\Gamma\left(s_1-\tfrac{i\nu_1}{2}\right)\Big|_{s_1 \rightarrow -(s_1+\tfrac{i\nu_1}{2})} &= \Gamma\left(-i\nu_1\right)\Gamma\left(i\nu_1+1\right)\left(-1\right)^{-s_1}\frac{\Gamma\left(-s_1\right)}{\Gamma\left(s_1+i\nu_1+1\right)},\\
    \Gamma\left(s_2+\tfrac{i\nu_2}{2}\right)\Gamma\left(s_2-\tfrac{i\nu_2}{2}\right)\Big|_{s_2 \rightarrow -(s_2+\tfrac{i\nu_2}{2})} &= \Gamma\left(-i\nu_2\right)\Gamma\left(i\nu_2+1\right)\left(-1\right)^{-s_2}\frac{\Gamma\left(-s_2\right)}{\Gamma\left(s_2+i\nu_2+1\right)}.
\end{align}
\end{subequations}

\section{Pole generation in multiple-Mellin-Barnes integrals}

\label{appendix::polegen}

Since late-time correlators are in general given by multiple Mellin-Barnes integrals, the entire pole structure of the integrand in a given Mellin variable may not be manifest. It can however in general be determined without the need to evaluate any of the Mellin integrals, as we describe in the following. There are two mechanisms through which poles in a given Mellin variable, say $u$, can be generated by an integral in a second Mellin variable, say $s$, with which it is entangled: 

\subsubsection*{1. Collision of poles}

Poles at the values of $u$ for which series of poles in the second Mellin variable $s$ collide. For a simple example of this mechanism consider the double Mellin-Barnes integral:
    \begin{equation}
    I_1=\int^{i\infty}_{-i\infty}\frac{du}{2\pi i}\frac{ds}{2\pi i}\, \Gamma (b+s) \Gamma (a-s+u) \Gamma\left(-u\right) \epsilon^u.
\end{equation}
We would like to determine the poles in the Mellin variable $u$. While the poles of the Gamma function $\Gamma\left(-u\right)$ are manifest, there may be poles generated by the Mellin integral in the variable $s$ since the two integrals are entangled through the Gamma function $\Gamma (a-s+u)$. Considering $u$ fixed, the $s$-integrand has the following two series of poles:
   \begin{subequations}
    \begin{align}
        s&=-\left(b+n\right), \qquad n \in \mathbb{N}_0, \\s&=a+u+m, \qquad m \in \mathbb{N}_0,
    \end{align}
   \end{subequations}
   which overlap by $n^\prime$ poles when $u=-\left(a+b+n^\prime\right)$. From this we can infer that the $u$-integrand has poles precisely at those values. This can be easily verified in the current example because the $s$ integral is simple to evaluate:
      \begin{equation}
    I_1=2^{-\left(a+b\right)}\int^{i\infty}_{-i\infty}\frac{du}{2\pi i}\,  \Gamma (a+b+u) \Gamma\left(-u\right) \left(\frac{\epsilon}{2}\right)^u,
\end{equation}
which can be obtained by closing the contour on either series of Gamma function poles. The poles at $u=-\left(a+b+n^\prime\right)$ for $n^\prime \in \mathbb{N}_0$ are now manifest as anticipated, where they are encoded in the Gamma function $\Gamma (a+b+u)$.

\subsubsection*{2. Divergences}   
   
Poles at the values of $u$ for which the integral in the second Mellin variable $s$ diverges. Consider the integral:
    \begin{equation}
   I_2=\int^{i\infty}_{-i\infty}\frac{du}{2\pi i}\frac{ds}{2\pi i} \frac{ \Gamma (-2 s) \Gamma (2 s-a) \Gamma (u-2 s)\Gamma\left(-u\right)}{\Gamma (a-2 s)} (-1)^{2 s}  \epsilon ^u.
\end{equation}
Using Stirling's formula, we have that:
\begin{equation}
    \frac{ \Gamma (-2 s) \Gamma (2 s-a) \Gamma (u-2 s)}{\Gamma (a-2 s)} (-1)^{2 s}  \sim |\mathfrak{Im}\left[s\right]|^{\mathfrak{Re}\left[u\right]-1-2a},
\end{equation}
as $\mathfrak{Im}\left[s\right] \rightarrow -\infty$, so the integral in $s$ diverges for $\mathfrak{Re}\left[u-2a\right] \ge 0$. This translates into a series of poles at $u=2a+n$ for $n \in \mathbb{N}_0$, as can be verified upon evaluating the $s$-integral:
    \begin{equation}
   I_2= \frac{e^{i \pi  a}}{2}\int^{i\infty}_{-i\infty}\frac{du}{2\pi i} \frac{\Gamma (1-a) \Gamma (2 a-u) \Gamma (-a+u+1)\Gamma\left(-u\right)}{\Gamma (a+1) \Gamma (a-u+1)} \epsilon ^u.
\end{equation}
where the anticipated poles are now manifest, where they are encoded in the Gamma function $\Gamma (2 a-u)$. Instead the poles at $u=a-n-1$ for $n \in \mathbb{N}_0$ are generated by the first mechanism -- i.e. due to the collision of poles in the Mellin variable $s$ at those values of $u$.

\bibliographystyle{JHEP}
\bibliography{refs}

\providecommand{\href}[2]{#2}\begingroup\raggedright\begin{thebibliography}{100}

\bibitem{Guth:1980zm}
A.~H. Guth, \emph{{The Inflationary Universe: A Possible Solution to the
  Horizon and Flatness Problems}},
  \href{https://doi.org/10.1103/PhysRevD.23.347}{\emph{Phys. Rev.} {\bfseries
  D23} (1981) 347}.

\bibitem{Linde:1981mu}
A.~D. Linde, \emph{{A New Inflationary Universe Scenario: A Possible Solution
  of the Horizon, Flatness, Homogeneity, Isotropy and Primordial Monopole
  Problems}}, \href{https://doi.org/10.1016/0370-2693(82)91219-9}{\emph{Phys.
  Lett.} {\bfseries 108B} (1982) 389}.

\bibitem{Albrecht:1982wi}
A.~Albrecht and P.~J. Steinhardt, \emph{{Cosmology for Grand Unified Theories
  with Radiatively Induced Symmetry Breaking}},
  \href{https://doi.org/10.1103/PhysRevLett.48.1220}{\emph{Phys. Rev. Lett.}
  {\bfseries 48} (1982) 1220}.

\bibitem{Starobinsky:1982ee}
A.~A. Starobinsky, \emph{{Dynamics of Phase Transition in the New Inflationary
  Universe Scenario and Generation of Perturbations}},
  \href{https://doi.org/10.1016/0370-2693(82)90541-X}{\emph{Phys. Lett.}
  {\bfseries 117B} (1982) 175}.

\bibitem{Chen:2010xka}
X.~Chen, \emph{{Primordial Non-Gaussianities from Inflation Models}},
  \href{https://doi.org/10.1155/2010/638979}{\emph{Adv. Astron.} {\bfseries
  2010} (2010) 638979} [\href{https://arxiv.org/abs/1002.1416}{{\ttfamily
  1002.1416}}].

\bibitem{Maldacena:2011nz}
J.~M. Maldacena and G.~L. Pimentel, \emph{{On graviton non-Gaussianities during
  inflation}}, \href{https://doi.org/10.1007/JHEP09(2011)045}{\emph{JHEP}
  {\bfseries 09} (2011) 045} [\href{https://arxiv.org/abs/1104.2846}{{\ttfamily
  1104.2846}}].

\bibitem{Mata:2012bx}
I.~Mata, S.~Raju and S.~Trivedi, \emph{{CMB from CFT}},
  \href{https://doi.org/10.1007/JHEP07(2013)015}{\emph{JHEP} {\bfseries 07}
  (2013) 015} [\href{https://arxiv.org/abs/1211.5482}{{\ttfamily 1211.5482}}].

\bibitem{Anninos:2014lwa}
D.~Anninos, T.~Anous, D.~Z. Freedman and G.~Konstantinidis, \emph{{Late-time
  Structure of the Bunch-Davies De Sitter Wavefunction}},
  \href{https://doi.org/10.1088/1475-7516/2015/11/048}{\emph{JCAP} {\bfseries
  1511} (2015) 048} [\href{https://arxiv.org/abs/1406.5490}{{\ttfamily
  1406.5490}}].

\bibitem{Ghosh:2014kba}
A.~Ghosh, N.~Kundu, S.~Raju and S.~P. Trivedi, \emph{{Conformal Invariance and
  the Four Point Scalar Correlator in Slow-Roll Inflation}},
  \href{https://doi.org/10.1007/JHEP07(2014)011}{\emph{JHEP} {\bfseries 07}
  (2014) 011} [\href{https://arxiv.org/abs/1401.1426}{{\ttfamily 1401.1426}}].

\bibitem{Kehagias:2015jha}
A.~Kehagias and A.~Riotto, \emph{{High Energy Physics Signatures from Inflation
  and Conformal Symmetry of de Sitter}},
  \href{https://doi.org/10.1002/prop.201500025}{\emph{Fortsch. Phys.}
  {\bfseries 63} (2015) 531}
  [\href{https://arxiv.org/abs/1501.03515}{{\ttfamily 1501.03515}}].

\bibitem{Arkani-Hamed:2015bza}
N.~Arkani-Hamed and J.~Maldacena, \emph{{Cosmological Collider Physics}},
  \href{https://arxiv.org/abs/1503.08043}{{\ttfamily 1503.08043}}.

\bibitem{Lee:2016vti}
H.~Lee, D.~Baumann and G.~L. Pimentel, \emph{{Non-Gaussianity as a Particle
  Detector}}, \href{https://doi.org/10.1007/JHEP12(2016)040}{\emph{JHEP}
  {\bfseries 12} (2016) 040}
  [\href{https://arxiv.org/abs/1607.03735}{{\ttfamily 1607.03735}}].

\bibitem{Arkani-Hamed:2017fdk}
N.~Arkani-Hamed, P.~Benincasa and A.~Postnikov, \emph{{Cosmological Polytopes
  and the Wavefunction of the Universe}},
  \href{https://arxiv.org/abs/1709.02813}{{\ttfamily 1709.02813}}.

\bibitem{Benincasa:2018ssx}
P.~Benincasa, \emph{{From the flat-space S-matrix to the Wavefunction of the
  Universe}},  \href{https://arxiv.org/abs/1811.02515}{{\ttfamily 1811.02515}}.

\bibitem{Li:2018wkt}
S.~Y. Li, Y.~Wang and S.~Zhou, \emph{{KLT-Like Behaviour of Inflationary
  Graviton Correlators}},
  \href{https://doi.org/10.1088/1475-7516/2018/12/023}{\emph{JCAP} {\bfseries
  1812} (2018) 023} [\href{https://arxiv.org/abs/1806.06242}{{\ttfamily
  1806.06242}}].

\bibitem{Farrow:2018yni}
J.~A. Farrow, A.~E. Lipstein and P.~McFadden, \emph{{Double copy structure of
  CFT correlators}}, \href{https://doi.org/10.1007/JHEP02(2019)130}{\emph{JHEP}
  {\bfseries 02} (2019) 130}
  [\href{https://arxiv.org/abs/1812.11129}{{\ttfamily 1812.11129}}].

\bibitem{Arkani-Hamed:2018kmz}
N.~Arkani-Hamed, D.~Baumann, H.~Lee and G.~L. Pimentel, \emph{{The Cosmological
  Bootstrap: Inflationary Correlators from Symmetries and Singularities}},
  \href{https://arxiv.org/abs/1811.00024}{{\ttfamily 1811.00024}}.

\bibitem{Goon:2018fyu}
G.~Goon, K.~Hinterbichler, A.~Joyce and M.~Trodden, \emph{{Shapes of gravity:
  Tensor non-Gaussianity and massive spin-2 fields}},
  \href{https://arxiv.org/abs/1812.07571}{{\ttfamily 1812.07571}}.

\bibitem{ToAppear}
C.~Sleight and M.~Taronna, \emph{{Bootstrapping Inflationary Correlators in
  Mellin Space}},  \href{https://arxiv.org/abs/19xx.xxxxx}{{\ttfamily
  19xx.xxxxx}}.

\bibitem{Coriano:2013jba}
C.~Coriano, L.~Delle~Rose, E.~Mottola and M.~Serino, \emph{{Solving the
  Conformal Constraints for Scalar Operators in Momentum Space and the
  Evaluation of Feynman's Master Integrals}},
  \href{https://doi.org/10.1007/JHEP07(2013)011}{\emph{JHEP} {\bfseries 07}
  (2013) 011} [\href{https://arxiv.org/abs/1304.6944}{{\ttfamily 1304.6944}}].

\bibitem{Bzowski:2013sza}
A.~Bzowski, P.~McFadden and K.~Skenderis, \emph{{Implications of conformal
  invariance in momentum space}},
  \href{https://doi.org/10.1007/JHEP03(2014)111}{\emph{JHEP} {\bfseries 03}
  (2014) 111} [\href{https://arxiv.org/abs/1304.7760}{{\ttfamily 1304.7760}}].

\bibitem{Bzowski:2015pba}
A.~Bzowski, P.~McFadden and K.~Skenderis, \emph{{Scalar 3-point functions in
  CFT: renormalisation, beta functions and anomalies}},
  \href{https://doi.org/10.1007/JHEP03(2016)066}{\emph{JHEP} {\bfseries 03}
  (2016) 066} [\href{https://arxiv.org/abs/1510.08442}{{\ttfamily
  1510.08442}}].

\bibitem{Bzowski:2017poo}
A.~Bzowski, P.~McFadden and K.~Skenderis, \emph{{Renormalised 3-point functions
  of stress tensors and conserved currents in CFT}},
  \href{https://doi.org/10.1007/JHEP11(2018)153}{\emph{JHEP} {\bfseries 11}
  (2018) 153} [\href{https://arxiv.org/abs/1711.09105}{{\ttfamily
  1711.09105}}].

\bibitem{Coriano:2018zdo}
C.~Corianò and M.~M. Maglio, \emph{{Renormalization, Conformal Ward Identities
  and the Origin of a Conformal Anomaly Pole}},
  \href{https://doi.org/10.1016/j.physletb.2018.04.003}{\emph{Phys. Lett.}
  {\bfseries B781} (2018) 283}
  [\href{https://arxiv.org/abs/1802.01501}{{\ttfamily 1802.01501}}].

\bibitem{Coriano:2018bbe}
C.~Corian\`o and M.~M. Maglio, \emph{{Exact Correlators from Conformal Ward
  Identities in Momentum Space and the Perturbative $TJJ$ Vertex}},
  \href{https://doi.org/10.1016/j.nuclphysb.2018.11.016}{\emph{Nucl. Phys.}
  {\bfseries B938} (2019) 440}
  [\href{https://arxiv.org/abs/1802.07675}{{\ttfamily 1802.07675}}].

\bibitem{Isono:2018rrb}
H.~Isono, T.~Noumi and G.~Shiu, \emph{{Momentum space approach to crossing
  symmetric CFT correlators}},
  \href{https://doi.org/10.1007/JHEP07(2018)136}{\emph{JHEP} {\bfseries 07}
  (2018) 136} [\href{https://arxiv.org/abs/1805.11107}{{\ttfamily
  1805.11107}}].

\bibitem{Bzowski:2018fql}
A.~Bzowski, P.~McFadden and K.~Skenderis, \emph{{Renormalised CFT 3-point
  functions of scalars, currents and stress tensors}},
  \href{https://doi.org/10.1007/JHEP11(2018)159}{\emph{JHEP} {\bfseries 11}
  (2018) 159} [\href{https://arxiv.org/abs/1805.12100}{{\ttfamily
  1805.12100}}].

\bibitem{Coriano:2018bsy}
C.~Corian\`o and M.~M. Maglio, \emph{{The general 3-graviton vertex ($TTT$) of
  conformal field theories in momentum space in $d=4$}},
  \href{https://doi.org/10.1016/j.nuclphysb.2018.10.007}{\emph{Nucl. Phys.}
  {\bfseries B937} (2018) 56}
  [\href{https://arxiv.org/abs/1808.10221}{{\ttfamily 1808.10221}}].

\bibitem{Maglio:2019grh}
C.~Corian\`o and M.~M. Maglio, \emph{{On Some Hypergeometric Solutions of the
  Conformal Ward Identities of Scalar 4-point Functions in Momentum Space}},
  \href{https://arxiv.org/abs/1903.05047}{{\ttfamily 1903.05047}}.

\bibitem{Isono:2019ihz}
H.~Isono, T.~Noumi and T.~Takeuchi, \emph{{Momentum space conformal three-point
  functions of conserved currents and a general spinning operator}},
  \href{https://arxiv.org/abs/1903.01110}{{\ttfamily 1903.01110}}.

\bibitem{Liu:1998th}
H.~Liu, \emph{{Scattering in anti-de Sitter space and operator product
  expansion}}, \href{https://doi.org/10.1103/PhysRevD.60.106005}{\emph{Phys.
  Rev.} {\bfseries D60} (1999) 106005}
  [\href{https://arxiv.org/abs/hep-th/9811152}{{\ttfamily hep-th/9811152}}].

\bibitem{Mack:2009gy}
G.~Mack, \emph{{D-dimensional Conformal Field Theories with anomalous
  dimensions as Dual Resonance Models}}, {\emph{Bulg. J. Phys.} {\bfseries 36}
  (2009) 214} [\href{https://arxiv.org/abs/0909.1024}{{\ttfamily 0909.1024}}].

\bibitem{Mack:2009mi}
G.~Mack, \emph{{D-independent representation of Conformal Field Theories in D
  dimensions via transformation to auxiliary Dual Resonance Models. Scalar
  amplitudes}},  \href{https://arxiv.org/abs/0907.2407}{{\ttfamily 0907.2407}}.

\bibitem{Penedones:2010ue}
J.~Penedones, \emph{{Writing CFT correlation functions as AdS scattering
  amplitudes}}, \href{https://doi.org/10.1007/JHEP03(2011)025}{\emph{JHEP}
  {\bfseries 03} (2011) 025} [\href{https://arxiv.org/abs/1011.1485}{{\ttfamily
  1011.1485}}].

\bibitem{Paulos:2011ie}
M.~F. Paulos, \emph{{Towards Feynman rules for Mellin amplitudes}},
  \href{https://doi.org/10.1007/JHEP10(2011)074}{\emph{JHEP} {\bfseries 10}
  (2011) 074} [\href{https://arxiv.org/abs/1107.1504}{{\ttfamily 1107.1504}}].

\bibitem{Fitzpatrick:2011ia}
A.~L. Fitzpatrick, J.~Kaplan, J.~Penedones, S.~Raju and B.~C. van Rees,
  \emph{{A Natural Language for AdS/CFT Correlators}},
  \href{https://doi.org/10.1007/JHEP11(2011)095}{\emph{JHEP} {\bfseries 11}
  (2011) 095} [\href{https://arxiv.org/abs/1107.1499}{{\ttfamily 1107.1499}}].

\bibitem{Fitzpatrick:2011hu}
A.~L. Fitzpatrick and J.~Kaplan, \emph{{Analyticity and the Holographic
  S-Matrix}}, \href{https://doi.org/10.1007/JHEP10(2012)127}{\emph{JHEP}
  {\bfseries 10} (2012) 127} [\href{https://arxiv.org/abs/1111.6972}{{\ttfamily
  1111.6972}}].

\bibitem{Chalmers:1998wu}
G.~Chalmers and K.~Schalm, \emph{{The Large N(c) limit of four point functions
  in N=4 superYang-Mills theory from Anti-de Sitter supergravity}},
  \href{https://doi.org/10.1016/S0550-3213(99)00275-8}{\emph{Nucl. Phys.}
  {\bfseries B554} (1999) 215}
  [\href{https://arxiv.org/abs/hep-th/9810051}{{\ttfamily hep-th/9810051}}].

\bibitem{Raju:2010by}
S.~Raju, \emph{{BCFW for Witten Diagrams}},
  \href{https://doi.org/10.1103/PhysRevLett.106.091601}{\emph{Phys. Rev. Lett.}
  {\bfseries 106} (2011) 091601}
  [\href{https://arxiv.org/abs/1011.0780}{{\ttfamily 1011.0780}}].

\bibitem{Raju:2012zr}
S.~Raju, \emph{{New Recursion Relations and a Flat Space Limit for AdS/CFT
  Correlators}}, \href{https://doi.org/10.1103/PhysRevD.85.126009}{\emph{Phys.
  Rev.} {\bfseries D85} (2012) 126009}
  [\href{https://arxiv.org/abs/1201.6449}{{\ttfamily 1201.6449}}].

\bibitem{Albayrak:2018tam}
S.~Albayrak and S.~Kharel, \emph{{Towards the higher point holographic momentum
  space amplitudes}},
  \href{https://doi.org/10.1007/JHEP02(2019)040}{\emph{JHEP} {\bfseries 02}
  (2019) 040} [\href{https://arxiv.org/abs/1810.12459}{{\ttfamily
  1810.12459}}].

\bibitem{Albayrak:2019asr}
S.~Albayrak, C.~Chowdhury and S.~Kharel, \emph{{New relation for AdS
  amplitudes}},  \href{https://arxiv.org/abs/1904.10043}{{\ttfamily
  1904.10043}}.

\bibitem{Spradlin:2001pw}
M.~Spradlin, A.~Strominger and A.~Volovich, \emph{{Les Houches lectures on de
  Sitter space}},  in \emph{{Unity from duality: Gravity, gauge theory and
  strings. Proceedings, NATO Advanced Study Institute, Euro Summer School, 76th
  session, Les Houches, France, July 30-August 31, 2001}}, pp.~423--453, 2001,
  \href{https://arxiv.org/abs/hep-th/0110007}{{\ttfamily hep-th/0110007}}.

\bibitem{Baumann:2009ds}
D.~Baumann, \emph{{Inflation}},  in \emph{{Physics of the large and the small,
  TASI 09, proceedings of the Theoretical Advanced Study Institute in
  Elementary Particle Physics, Boulder, Colorado, USA, 1-26 June 2009}},
  pp.~523--686, 2011, \href{https://arxiv.org/abs/0907.5424}{{\ttfamily
  0907.5424}}, \href{https://doi.org/10.1142/9789814327183_0010}{DOI}.

\bibitem{Anninos:2012qw}
D.~Anninos, \emph{{De Sitter Musings}},
  \href{https://doi.org/10.1142/S0217751X1230013X}{\emph{Int. J. Mod. Phys.}
  {\bfseries A27} (2012) 1230013}
  [\href{https://arxiv.org/abs/1205.3855}{{\ttfamily 1205.3855}}].

\bibitem{Akhmedov:2013vka}
E.~T. Akhmedov, \emph{{Lecture notes on interacting quantum fields in de Sitter
  space}}, \href{https://doi.org/10.1142/S0218271814300018}{\emph{Int. J. Mod.
  Phys.} {\bfseries D23} (2014) 1430001}
  [\href{https://arxiv.org/abs/1309.2557}{{\ttfamily 1309.2557}}].

\bibitem{Chen:2017ryl}
X.~Chen, Y.~Wang and Z.-Z. Xianyu, \emph{{Schwinger-Keldysh Diagrammatics for
  Primordial Perturbations}},
  \href{https://doi.org/10.1088/1475-7516/2017/12/006}{\emph{JCAP} {\bfseries
  1712} (2017) 006} [\href{https://arxiv.org/abs/1703.10166}{{\ttfamily
  1703.10166}}].

\bibitem{doi:10.1063/1.1665471}
F.~Schwarz, \emph{Unitary irreducible representations of the groups so0(n, 1)},
  \href{https://doi.org/10.1063/1.1665471}{\emph{Journal of Mathematical
  Physics} {\bfseries 12} (1971) 131}
  [\href{https://arxiv.org/abs/https://doi.org/10.1063/1.1665471}{{\ttfamily
  https://doi.org/10.1063/1.1665471}}].

\bibitem{Dobrev:1977qv}
V.~K. Dobrev, G.~Mack, V.~B. Petkova, S.~G. Petrova and I.~T. Todorov,
  \emph{{Harmonic Analysis on the n-Dimensional Lorentz Group and Its
  Application to Conformal Quantum Field Theory}},
  \href{https://doi.org/10.1007/BFb0009678}{\emph{Lect. Notes Phys.} {\bfseries
  63} (1977) 1}.

\bibitem{Joung:2006gj}
E.~Joung, J.~Mourad and R.~Parentani, \emph{{Group theoretical approach to
  quantum fields in de Sitter space. I. The Principle series}},
  \href{https://doi.org/10.1088/1126-6708/2006/08/082}{\emph{JHEP} {\bfseries
  08} (2006) 082} [\href{https://arxiv.org/abs/hep-th/0606119}{{\ttfamily
  hep-th/0606119}}].

\bibitem{Joung:2007je}
E.~Joung, J.~Mourad and R.~Parentani, \emph{{Group theoretical approach to
  quantum fields in de Sitter space. II. The complementary and discrete
  series}}, \href{https://doi.org/10.1088/1126-6708/2007/09/030}{\emph{JHEP}
  {\bfseries 09} (2007) 030} [\href{https://arxiv.org/abs/0707.2907}{{\ttfamily
  0707.2907}}].

\bibitem{Basile:2016aen}
T.~Basile, X.~Bekaert and N.~Boulanger, \emph{{Mixed-symmetry fields in de
  Sitter space: a group theoretical glance}},
  \href{https://arxiv.org/abs/1612.08166}{{\ttfamily 1612.08166}}.

\bibitem{Burges:1984qm}
C.~J.~C. Burges, \emph{{The De Sitter Vacuum}},
  \href{https://doi.org/10.1016/0550-3213(84)90562-5}{\emph{Nucl. Phys.}
  {\bfseries B247} (1984) 533}.

\bibitem{Mottola:1984ar}
E.~Mottola, \emph{{Particle Creation in de Sitter Space}},
  \href{https://doi.org/10.1103/PhysRevD.31.754}{\emph{Phys. Rev.} {\bfseries
  D31} (1985) 754}.

\bibitem{Allen:1985ux}
B.~Allen, \emph{{Vacuum States in de Sitter Space}},
  \href{https://doi.org/10.1103/PhysRevD.32.3136}{\emph{Phys. Rev.} {\bfseries
  D32} (1985) 3136}.

\bibitem{Gibbons:1977mu}
G.~W. Gibbons and S.~W. Hawking, \emph{{Cosmological Event Horizons,
  Thermodynamics, and Particle Creation}},
  \href{https://doi.org/10.1103/PhysRevD.15.2738}{\emph{Phys. Rev.} {\bfseries
  D15} (1977) 2738}.

\bibitem{Streater:1989vi}
R.~F. Streater and A.~S. Wightman, \emph{{PCT, spin and statistics, and all
  that}}. 1989.

\bibitem{PhysRevLett.73.1746}
J.~Bros, J.-P. Gazeau and U.~Moschella, \emph{Quantum field theory in the de
  sitter universe},
  \href{https://doi.org/10.1103/PhysRevLett.73.1746}{\emph{Phys. Rev. Lett.}
  {\bfseries 73} (1994) 1746}.

\bibitem{Bros:1995js}
J.~Bros and U.~Moschella, \emph{{Two point functions and quantum fields in de
  Sitter universe}},
  \href{https://doi.org/10.1142/S0129055X96000123}{\emph{Rev. Math. Phys.}
  {\bfseries 8} (1996) 327}
  [\href{https://arxiv.org/abs/gr-qc/9511019}{{\ttfamily gr-qc/9511019}}].

\bibitem{doi:10.1063/1.1703727}
J.~Schwinger, \emph{Brownian motion of a quantum oscillator},
  \href{https://doi.org/10.1063/1.1703727}{\emph{Journal of Mathematical
  Physics} {\bfseries 2} (1961) 407}.

\bibitem{kadanoff1962quantum}
L.~Kadanoff and G.~Baym, \emph{Quantum statistical mechanics: Green's function
  methods in equilibrium and nonequilibrium problems}, Frontiers in physics.
  W.A. Benjamin, 1962.

\bibitem{Keldysh:1964ud}
L.~V. Keldysh, \emph{{Diagram technique for nonequilibrium processes}},
  {\emph{Zh. Eksp. Teor. Fiz.} {\bfseries 47} (1964) 1515}.

\bibitem{Maldacena:2002vr}
J.~M. Maldacena, \emph{{Non-Gaussian features of primordial fluctuations in
  single field inflationary models}},
  \href{https://doi.org/10.1088/1126-6708/2003/05/013}{\emph{JHEP} {\bfseries
  05} (2003) 013} [\href{https://arxiv.org/abs/astro-ph/0210603}{{\ttfamily
  astro-ph/0210603}}].

\bibitem{Bernardeau:2003nx}
F.~Bernardeau, T.~Brunier and J.-P. Uzan, \emph{{High order correlation
  functions for self interacting scalar field in de Sitter space}},
  \href{https://doi.org/10.1103/PhysRevD.69.063520}{\emph{Phys. Rev.}
  {\bfseries D69} (2004) 063520}
  [\href{https://arxiv.org/abs/astro-ph/0311422}{{\ttfamily
  astro-ph/0311422}}].

\bibitem{Weinberg:2005vy}
S.~Weinberg, \emph{{Quantum contributions to cosmological correlations}},
  \href{https://doi.org/10.1103/PhysRevD.72.043514}{\emph{Phys. Rev.}
  {\bfseries D72} (2005) 043514}
  [\href{https://arxiv.org/abs/hep-th/0506236}{{\ttfamily hep-th/0506236}}].

\bibitem{Hartman:2006dy}
T.~Hartman and L.~Rastelli, \emph{{Double-trace deformations, mixed boundary
  conditions and functional determinants in AdS/CFT}},
  \href{https://doi.org/10.1088/1126-6708/2008/01/019}{\emph{JHEP} {\bfseries
  01} (2008) 019} [\href{https://arxiv.org/abs/hep-th/0602106}{{\ttfamily
  hep-th/0602106}}].

\bibitem{Giombi:2011ya}
S.~Giombi and X.~Yin, \emph{{On Higher Spin Gauge Theory and the Critical O(N)
  Model}}, \href{https://doi.org/10.1103/PhysRevD.85.086005}{\emph{Phys. Rev.}
  {\bfseries D85} (2012) 086005}
  [\href{https://arxiv.org/abs/1105.4011}{{\ttfamily 1105.4011}}].

\bibitem{Costa:2014kfa}
M.~S. Costa, V.~Gon\c{c}alves and J.~Penedones, \emph{{Spinning AdS
  Propagators}}, \href{https://doi.org/10.1007/JHEP09(2014)064}{\emph{JHEP}
  {\bfseries 09} (2014) 064} [\href{https://arxiv.org/abs/1404.5625}{{\ttfamily
  1404.5625}}].

\bibitem{Bekaert:2014cea}
X.~Bekaert, J.~Erdmenger, D.~Ponomarev and C.~Sleight, \emph{{Towards
  holographic higher-spin interactions: Four-point functions and higher-spin
  exchange}}, \href{https://doi.org/10.1007/JHEP03(2015)170}{\emph{JHEP}
  {\bfseries 03} (2015) 170} [\href{https://arxiv.org/abs/1412.0016}{{\ttfamily
  1412.0016}}].

\bibitem{Bekaert:2015tva}
X.~Bekaert, J.~Erdmenger, D.~Ponomarev and C.~Sleight, \emph{{Quartic AdS
  Interactions in Higher-Spin Gravity from Conformal Field Theory}},
  \href{https://doi.org/10.1007/JHEP11(2015)149}{\emph{JHEP} {\bfseries 11}
  (2015) 149} [\href{https://arxiv.org/abs/1508.04292}{{\ttfamily
  1508.04292}}].

\bibitem{Sleight:2016hyl}
C.~Sleight, \emph{{Interactions in Higher-Spin Gravity: a Holographic
  Perspective}}, \href{https://doi.org/10.1088/1751-8121/aa820c}{\emph{J.
  Phys.} {\bfseries A50} (2017) 383001}
  [\href{https://arxiv.org/abs/1610.01318}{{\ttfamily 1610.01318}}].

\bibitem{Chen:2017yia}
H.-Y. Chen, E.-J. Kuo and H.~Kyono, \emph{{Anatomy of Geodesic Witten
  Diagrams}}, \href{https://doi.org/10.1007/JHEP05(2017)070}{\emph{JHEP}
  {\bfseries 05} (2017) 070}
  [\href{https://arxiv.org/abs/1702.08818}{{\ttfamily 1702.08818}}].

\bibitem{Sleight:2017fpc}
C.~Sleight and M.~Taronna, \emph{{Spinning Witten Diagrams}},
  \href{https://doi.org/10.1007/JHEP06(2017)100}{\emph{JHEP} {\bfseries 06}
  (2017) 100} [\href{https://arxiv.org/abs/1702.08619}{{\ttfamily
  1702.08619}}].

\bibitem{Tamaoka:2017jce}
K.~Tamaoka, \emph{{Geodesic Witten diagrams with antisymmetric tensor
  exchange}}, \href{https://doi.org/10.1103/PhysRevD.96.086007}{\emph{Phys.
  Rev.} {\bfseries D96} (2017) 086007}
  [\href{https://arxiv.org/abs/1707.07934}{{\ttfamily 1707.07934}}].

\bibitem{Giombi:2017hpr}
S.~Giombi, C.~Sleight and M.~Taronna, \emph{{Spinning AdS Loop Diagrams: Two
  Point Functions}}, \href{https://doi.org/10.1007/JHEP06(2018)030}{\emph{JHEP}
  {\bfseries 06} (2018) 030}
  [\href{https://arxiv.org/abs/1708.08404}{{\ttfamily 1708.08404}}].

\bibitem{Yuan:2017vgp}
E.~Y. Yuan, \emph{{Loops in the Bulk}},
  \href{https://arxiv.org/abs/1710.01361}{{\ttfamily 1710.01361}}.

\bibitem{Giombi:2018vtc}
S.~Giombi, V.~Kirilin and E.~Perlmutter, \emph{{Double-Trace Deformations of
  Conformal Correlations}},
  \href{https://doi.org/10.1007/JHEP02(2018)175}{\emph{JHEP} {\bfseries 02}
  (2018) 175} [\href{https://arxiv.org/abs/1801.01477}{{\ttfamily
  1801.01477}}].

\bibitem{Yuan:2018qva}
E.~Y. Yuan, \emph{{Simplicity in AdS Perturbative Dynamics}},
  \href{https://arxiv.org/abs/1801.07283}{{\ttfamily 1801.07283}}.

\bibitem{Nishida:2018opl}
M.~Nishida and K.~Tamaoka, \emph{{Fermions in Geodesic Witten Diagrams}},
  \href{https://doi.org/10.1007/JHEP07(2018)149}{\emph{JHEP} {\bfseries 07}
  (2018) 149} [\href{https://arxiv.org/abs/1805.00217}{{\ttfamily
  1805.00217}}].

\bibitem{Costa:2018mcg}
M.~S. Costa and T.~Hansen, \emph{{AdS Weight Shifting Operators}},
  \href{https://doi.org/10.1007/JHEP09(2018)040}{\emph{JHEP} {\bfseries 09}
  (2018) 040} [\href{https://arxiv.org/abs/1805.01492}{{\ttfamily
  1805.01492}}].

\bibitem{Carmi:2018qzm}
D.~Carmi, L.~Di~Pietro and S.~Komatsu, \emph{{A Study of Quantum Field Theories
  in AdS at Finite Coupling}},
  \href{https://doi.org/10.1007/JHEP01(2019)200}{\emph{JHEP} {\bfseries 01}
  (2019) 200} [\href{https://arxiv.org/abs/1810.04185}{{\ttfamily
  1810.04185}}].

\bibitem{Jepsen:2019svc}
C.~B. Jepsen and S.~Parikh, \emph{{Propagator identities, holographic conformal
  blocks, and higher-point AdS diagrams}},
  \href{https://arxiv.org/abs/1906.08405}{{\ttfamily 1906.08405}}.

\bibitem{Polyakov:2007mm}
A.~M. Polyakov, \emph{{De Sitter space and eternity}},
  \href{https://doi.org/10.1016/j.nuclphysb.2008.01.002}{\emph{Nucl. Phys.}
  {\bfseries B797} (2008) 199}
  [\href{https://arxiv.org/abs/0709.2899}{{\ttfamily 0709.2899}}].

\bibitem{Leonhardt:2003qu}
T.~Leonhardt, R.~Manvelyan and W.~Ruhl, \emph{{The Group approach to AdS space
  propagators}},
  \href{https://doi.org/10.1016/j.nuclphysb.2003.07.007}{\emph{Nucl. Phys.}
  {\bfseries B667} (2003) 413}
  [\href{https://arxiv.org/abs/hep-th/0305235}{{\ttfamily hep-th/0305235}}].

\bibitem{Moschella:2007zza}
U.~Moschella and R.~Schaeffer, \emph{{Quantum theory on Lobatchevski spaces}},
  \href{https://doi.org/10.1088/0264-9381/24/14/003}{\emph{Class. Quant. Grav.}
  {\bfseries 24} (2007) 3571}
  [\href{https://arxiv.org/abs/0709.2795}{{\ttfamily 0709.2795}}].

\bibitem{Witten:1998qj}
E.~Witten, \emph{{Anti-de Sitter space and holography}},
  \href{https://doi.org/10.4310/ATMP.1998.v2.n2.a2}{\emph{Adv. Theor. Math.
  Phys.} {\bfseries 2} (1998) 253}
  [\href{https://arxiv.org/abs/hep-th/9802150}{{\ttfamily hep-th/9802150}}].

\bibitem{Penedones:2007ns}
J.~Penedones, \emph{{High Energy Scattering in the AdS/CFT Correspondence}},
  Ph.D. thesis, Porto U., 2007.
\newblock \href{https://arxiv.org/abs/0712.0802}{{\ttfamily 0712.0802}}.

\bibitem{Bros:1994dn}
J.~Bros, U.~Moschella and J.~P. Gazeau, \emph{{Quantum field theory in the de
  Sitter universe}},
  \href{https://doi.org/10.1103/PhysRevLett.73.1746}{\emph{Phys. Rev. Lett.}
  {\bfseries 73} (1994) 1746}.

\bibitem{Gubser:1998bc}
S.~S. Gubser, I.~R. Klebanov and A.~M. Polyakov, \emph{{Gauge theory
  correlators from noncritical string theory}},
  \href{https://doi.org/10.1016/S0370-2693(98)00377-3}{\emph{Phys. Lett.}
  {\bfseries B428} (1998) 105}
  [\href{https://arxiv.org/abs/hep-th/9802109}{{\ttfamily hep-th/9802109}}].

\bibitem{MellinBook}
R.~Paris and D.~Kaminski, \emph{Asymptotics and Mellin-Barnes integrals},
  no.~85 in Encyclopedia of Mathematics and its Applications. Cambridge
  University Press, United Kingdom, 2001.

\bibitem{watson1944treatise}
G.~Watson, \emph{A Treatise on the Theory of Bessel Functions}. Cambridge
  University Press, 1944.

\bibitem{Bzowski:2015yxv}
A.~Bzowski, P.~McFadden and K.~Skenderis, \emph{{Evaluation of conformal
  integrals}}, \href{https://doi.org/10.1007/JHEP02(2016)068}{\emph{JHEP}
  {\bfseries 02} (2016) 068}
  [\href{https://arxiv.org/abs/1511.02357}{{\ttfamily 1511.02357}}].

\bibitem{appell1880series}
P.~Appell, \emph{Sur les s{\'e}ries hyperg{\'e}ometriques de deux variables et
  sur des {\'e}quations diff{\'e}rentielles lin{\'e}aires aux d{\'e}riv{\'e}es
  partielles}, {\emph{Comptes Rendus} {\bfseries 90} (1880) 296}.

\bibitem{AppelletKampe}
P.~Appell and J.~Kamp\'e~de F\'eriet, \emph{Fonctions hyperge\'om\'etriques et
  hypersph\'eriques : Polynomes d'hermite}. Gauthier-Villars, 1926.

\bibitem{Davydychev:1992xr}
A.~I. Davydychev, \emph{{Recursive algorithm of evaluating vertex type Feynman
  integrals}}, {\emph{J. Phys.} {\bfseries A25} (1992) 5587}.

\bibitem{Antoniadis:2011ib}
I.~Antoniadis, P.~O. Mazur and E.~Mottola, \emph{{Conformal Invariance, Dark
  Energy, and CMB Non-Gaussianity}},
  \href{https://doi.org/10.1088/1475-7516/2012/09/024}{\emph{JCAP} {\bfseries
  1209} (2012) 024} [\href{https://arxiv.org/abs/1103.4164}{{\ttfamily
  1103.4164}}].

\bibitem{Creminelli:2011mw}
P.~Creminelli, \emph{{Conformal invariance of scalar perturbations in
  inflation}}, \href{https://doi.org/10.1103/PhysRevD.85.041302}{\emph{Phys.
  Rev.} {\bfseries D85} (2012) 041302}
  [\href{https://arxiv.org/abs/1108.0874}{{\ttfamily 1108.0874}}].

\bibitem{Boyanovsky:2011xn}
D.~Boyanovsky and R.~Holman, \emph{{On the Perturbative Stability of Quantum
  Field Theories in de Sitter Space}},
  \href{https://doi.org/10.1007/JHEP05(2011)047}{\emph{JHEP} {\bfseries 05}
  (2011) 047} [\href{https://arxiv.org/abs/1103.4648}{{\ttfamily 1103.4648}}].

\bibitem{Falk:1992sf}
T.~Falk, R.~Rangarajan and M.~Srednicki, \emph{{The Angular dependence of the
  three point correlation function of the cosmic microwave background radiation
  as predicted by inflationary cosmologies}},
  \href{https://doi.org/10.1086/186707}{\emph{Astrophys. J.} {\bfseries 403}
  (1993) L1} [\href{https://arxiv.org/abs/astro-ph/9208001}{{\ttfamily
  astro-ph/9208001}}].

\bibitem{Zaldarriaga:2003my}
M.~Zaldarriaga, \emph{{Non-Gaussianities in models with a varying inflaton
  decay rate}}, \href{https://doi.org/10.1103/PhysRevD.69.043508}{\emph{Phys.
  Rev.} {\bfseries D69} (2004) 043508}
  [\href{https://arxiv.org/abs/astro-ph/0306006}{{\ttfamily
  astro-ph/0306006}}].

\bibitem{Seery:2008qj}
D.~Seery, K.~A. Malik and D.~H. Lyth, \emph{{Non-gaussianity of inflationary
  field perturbations from the field equation}},
  \href{https://doi.org/10.1088/1475-7516/2008/03/014}{\emph{JCAP} {\bfseries
  0803} (2008) 014} [\href{https://arxiv.org/abs/0802.0588}{{\ttfamily
  0802.0588}}].

\bibitem{Chen:2006nt}
X.~Chen, M.-x. Huang, S.~Kachru and G.~Shiu, \emph{{Observational signatures
  and non-Gaussianities of general single field inflation}},
  \href{https://doi.org/10.1088/1475-7516/2007/01/002}{\emph{JCAP} {\bfseries
  0701} (2007) 002} [\href{https://arxiv.org/abs/hep-th/0605045}{{\ttfamily
  hep-th/0605045}}].

\bibitem{Holman:2007na}
R.~Holman and A.~J. Tolley, \emph{{Enhanced Non-Gaussianity from Excited
  Initial States}},
  \href{https://doi.org/10.1088/1475-7516/2008/05/001}{\emph{JCAP} {\bfseries
  0805} (2008) 001} [\href{https://arxiv.org/abs/0710.1302}{{\ttfamily
  0710.1302}}].

\bibitem{LopezNacir:2011kk}
D.~Lopez~Nacir, R.~A. Porto, L.~Senatore and M.~Zaldarriaga, \emph{{Dissipative
  effects in the Effective Field Theory of Inflation}},
  \href{https://doi.org/10.1007/JHEP01(2012)075}{\emph{JHEP} {\bfseries 01}
  (2012) 075} [\href{https://arxiv.org/abs/1109.4192}{{\ttfamily 1109.4192}}].

\bibitem{Flauger:2013hra}
R.~Flauger, D.~Green and R.~A. Porto, \emph{{On squeezed limits in single-field
  inflation. Part I}}, \href{https://doi.org/10.1088/1475-7516/2013/08/032,
  10.1088/1475-7516/2013/08/032/}{\emph{JCAP} {\bfseries 1308} (2013) 032}
  [\href{https://arxiv.org/abs/1303.1430}{{\ttfamily 1303.1430}}].

\bibitem{Aravind:2013lra}
A.~Aravind, D.~Lorshbough and S.~Paban, \emph{{Non-Gaussianity from Excited
  Initial Inflationary States}},
  \href{https://doi.org/10.1007/JHEP07(2013)076}{\emph{JHEP} {\bfseries 07}
  (2013) 076} [\href{https://arxiv.org/abs/1303.1440}{{\ttfamily 1303.1440}}].

\bibitem{Raju:2012zs}
S.~Raju, \emph{{Four Point Functions of the Stress Tensor and Conserved
  Currents in AdS$_4$/CFT$_3$}},
  \href{https://doi.org/10.1103/PhysRevD.85.126008}{\emph{Phys. Rev.}
  {\bfseries D85} (2012) 126008}
  [\href{https://arxiv.org/abs/1201.6452}{{\ttfamily 1201.6452}}].

\bibitem{meijer1941multiplikationstheoreme}
C.~S. Meijer, \emph{Multiplikationstheoreme f{\"u}r die Funktion Gmnpq (z)}.
  Noord-Hollandsche Uitgevers Maatschappij, 1941.

\bibitem{Sleight:2016dba}
C.~Sleight and M.~Taronna, \emph{{Higher Spin Interactions from Conformal Field
  Theory: The Complete Cubic Couplings}},
  \href{https://doi.org/10.1103/PhysRevLett.116.181602}{\emph{Phys. Rev. Lett.}
  {\bfseries 116} (2016) 181602}
  [\href{https://arxiv.org/abs/1603.00022}{{\ttfamily 1603.00022}}].

\bibitem{Sleight:2017krf}
C.~Sleight, \emph{{Lectures on Higher Spin Holography}}, {\emph{PoS} {\bfseries
  Modave2016} (2017) 003} [\href{https://arxiv.org/abs/1701.08360}{{\ttfamily
  1701.08360}}].

\bibitem{Castro:2017hpx}
A.~Castro, E.~Llabr\'es and F.~Rejon-Barrera, \emph{{Geodesic Diagrams,
  Gravitational Interactions \& OPE Structures}},
  \href{https://doi.org/10.1007/JHEP06(2017)099}{\emph{JHEP} {\bfseries 06}
  (2017) 099} [\href{https://arxiv.org/abs/1702.06128}{{\ttfamily
  1702.06128}}].

\bibitem{Chu:2018kec}
S.~K. Chu, Y.~Wang and S.~Zhou, \emph{{Operator Method and Recursion Relations
  for Inflationary Correlator}},
  \href{https://arxiv.org/abs/1812.00322}{{\ttfamily 1812.00322}}.

\bibitem{Noumi:2012vr}
T.~Noumi, M.~Yamaguchi and D.~Yokoyama, \emph{{Effective field theory approach
  to quasi-single field inflation and effects of heavy fields}},
  \href{https://doi.org/10.1007/JHEP06(2013)051}{\emph{JHEP} {\bfseries 06}
  (2013) 051} [\href{https://arxiv.org/abs/1211.1624}{{\ttfamily 1211.1624}}].

\bibitem{barnes1}
E.~W. Barnes, \emph{A new development of the theory of the hypergeometric
  functions}, \href{https://doi.org/10.1112/plms/s2-6.1.141}{\emph{Proceedings
  of the London Mathematical Society} {\bfseries s2-6} (1908) 141}.

\bibitem{barnes2}
E.~W. {Barnes}, \emph{{A transformation of generalised hypergeometric
  series.}}, {\emph{{Quart. J.}} {\bfseries 41} (1910) 136}.

\bibitem{Baileybook}
W.~N. Bailey, \emph{{Generalized Hypergeometric Series}}. Cambridge University
  Press, 1935.

\bibitem{gelfand1986general}
I.~Gelfand, \emph{General theory of hypergeometric functions},  in \emph{Soviet
  Math. Dokl.}, vol.~33, pp.~573--577, 1986.

\end{thebibliography}\endgroup

\end{document}